\documentclass[12pt,preprint]{aastex}

\usepackage{graphicx}
\usepackage{natbib}
\shorttitle{SMA Observations of Protobinary Systems}%
\shortauthors{X.~Chen et al.}%

\begin{document}

\title{SMA Observations of Class\,0 Protostars: A High-Angular Resolution Survey of Protostellar Binary Systems}

\author{Xuepeng~Chen\altaffilmark{1,2}, H\'{e}ctor~G.~Arce\altaffilmark{2}, Qizhou~Zhang\altaffilmark{3}, Tyler~L.~Bourke\altaffilmark{3},
Ralf~Launhardt\altaffilmark{4}, Jes~K.~J{\o}rgensen\altaffilmark{5}, Chin-Fei Lee\altaffilmark{6}, Jonathan~B.~Foster\altaffilmark{7},
Michael~M.~Dunham\altaffilmark{2}, Jaime~E.~Pineda\altaffilmark{8,9}, and Thomas~Henning\altaffilmark{4}}

\affil{$^1$Purple Mountain Observatory, Chinese Academy of Sciences, 2 West Beijing Road, Nanjing 210008, China; xpchen@pmo.ac.cn}%
\affil{$^2$Department of Astronomy, Yale University, Box 208101, New Haven, CT 06520-8101, USA; xuepeng.chen@yale.edu}%
\affil{$^3$Harvard-Smithsonian Center for Astrophysics, 60 Garden Street., Cambridge, MA 02138, USA}%
\affil{$^4$Max Planck Institute for Astronomy, K\"{o}nigstuhl 17, D-69117 Heidelberg, Germany}%
\affil{$^5$Niels Bohr Institute and Centre for Star and Planet Formation, Copenhagen University, Juliane Maries Vej 30, DK-2100 Copenhagen {\O}, Denmark}%
\affil{$^6$Academia Sinica Institute of Astronomy and Astrophysics, P.O. Box 23-141, Taipei 106, Taiwan}%
\affil{$^7$Institute for Astrophysical Research, Boston University, Boston, MA 02215, USA}%
\affil{$^8$ESO, Karl Schwarzschild Str. 2, 85748 Garching bei Munchen, Germany}%
\affil{$^9$Jodrell Bank Centre for Astrophysics, School of Physics and Astronomy, University of Manchester, Manchester, M13 9PL, UK}%

\begin{abstract}

We present high angular resolution 1.3\,mm and 850\,$\mu$m dust continuum data obtained with the Submillimeter 
Array toward 33 Class\,0 protostars in nearby clouds (distance $<$\,500\,pc), which represents so far the largest 
survey toward protostellar binary/multiple systems. The median angular resolution in the survey is 2\farcs5, while the 
median linear resolution is approximately 600\,AU. Compact dust continuum emission is observed from all sources 
in the sample. Twenty-one sources in the sample show signatures of binarity/multiplicity, with separations ranging 
from 50\,AU to 5000\,AU. The numbers of singles, binaries, triples, and quadruples in the sample are 12, 14, 5, and 
2, respectively. The derived multiplicity frequency (MF) and companion star fraction (CSF) for Class\,0 protostars are 
0.64\,$\pm$\,0.08 and 0.91\,$\pm$\,0.05, respectively, with no correction for completeness. The derived MF and CSF 
in this survey are approximately two times higher than the values found in the binary surveys toward Class\,I young 
stellar objects, and approximately three (for MF) and four (for CSF) times larger than the values found among main 
sequence stars, with a similar range of separations. Furthermore, the observed fraction of high order multiple systems 
to binary systems in Class\,0 protostars (0.50\,$\pm$\,0.09) is also larger than the fractions found in Class\,I young 
stellar objects (0.31\,$\pm$\,0.07) and main sequence stars ($\leq$\,0.2). These results suggest that binary properties 
evolve as protostars evolve, as predicted by numerical simulations. The distribution of separations for Class\,0 protostellar 
binary/multiple systems shows a general trend in which companion star fraction increases with decreasing companion 
separation. We find that 67\%\,$\pm$\,8\% of the protobinary systems have circumstellar mass ratios below 0.5, implying 
that unequal-mass systems are preferred in the process of binary star formation. We suggest an empirical sequential 
fragmentation picture for binary star formation, based on this work and existing lower resolution single-dish observations.

\end{abstract}


\clearpage

\section{INTRODUCTION}

Over the past two decades, our knowledge of the formation and evolution of low-mass stars 
has made significant progress (see, e.g., Reipurth et al. 2007 for several reviews). It is widely 
accepted that low-mass stars form from the gravitational collapse of molecular cloud cores 
(see, e.g., Shu et al. 1987; McKee \& Ostriker 2007). Initially, these cores, generally referred 
to as prestellar cores, are cold dense condensations with infall motions that have not yet formed 
central stellar objects (Andr\'{e} et al. 2000; 2009). Resulting from the collapse of prestellar 
cores, Class\,0 objects are the youngest accreting protostars observed right after point mass 
formation, when most of the mass of the system is still in the surrounding dense core/envelope 
(Andr\'{e} et al. 2000). As the protostellar envelope dissipates through accretion and ejection 
of circumstellar material, Class\,0 protostars evolve into Class\,I young stellar objects (YSOs; 
Lada 1987). Class\,I YSOs subsequently evolve into pre-main sequence (PMS) stars surrounded 
by circumstellar disks, but without dense envelopes. Eventually, PMS stars evolve into main 
sequence (MS) stars.

One of the major puzzles in our understanding of star formation concerns the origin of binary 
stars (Tohline 2002). A series of surveys toward solar-type MS stars have found that about 
40\%--60\% of MS stars are actually binary systems, with periods from less than one day to over 
10 million years (e.g., Duquennoy \& Mayor 1991; Fischer \& Marcy 1992). The binary frequency 
found among PMS stars is even higher than that among the MS stars (e.g., Mathieu 1994). For 
example, in the survey toward PMS stars in the Taurus star formation region, K\"{o}hler \& Leinert 
(1998) found that for systems with separations from 18\,AU to 1800\,AU the binary frequency is 
48.9\%\,$\pm$\,5.3\%, which is larger than a factor of two compared to the binary frequency in 
solar-type MS stars with the same separation range. Recent surveys toward Class\,I YSOs 
show that binary systems are also common in the Class\,I phase (see, e.g., Haisch et al. 2004; 
Duch\^{e}ne et al. 2004, 2007; Connelley et al. 2008a, b). For example, the large survey by 
Connelley et al. (2008a, b) found that for Class\,I YSOs the binary frequency for separations from 
100\,AU to 4500\,AU is about 43\%. The high frequency of binary systems found in different stellar 
evolution stages has led to a widely held opinion that most stars form in binary/multiple systems 
(e.g., Mathieu 1994; Goodwin et al. 2007). In contrast, however, Lada (2006) argued that the 
most common outcome of the star formation process is a single star rather than a binary star, 
based on two observational facts: (1) that most low-mass stars (stellar mass $<$\,0.5\,$M_\odot$) 
are single objects (e.g., Delfosse et al. 2004; Siegler et al. 2005), and (2) that these low-mass 
stars represent the majority of the field stars (Kroupa 2002; Chabrier 2005).

Do most stars form as singles or binaries? Before trying to answer this question still in debate, 
it must be noted that multiple stellar systems undergo dynamical evolution at the beginning of 
their formation phase (e.g., Reipurth \& Clarke 2001; Reipurth et al. 2010). Dynamical evolution 
will modify the binary population by reducing the overall binary fraction and by hardening the 
remaining binaries on a timescale of less than 10$^5$ yr (see Goodwin et al. 2007). Therefore, 
neither observations of PMS/Class\,I young stars nor observations of field dwarfs can constrain 
the {\it original} properties of binary/multiple systems. Clearly, the most effective way to answer 
this question is to extensively observe the earliest stages of the star formation process.

Class\,0 objects, which represent the youngest protostars at an age of a few 10$^4$$-$10$^5$~yr 
(Andr\'{e}  et al. 2000; Evans et al. 2009), provide an important opportunity to probe the earliest 
phase of the star formation process, where most of the initial information of (binary) star formation 
is expected to still be preserved. To understand the formation of binary stars, it is therefore critical 
to observe large samples of Class\,0 protostars in nearby molecular clouds to answer a number of 
key questions, e.g., how common is binarity/multiplicity in the Class\,0 protostellar phase? What are 
the initial distributions of separations and mass ratios? Unfortunately, direct observations of the 
Class\,0 protostellar stages, when the main collapse has started but no optical or near-infrared 
emission emerges from the protostar through the opaque infalling envelope, were long hampered 
by the low angular resolution of single-dish (sub-)millimeter telescopes. Only the high angular 
resolution attained with current millimeter interferometers enables us to directly observe the formation 
phase of binary stars. In the past decade, a handful of Class\,0 protostellar binary (protobinary) 
systems have been found using different interferometric arrays (e.g., Looney et al. 2000; Reipurth 
et al. 2002; Launhardt 2004), and there are increasing interferometric studies of binarity in the 
protostellar phase (e.g., Volgenau et al. 2006; Maury et al. 2010). However, these studies were 
limited to small samples of objects and no statistically strong conclusions could be drawn from them.

On the theoretical side, numerous simulations support the hypothesis that the fragmentation of 
molecular cloud cores is the main mechanism for the formation of binary/multiple stellar systems 
(see reviews by Bodenheimer et al. 2000, Tohline 2002, and Goodwin et al. 2007). Yet, many key 
questions concerning this fragmentation process (e.g., exactly {\it when}, {\it where}, {\it why}, and 
{\it how}) are still under debate. 
In general, fragmentation models can be classified depending on when the fragmentation takes 
place during the different collapse phases of the core\footnote{See, e.g., Figure~2 in Tohline 2002 
or Figure~9 in Andr\'{e} et al. 2009 for a discussion of the different evolutionary phases of core 
collapse.}. Initial clump/filament fragmentation takes place prior to the collapse of individual cores, and 
results in cores with separations of $\sim$\,10$^3$$-$10$^4$\,AU (see, e.g., Kauffmann et al. 2008; 
Launhardt et al. 2010), while prompt (isothermal) fragmentation takes place at the end of the core's 
isothermal collapse phase, and results in fragments with separations of $\sim$\,10$^2$$-$10$^3$\,AU 
(see Tohline 2002). On the other hand, adiabatic fragmentation occurs when the collapse evolves into
the adiabatic phase, and results in fragments with separations of $\sim$\,3$-$300\,AU (Machida et al. 
2005; 2008). Finally, secondary fragmentation takes place during the core's secondary collapse phase, 
and results in fragments with separations of 0.01 to 0.1\,AU (Machida et al. 2008). In all these cases, 
fragmentation takes place when there is some angular momentum in the core (see the review by 
Goodwin et al. 2007). The source of the angular momentum may be ordered rotational motions (i.e., 
rotational fragmentation) or turbulence (i.e., turbulent fragmentation).

These different possible models of fragmentation result in different binary/multiple systems with a wide range 
of properties. These properties are largely determined by the accretion process which, in turn, strongly depends 
on the initial conditions of the cloud cores, e.g., the initial distributions of mass and angular momentum (see 
Bate \& Bonnell 1997). It must be noted that the constraints used on current binary formation models mostly come 
from the observations of MS and PMS stars. These observations cannot provide information on the initial conditions 
of fragmenting cores. Hence, it is critical to observe the early phase of core fragmentation and study in detail their 
properties (e.g., mass and separation distributions), in order to put direct constraints on different fragmentation 
models.

To achieve a comprehensive understanding of binary star formation, we have started a systematic program 
to observe at different (sub-)\,millimeter interferometers a large sample of low-mass prestellar and protostellar 
cores. The early survey was conducted at the Owens Valley Radio Observatory (OVRO) millimeter array (Chen et 
al. 2007; R. Launhardt et al. in preparation), and was then continued with the Australia Telescope Compact Array
(ATCA; Chen et al. 2008a), the IRAM Plateau de Bure Interferometer (PdBI; Chen et al. 2009; X.~Chen et al. in 
preparation), and the Submillimeter Array (SMA; Chen et al. 2008b; Chen \& Arce 2010). As part of this program,
we present in this paper high angular resolution (sub-)\,millimeter continuum observations toward 33 Class\,0 
protostars using the SMA, which represents so far the largest survey of protobinary systems. 
High angular resolution (sub-)\,millimeter continuum emission traces the circumstellar dust from the inner envelopes 
and disks surrounding protostars, which we used to resolve binary/multiple systems, and measure their masses, 
sizes, and spatial distribution. 
In Section~2 we describe the target list, observations, and data reduction. Observational results are presented 
in Section~3 and discussed in Section~4. The main conclusions of this study are summarized in Section~5. We 
give a detailed description of the individual sources in Appendix~A. 

\section{SAMPLE, OBSERVATIONS, AND DATA REDUCTION}

\subsection{Sample}

The sources in this study were collected from the SMA science archive. A total of 33 sources were selected 
by using the following three criteria: (1) available in the SMA archive\footnote{All the SMA data used in this 
work are publicly available from the SMA archive. Some of the data are unpublished and in Tables~2-3 
we list the name of the principle investigator (PI) responsible for the observations. For data that had been 
published we list the paper that first presented the data in Tables 2-3.}; (2) confirmed classification 
as a Class\,0 object by other studies; and (3) known distance less than 500\,pc. 
These sources were observed at the SMA between 2004 and 2009 with different science goals (e.g., for studying 
outflow, kinematics, or chemistry in the Class\,0 protostars). Only for a few sources (i.e., IRAS\,03282, NGC\,1333 
SVS\,13, and CG\,30), the proposed science goal was to study the formation of binary stars.
The target list and their basic properties (i.e., bolometric luminosity $L_{\rm bol}$, bolometric temperature 
$T_{\rm bol}$, and envelope mass $M_{\rm env}$) are summarized in Table\,1.
Most sources in Table~1 were cataloged by both Andr\'{e} et al. (2000) and Froebrich  (2005). We refer readers 
to these two papers and references therein for more details about the basic properties of these sources. Note 
that several sources may represent transition objects between the Class\,0 and Class\,I phases (see Froebrich 
2005 for more detailed discussion). 

Distances to the different sources were obtained from the literature. Early estimates of the distance 
to Perseus indicated that this region is 350\,pc away (e.g., Herbig \& Jones 1983). However, more recent studies 
indicate that this region is significantly closer. Photometric distances towards the western part of Perseus place 
this cloud at 220\,pc ({\v C}ernis 1990; {\v C}ernis \& Strai{\v z}ys 2003), consistent with recent VLBI observations 
that give a distance to NGC\,1333 of 240\,$\pm$\,20\,pc (Hirota et al. 2008, 2011). We therefore adopt a distance 
of 240\,pc for all objects in Perseus, in agreement with other recent studies of this cloud (e.g., Enoch et al. 2006; 
Evans et al. 2009; Arce et al. 2010, 2011). The basic properties (bolometric luminosity and envelope mass) of the 
Perseus sources listed in Andr\'{e} et al. (2000) and Froebrich (2005) were modified using the new distance of 
240\,pc (see Table~1). For sources in the Taurus star forming region, we adopt a distance of 140\,pc (Elias 1978; 
Kenyon et al. 1994; Loinard et al. 2007b). For sources in the Orion region, we adopt a distance of 440\,pc, obtained 
by the VLBI observations of Hirota et al. (2007). For sources in the Ophuichus cloud, we use a distance of 120\,pc, 
derived from recent VLBA observations by Loinard et al. (2008). For the Lupus region, the distance is estimated to 
be 130\,pc (Murphy et al. 1986). For isolated objects, the distance references are as follows: CB\,17 (Launhardt et 
al. 2010), CG\,30 (Knude et al. 1999), L483 (Dame \& Thaddeus 1985), L723 (Goldsmith et al. 1984), B335 (Tomita 
et al. 1979), L1157 (Strai{\v z}ys et al. 1992), and L1251B (Kun \& Prusti 1993).

\begin{deluxetable}{lcccccccc}
\tabletypesize{\scriptsize}\tablecaption{\footnotesize Basic
properties of target sources\label{tbl-1}} \tablewidth{0pt}
\tablehead{\colhead{Source}&\colhead{Other (or short)}& \colhead{Association}
&\colhead{Distance}&\colhead{$L_{\rm bol}$}&\colhead{$T_{\rm bol}$}
& \colhead{$M_{\rm env}$} &\colhead{Class}&\colhead{References$^b$}\\
\colhead{Name}&\colhead{Name$^a$}&\colhead{Region}&\colhead{[pc]}&\colhead{[L$_{\odot}$]}&
\colhead{[K]}&\colhead{[M$_{\odot}$]}&\colhead{}&\colhead{}}
\startdata

L1448\,IRS2                 &  IRAS\,03222+3034  &  Perseus  & 240  & 1.7-4.1   & 43-63  & 0.6-1.4  & 0   &  1, 2, 3, 4 \\

L1448\,N                      &  IRAS\,03225+3034  &  Perseus  & 240  & 4.3-17    & 53-90  & 1.5-4.5  & 0/I   &  1, 2, 3, 4 \\

L1448\,C                      &  L1448-mm          &  Perseus  & 240  & 4.4-6.3   & 49-69 & 0.7-1.9 & 0   &  1, 2, 3, 4 \\

PER\,065                     &  Per-Bolo\,30      &  Perseus  & 240  & 0.03-0.22 & 31-37 & 0.4-2.1  & 0   & 3, 4  \\

NGC\,1333~IRAS\,2A  &  IRAS\,2A  &  Perseus  & 240 &  19-23 & 50-69 &  1.0-2.8 & 0   & 1, 2, 3, 4  \\

NGC\,1333~IRAS\,2B  &  IRAS\,2B  &  Perseus  & 240 & 19-23  & 50-69 & 1.0-2.8 & 0    & 1, 2, 3, 4   \\

NGC\,1333~SVS\,13   &  SVS\,13 &  Perseus  & 240 & 7-33  & 30-180 & 1.3-3.3  & 0/I   & 1, 2, 3, 4  \\

NGC\,1333~IRAS\,4A  &  IRAS\,4A &  Perseus  & 240 & 4.2-8.5  & 34-51  &  2.7-7.8  & 0   & 1, 2, 3, 4  \\

NGC\,1333~IRAS\,4B  &  IRAS\,4B  &  Perseus  & 240 & 1.6-7.6  & 36-54  &  1.4-3.7  & 0   & 1, 2, 3, 4 \\

IRAS\,03282+3035     &  IRAS\,03282  &  Perseus & 240 & 0.7-1.2  & 33-60 & 0.4-1.0  & 0   & 1, 2, 3, 4  \\

Per~B1-c                    &  B1-c  &  Perseus & 240 & 1.8-3.7  & 53-76 & 2.1-18  & 0/I   &  3, 4, 5, 6 \\

Per~B1-b                   &   B1-b  &  Perseus & 240 &  1.2-2.5  & $<$\,25 &  2.8-26  & 0   &  3, 5, 7 \\

HH\,211-MM              &    &  Perseus & 240 & 1.5-4.5  & 24-31 & 0.5-3.0 & 0   & 1, 2, 3, 4  \\

IC\,348~MMS            &    &  Perseus  & 240 &  0.5-1.9 & 49-59 & 1.8-10  & 0   & 3, 4  \\

CB\,17 IRS                & L1389 & Isolated & 250 & 0.6 & $>$\,55 & $<$\,4 & 0/I & 8 \\

IRAM\,04191+1522   &  IRAM\,04191  &  Taurus &  140 & 0.12-0.28  & 18-36  & 0.5  & 0   & 1, 2, 9, 10 \\

L1521F-IRS              &   L1521F/MC\,27  &  Taurus & 140 & 0.1-0.36 & 20 &  0.7-4.8 & 0   & 11, 12, 13 \\

L1527 IRS                &  IRAS\,04368+2557   &  Taurus & 140 & 1.9-2.2  &  36-60  & 0.4-0.9 & 0/I   & 1, 2, 14  \\

HH\,114~MMS         &    &  Orion & 440 & $<$\,25  & 40-84  & 1.4-2.8  & 0  & 1, 2  \\

OMC\,3~MM\,6       &    &  Orion &  440 & $<$\,60 & 30-72 & 6-12 & 0   &  1, 2  \\

NGC\,2024-FIR\,5  &    &  Orion & 440  & $\geq$\,10  & 20 & 8.2-15  & 0   & 1, 2  \\

HH\,212-MM          &    &  Orion &  440 &  7.7-14  & $<$\,56 & 0.3-1.2  & 0  &  1, 2   \\

HH\,25~MMS          &    &  Orion & 440  &  6.0-7.2 &   34-41 & 0.5-1.2  & 0   & 1, 2  \\

CG\,30                   &  BHR\,12  &  Isolated & 200 & 10-17  & 74-117 &  1.1-2.3  & 0/I   & 1, 2  \\

B228                      &  IRAS\,15398--3359  &  Lupus & 130  & 0.9-1.2  & 48-61 & 0.3-0.4  & 0   & 2, 14 \\

VLA\,1623             &    &  Ophuichus & 120 &  0.4-1.2  & $<$\,59 &  0.2-0.5  & 0   & 1, 2, 4  \\

Oph~MMS\,126    & IRAS\,16253--2429   &  Ophuichus &  120 &  0.25 & 35  &  0.51   & 0   &  4 \\

IRAS\,16293--2422 &  IRAS\,16293  &  Ophuichus & 120  & 6.9-14 &  41-54 & 1.2-2.8  & 0   & 1, 2, 4 \\

L483 &  IRAS\,18148--0440    &  Isolated & 200  & 9-13  & 50-54 &  0.3-1.8 & 0/I   &  1, 2, 14\\

L723\,VLA2   & IRAS\,19156+1906     &  Isolated & 300 & 3-3.3 & 47-50 & 0.6-1.6  & 0   & 1, 14  \\

B335  & IRAS\,19345+0727     &  Isolated & 250 &  3.0-3.1 & 28-45  & 0.7-1.2 & 0   &  1, 2, 14 \\

L1157  &   IRAS\,20386+6751  &  Isolated & 325 &  5.8-11  & 42-60 & 0.5-2.6  & 0   &  1, 2, 14  \\

L1251\,B~IRS1 &  IRAS\,22376+7455  &  Isolated & 300 &  $\leq$\,10  & 83-87 & 2.0-2.1  & 0/I   & 2, 15, 16 \\

\enddata
\tablenotetext{a}{The IRAS source is not always associated with the mm source.}
\tablenotetext{b}{References for the basic properties of the objects. --- (1) Andr\'{e} et al. (2000); (2) Froberich (2005); (3) Hatchell et al. (2007); 
(4) Enoch et al. (2009); (5) Matthews \& Wilson (2002); (6) Matthews et al. (2006); (7) Hirano et al. (1999); (8) Launhardt et al. (2010); (9) Andr{\'e} 
et al. (1999); (10) Dunham et al. (2006); (11) Crapsi et al. (2004); (12) Bourke et al. (2006); (13) Terebey et al. (2009); (14) Shirley et al. (2000); 
(15) J.~Lee et al. (2006); (16) J.~Lee et al. (2007).}
\end{deluxetable}


\begin{deluxetable}{lccccccc}
\rotate
\tabletypesize{\scriptsize} \tablecaption{SMA observation log: 1.3\,mm dust continuum\label{tbl-2}} \tablewidth{0pt}
\tablehead{\colhead{Source}  &\colhead{R.A. \&
Dec. (J2000)$^{a}$} &\colhead{Observing Dates} &\colhead{Array} &\colhead{UV coverage}&\colhead{HPBW$^{b}$} &\colhead{rms$^{c}$}&\colhead{PI or Referred Paper}\\
\colhead{Name} &\colhead{[h\,:\,m\,:\,s,
$^{\circ}:\,':\,''$]} &\colhead{[day/mon/yr]} &\colhead{configurations}
&\colhead{[k$\lambda$]}&\colhead{[arcsecs]}
&\colhead{[mJy/beam]}&\colhead{}}\startdata

L1448\,IRS2                        & 03:25:22.40, +30:45:12.0  & 04/11/07, 06/11/07 &  Com      & 7--53 & 3.4$\times$2.5 & 0.68/2.1 & J.~Foster \\

L1448\,N                             & 03:25:36.00, +30:45:20.0  & 21/11/07                 &  Com      & 8--59 & 2.8$\times$2.4 & 0.74/7.0 & J{\o}rgensen et al. (2007)\\

L1448\,C                             & 03:25:38.80, +30:44:05.0  & 07/11/04                 &  Com-N  & 8--107 & 2.7$\times$1.9 & 1.1/4.0 & J{\o}rgensen et al. (2007)\\

PER\,065                            & 03:28:39.10, +31:06:01.8  & 20/01/09, 28/01/09  & Com      & 6--53 & 3.7$\times$2.7 & 0.45/5.5 & T.-H.~Hsieh \\

NGC\,1333~IRAS2\,(A,\,B)  & 03:28:55.70, +31:14:37.0 & 07/11/04                  &  Com-N  & 8--107 & 2.8$\times$1.9 & 1.4/1.8 & J{\o}rgensen et al. (2007)\\

NGC\,1333~SVS\,13           & 03:29:03.07, +31:15:52.0 & 07/12/08                  &  Com      & 9--59 & 2.8$\times$2.6 & 0.40/2.5 & X.~Chen \\

NGC\,1333~IRAS4\,(A,\,B)  & 03:29:10.50, +31:13:31.0 & 22/11/04, 17/01/06  &  Com-N, Com  & 9--140 & 1.5$\times$1.4 & 1.4/12 & J{\o}rgensen et al. (2007)\\

IRAS\,03282                       & 03:31:21.00, +30:45:30.0  & 08/12/08 &  Com & 6--59 & 2.5$\times$2.2 & 0.37/1.8 & X.~Chen \\

Per-B1c                              & 03:33:17.88, +31:09:32.0  & 17/10/05 &  Com  & 6--53 & 3.4$\times$2.9 & 1.6/2.7 & B.~Matthews \\

Per-B1b                              & 03:33:21.14, +31:07:35.3 & 10/09/07,11/09/07 &  Subcom, Com & 5--52 & 7.6$\times$4.5 & 0.91/4.2 & N.~Hirano \\

IC\,348 MMS                      & 03:43:57.30, +32:03:09.0 & 20/11/05                &  Com & 5-53  & 3.3$\times$3.0 & 3.3/4.0 &  M.~Reid\\

CB\,17 IRS                         & 04:04:35.85, +56:56:03.1 & 30/11/08                 & Com & 6--53 & 3.1$\times$2.8 & 0.44/0.5 & X.~Chen\\

IRAM\,04191                      & 04:21:56.90, +15:29:46.4 & 02/11/07, 03/11/07 &  Com & 9-54  & 3.7$\times$3.1 & 0.60/0.7 & J.~Karr \\

L1521F-IRS                       & 04:28:38.95, +26:51:35.1 & 03/01/07 &  Com & 9--53 & 3.4$\times$2.4 & 0.98/1.5 & N.~Ohashi\\

L1527                                & 04:39:53.90, +26:03:10.0 & 08/11/04 &  Com-N & 11--108 & 2.7$\times$1.8 & 0.91/1.4 & J{\o}rgensen et al. (2007)\\

HH\,114 MMS                    & 05:18:15.20, +07:12:02.0 & 28/11/05, 01/12/05 &  Com & 8--52 & 3.4$\times$3.1 & 1.3/5.0 & H.~Arce\\

HH\,212 MM                      & 05:43:51.30, $-$01:02:53.0 & 30/10/06, 25/11/06 & Ext, V-Ext & 22-390  &  0.8$\times$0.7 & 0.63/1.1 & C.-F.~Lee\\

CG\,30                              & 08:09:33.00, $-$36:05:08.0 & 07/03/08 & Com & 6--53 & 5.0$\times$2.0 & 1.3/4.0 & X.~Chen\\

B\,228                               & 15:43:02.16, $-$34:09:09.0 &  29/04/09 &  Com & 5--87  & 3.9$\times$2.4 & 1.4/1.5 & A.~Hedden \\

B\,228 (1.1\,mm)               & 15:43:02.16, $-$34:09:09.0 & 12/05/09 &  Com & 8--109  & 2.5$\times$1.9 & 1.4/1.6 & A.~Hedden \\

VLA\,1623                         & 16:26:26.36, $-$24:24:30.3 & 25/05/07, 11/06/07 & V-Ext &  22--390  & 0.6$\times$0.3 & 0.38/1.0 & N.~Ohashi\\

Oph MMS\,126                 & 16:28:21.60, $-$24:36:23.4 & 02/05/08 & Com & 7--93  & 4.8$\times$2.2 & 2.0/2.1 & K.~Stapelfeldt \\

IRAS\,16293                     & 16:32:22.91, $-$24:28:35.5 & 22/06/04, 18/02/05 &  Com-N, Com & 7--146 & 1.5$\times$1.2 & 1.2/19 & T.~Bourke \\

L723\,VLA2                      & 19:17:53.40, +19:12:19.4   & 07/08/04, 22/10/04 & Com & 11--106 & 2.9$\times$2.1 & 0.81/1.2 & Girart et al. (2009)\\

B335                                & 19:37:00.89, +07:34:10.0   & 24/06/05 &  Com & 6--54  & 3.4$\times$3.1 & 0.98/2.9 & J{\o}rgensen et al. (2007)\\

L1157                              & 20:39:06.19, +68:02:15.9   & 06/07/05 &  Com & 5--52  & 4.3$\times$2.7 & 1.2/4.8 & J{\o}rgensen et al. (2007) \\

L1251\,B                         & 22:38:47.20, +75:11:28.7    & 25/09/05 &  Com &  5--42  & 4.1$\times$3.1 & 3.0/3.2 & J.~Lee et al. (2007)\\

\enddata
\tablenotetext{a}{Phase center of the observations (if more than two pointings in the observations, 
the final figure center of the mosaic image is listed here).} 
\tablenotetext{b}{Synthesized FWHM beam size with robust weighting (weighting parameter 1.0).}
\tablenotetext{c}{1\,$\sigma$ theoretical noise (robust weighting 1.0)\,/\,1\,$\sigma$ rms noise in the cleaned map.}
\end{deluxetable}

\begin{deluxetable}{lccccccc}
\rotate
\tabletypesize{\scriptsize} \tablecaption{SMA observation log: 850\,$\mu$m dust continuum\label{tbl-3}} \tablewidth{0pt}
\tablehead{\colhead{Source}  &\colhead{R.A. \&
Dec. (J2000)$^{a}$} &\colhead{Observing Dates} &\colhead{Array} &\colhead{UV coverage}&\colhead{HPBW$^{b}$} &\colhead{rms$^{c}$}&\colhead{PI or Referred Paper}\\
\colhead{Name} &\colhead{[h\,:\,m\,:\,s,
$^{\circ}:\,':\,''$]} &\colhead{[day/mon/yr]} &\colhead{configurations} &\colhead{[k$\lambda$]}&\colhead{[arcsecs]}
&\colhead{[mJy/beam]}&\colhead{}}\startdata

L1448\,C                    & 03:25:38.55, +30:44:13.4  & 05/12/06, 25/12/06 & Com, Ext & 10--265 & 0.97$\times$0.65 & 1.6/5.5 & Hirano et al. (2010)\\

NGC\,1333~IRAS2A  & 03:28:55.57, +31:14:37.2  & 17/10/04, 08/01/06 & Com, Ext  & 17--265 & 0.75$\times$0.61 & 1.7/4.5 & J{\o}rgensen et al. (2007)\\

NGC\,1333~IRAS4A  & 03:29:10.50, +31:13:31.0 & 11/11/04, 20/11/04  &  Com-N, Com & 14--143 & 1.8$\times$0.86 & 2.8/24 & J{\o}rgensen et al. (2007) \\

IRAS\,03282              & 03:31:21.00, +30:45:30.0 & 24/08/09, 25/10/09  & Com, Ext & 17-265 & 0.86$\times$0.76 & 1.1/4.0 & X.~Chen \\

HH\,211-MM              & 03:43:56.80, +32:00:50.3 & 23/01/08, 18/08/08 &  Ext, V-Ext & 60--600 & 0.20$\times$0.15 & 2/1.4 & Lee et al. (2009, 2010) \\

HH\,114 MMS            & 05:18:15.20, +07:12:02.0 & 31/12/05 & Com & 10--78 & 2.2$\times$1.9 & 3.1/16 & H.~Arce\\

OMC3 MMS6             & 05:35:23.49, $-$05:01:32.2 & 15/11/07, 15/12/07, 15/02/09 &  Com, Ext & 15--210 & 1.4$\times$1.1 & 1.5/20 & Y.-W.~Tang\\

NGC\,2024-FIR5       & 05:41:44.25, $-$01:55:40.8  & 24/11/07, 19/12/07 & Com & 15--88  & 2.2$\times$1.3 & 2.8/14 & Alves et al. (2011)\\

HH\,212 MM              & 05:43:51.41, $-$01:02:53.1 & 02/12/05, 14/01/06, 18/11/06 & Com, Ext, V-Ext & 10-550 & 0.35$\times$0.32 & 2.3/2.8 & Lee et al. (2007a, 2008)\\

HH\,25 MMS              & 05:46:07.56, $-$00:13:41.6 & 21/12/06, 19/02/07, 26/08/08 &  Subcom, Com, Com-N & 8--165 & 2.3$\times$1.9 & 1.2/6.8 & Q.~Zhang\\

IRAS\,16293              & 16:32:22.90, $-$24:28:35.5 & 22/03/09 &  eSMA & 30--900  & 0.35$\times$0.30 & 1.3/18 & S.~Bottinelli \\

L483                          & 18:17:29.86, $-$04:39:38.8 & 18/06/05, 10/07/05 &  Com & 13--82  & 2.6$\times$1.9 & 1.7/3.2 & J{\o}rgensen et al. (2007)\\

\enddata
\tablenotetext{a}{Phase center of the observations (if more than two pointings in the observations, 
the final figure center of the mosaic image is listed here).} 
\tablenotetext{b}{Synthesized FWHM beam size with robust weighting (weighting parameter 1.0).}
\tablenotetext{c}{1\,$\sigma$ theoretical noise (robust weighting 1.0)\,/\,1\,$\sigma$ rms noise in the cleaned map.}
\end{deluxetable}


\subsection{Observations}

Observations were carried out with the SMA\footnote{The Submillimeter Array is a joint project between 
the Smithsonian Astrophysical Observatory and the Academia Sinica Institute of Astronomy and 
Astrophysics and is funded by the Smithsonian Institution and the Academia Sinica.} (Ho et al. 2004) between
2004 and 2009. Five different array configurations, Subcompact (Subcom), Compact (Com), Compact-North 
(Com-N; for several equatorial and southern sources), Extended (Ext), and Very Extended (V-Ext), were used 
in the observations, with baselines ranging from about 10\,m to 500\,m. Only one source, IRAS\,16293--2422, 
was also observed with the Extended Submillimeter Array (eSMA\footnote{See http://www.jach.hawaii.edu/JCMT/eSMA/
for more details.}).

The 230\,GHz and 345\,GHz receivers were used in the observations. Both receivers have two sidebands 
(lower and upper), with 2\,GHz bandwidth each (separated by 10\,GHz). The receiver setups were typically 
tuned to the CO\,(2--1) line in the 230\,GHz observations (see, e.g., Chen et al. 2008b; 2010) and to the 
CO\,(3--2) line in the 345\,GHz observations (see, e.g., J{\o}rgenson et al. 2007; Lee et al. 2009). The total 
continuum bandwidth, combining the line-free portions of the two sidebands, are about 3.0--3.8\,GHz for the 
1.3\,mm and 850\,$\mu$m dust continuum. We present only the continuum results in this paper. At 230\,GHz, 
system temperatures generally range from 80 to 300\,K (depending on elevation), with a typical value of 150\,K. 
At 350\,GHz, system temperatures range from 150 to 500\,K, with a typical value of 250\,K. The SMA primary 
beams are $\sim$\,55$''$ at 230\,GHz and $\sim$\,36$''$ at 345\,GHz.

The visibility data were reduced using the MIR package (Qi 2005). During the reduction, visibility points 
with clearly deviating phases and/or amplitudes were flagged. The passband (spectral response) was 
calibrated through observations of available planets (e.g., Saturn) and strong quasars (e.g., 3C\,454.3) at 
the beginning and/or end of each track. Time variation of amplitude and phase was calibrated through 
frequent observations of quasars near to each source. The flux density scale was calibrated by observing 
planets such as Uranus and Ceres. For a few tracks, strong quasars (e.g., 3C\,273) with stable flux densities 
during the observing dates, were also used as flux calibrators. We estimate a typical flux accuracy of about
20\%--30\%. The calibrated visibility data were then imaged using the MIRIAD toolbox (Sault et al. 1995).
Each continuum map was deconvolved down to the 1\,$\sigma$ theoretical rms noise level using the MIRIAD 
$clean$ routine. The synthesized beam sizes (with robust weighting 1.0) and rms noise levels are summarized 
in Tables~2 and 3. Figure~1 shows the beam sizes plotted as a function of the corresponding linear resolutions. 
The median angular resolution in this survey is 3\farcs1 at $\lambda$\,1.3\,mm and 1\farcs0 at $\lambda$\,850\,$\mu$m, 
corresponding to the median linear resolutions of approximately 640\,AU and 220\,AU, respectively. Combining
the observations at the two wavelengths, the median angular and linear resolutions in this survey are 2\farcs5 
and 600\,AU, respectively. Figure~2 shows the distribution of the rms noise levels. The median 1\,$\sigma$ theoretical 
noise is 1.0\,mJy\,beam$^{-1}$ at $\lambda$\,1.3\,mm and 1.7\,mJy\,beam$^{-1}$ at $\lambda$\,850\,$\mu$m, 
while in the cleaned images, the median 1\,$\sigma$ noise is 2.3\,mJy\,beam$^{-1}$ and 5.5\,mJy\,beam$^{-1}$, 
respectively.

\section{RESULTS: SMA SUBMILLIMETER AND MILLIMETER MAPS}

Figures~3-17 present the SMA submillimeter and millimeter maps for the sources listed in Table~1. 
Most sources in Table~1 were observed with SCUBA on the James Clerk Maxwell Telescope (JCMT). 
Complementary SCUBA submillimeter images are also shown for selected regions and sources, to illustrate 
the relative locations of the protostellar cores within the clouds and/or the overall morphology of the 
sources. The SCUBA data for the sources in the Perseus and Ophuichus clouds are available from 
the COMPLETE database\footnote{See Ridge et al. (2006) and http://www.cfa.harvard.edu/COMPLETE/ 
for more details.}. 
The data for other objects were retrieved from the JCMT science archive\footnote{http://cadcwww.dao.nrc.ca/jcmt/. 
The JCMT archive at the Canadian Astronomy Data Center is operated by the Herzberg Institute of 
Astrophysics, National Research Council of Canada.}.

In the SMA observations, millimeter and/or submillimeter dust continuum emission is detected from all the 
sources in the sample. It must be noted that in this work we regard a local density peak or enhancement as 
a source if its emission is $\geq$\,5\,$\sigma$ levels with respect to its immediate surrounding. Except several 
very low luminosity sources such as PER\,065 (see Figure~4), L1521F-IRS (see Figure~9c), and Oph-MMS\,126 
(see Figure~15c), most sources in the sample show centrally peaked emission in the SMA dust continuum images, 
suggesting the existence of a compact inner envelope and/or circumstellar disk in these sources. 

As shown in the SCUBA images (angular resolution of $\sim$\,14$''$ at 850\,$\mu$m), most sources in the sample 
appear to be single cores (except CG\,30). At the high angular resolution obtained with the SMA (median angular
resolution of 3\farcs1 at 1.3\,mm and of 1\farcs0 at 850\,$\mu$m), 22 sources in the sample show signatures 
of binarity/multiplicity, with angular separations from 0\farcs3 (HH\,211 MMS) to 22$''$ (CG\,30). 
Detailed descriptions of the individual sources are presented in Appendix~A.

Tables~4 and 5 list the derived parameters (e.g., positions, integrated fluxes, and deconvolved FWHM 
sizes) for each of the sources/components ($\geq$\,5\,$\sigma$ level) from circular Gaussian fits to the 
visibilities. The statistical uncertainties on the derived parameters are generally small and are mainly 
determined by the rms noise levels of the images (see Tables 2 and 3). On the other hand, the calibration 
uncertainties of the fluxes are 20\%--30\%, and thus dominate the uncertainties of the derived fluxes.
For a few faint components in the systems such as IRAS\,03282, NGC\,1333~IRAS\,2A, and HH\,212~MMS, 
their fluxes are uncertain to within a factor of two, due to uncertainties in separating their emission from 
the surrounding envelope emission.

Figure~18 shows the distribution of fluxes derived in this work. The 1.3\,mm fluxes range from 1.1\,mJy to 
3.0\,Jy, with a median value of 100\,mJy, while the 850\,$\mu$m fluxes range from 20\,mJy to 3.2\,Jy, with 
a median value of 430\,mJy.
Compared with the total fluxes obtained from the single-dish IRAM-30\,m images at 1.3\,mm (see, e.g., 
Kauffmann et al. 2008; Launhardt et al. 2010) and the SCUBA images at 850\,$\mu$m (see, e.g., 
Chandler \& Richer 2000; Shirley et al. 2000; Sandell \& Knee 2001), the SMA resolved out a significant 
fraction (generally 80\%--90\%) of the extended emission and detected only the compact emission in the 
center of protostellar cores (i.e., inner envelopes and circumstellar disks).

Assuming that the 1.3\,mm and 850\,$\mu$m continuum emission is optically thin, the total gas mass ($M_{\rm gas}$) 
of the individual sources can be derived from the (sub-)\,millimeter fluxes using the formula:

\begin{equation}
M_{\rm gas} = \frac{F_{\nu} D^{2}}
            {\kappa_{\rm m}(\nu)\, B_{\nu} (T_{\rm d})}
\end{equation}

\noindent where $F_{\nu}$ is the flux of the source, $D$\ is the distance to the source, $T_{\rm d}$\ is the dust 
temperature, $\kappa_{\rm m}(\nu)$\ is the dust opacity (mass absorption coefficient per gram of dust, e.g., 
1.75\,${\rm cm}^2\,{\rm g}^{-1}$ at 850\,$\mu$m for OH5 dust; see Ossenkopf \& Henning 1994), and $B_{\nu} (T_{\rm d})$ 
is the Planck function at frequency $\nu$ and dust temperature $T_{\rm d}$.
In practice, we use the same method as in J{\o}rgensen et al. (2007) to calculate the masses (adopting a dust 
temperature of 30\,K for all sources),

\begin{equation}
M_{\rm 1.3\,mm} = 1.3\,M_\odot\,\left(\frac{F_{\rm 1.3mm}}{1~\rm Jy}\right)\left(\frac{D}{200~\rm pc}\right)^2\,\times\,\{\rm exp [0.36\left(\frac{30~\rm K}{T}\right)] - 1\},
\end{equation}

\begin{equation}
M_{\rm 0.85\,mm} = 0.18\,M_\odot\,\left(\frac{F_{\rm 0.85\,mm}}{1~\rm Jy}\right)\left(\frac{D}{200~\rm pc}\right)^2\,\times\,\{\rm exp [0.55\left(\frac{30~\rm K}{T}\right)] - 1\}.
\end{equation}

\noindent Column~5 of Tables~4 and 5 lists the total gas masses for each sources, calculated from the 1.3\,mm 
and 850$\,\mu$m fluxes. The relative uncertainty of the derived masses due to calibration errors of the 
(sub-)\,millimeter fluxes is about 20\%--30\% for all sources. Moreover, due to the uncertainties in the dust 
opacity and temperature, the masses obtained this way may contain uncertainties of up to a factor of two.

Figure~18 shows the distribution of gas masses for the objects in this sample. The masses derived from the
1.3\,mm observations range from 0.001\,$M_\odot$ to 2.4\,$M_\odot$, with a median value of 0.06\,$M_\odot$, 
while the masses derived from the 850\,$\mu$m observations range from 0.005\,$M_\odot$ to 1.7\,$M_\odot$, 
with a median value of 0.12\,$M_\odot$. Compared with the masses derived from single-dish observations, e.g., 
a median value of $\sim$\,0.5\,$M_\odot$ derived from IRAM-30m 1.3\,mm observations for protostars (see 
Kauffmann et al. 2008), the masses derived from this SMA survey are relatively small and represent only lower 
limits to the core masses. As indicated above, most of the (sub-)\,millimeter continuum emission is resolved 
out in the SMA images, and it is very likely that detected emission arises from the inner envelope and circumstellar 
disk. Therefore, the derived masses may also be regarded as upper limits to the disk masses (see, e.g., 
J{\o}rgensen et al. 2009 for further discussion).


\begin{deluxetable}{lcccccc}
\tabletypesize{\scriptsize} \tablecaption{SMA 1.3\,mm dust continuum results.\label{tbl-4}} \tablewidth{0pt}
\tablehead{\colhead{Source} & \colhead{R.A.$^{a}$} & \colhead{Dec.$^{a}$} & \colhead{$S_{\nu}$$^{a}$} & \colhead{FWHM$^{a}$} & \colhead{$M_{\rm gas}$$^b$}\\
\colhead{} & \colhead{(J2000)} & \colhead{(J2000)} & \colhead{[mJy]} & \colhead{(maj.$''$\,$\times$\,min.$''$)} & \colhead{[$M_{\odot}$]}}

\startdata

L1448 IRS2          & 03:25:22.38 & +30:45:13.3 & 180$\pm$36 & 3.9$\times$2.3   & 0.15  \\

L1448N\,A            & 03:25:36.48 & +30:45:21.8 & 140$\pm$30 & 3.0$\times$1.4   & 0.11  \\

L1448N\,B            & 03:25:36.33 & +30:45:14.8 & 660$\pm$130 & 2.3$\times$1.6   & 0.54    \\

L1448N\,NW        & 03:25:35.66 & +30:45:34.2 & 70$\pm$14 & 3.5$\times$2.5  & 0.057  \\

L1448C\,N           & 03:25:38.88 & +30:44:05.5 & 200$\pm$40 & 1.6$\times$0.9   &  0.16    \\

L1448C\,S           & 03:25:39.05 & +30:43:59.3 & 20$\pm$5 & 5.7$\times$1.9   &  0.016   \\

PER\,065\,A          & 03:28:39.17 & +31:05:55.6 & 4.4$\pm$1 & 8.1$\times$6.5   &  0.004   \\

PER\,065\,B          & 03:28:38.86 & +31:05:56.8 & 1.1$\pm$0.2 & 2.3$\times$1.3   &  0.001   \\

PER\,065\,C          & 03:28:38.96 & +31:06:04.2 & 1.5$\pm$0.3 & 2.3$\times$2.1   &  0.001   \\

NGC\,1333\,IRAS\,2A             & 03:28:55.58 & +31:14:37.2 & 330$\pm$66 & 1.8$\times$1.3   &  0.27  \\

NGC\,1333\,IRAS\,2B             & 03:28:57.38 & +31:14:15.8 & 50$\pm$10 & 3.8$\times$2.0   &  0.04   \\

NGC\,1333\,SVS\,13\,A          & 03:29:03.75 & +31:16:03.6 & 450$\pm$90 & 2.7$\times$2.0   &  0.37   \\

NGC\,1333\,SVS\,13\,VLA3    & 03:29:03.41 & +31:16:01.5 & 22$\pm$5 & (unresolved)   &  0.02   \\

NGC\,1333\,SVS\,13\,B          & 03:29:03.07 & +31:15:51.8 & 460$\pm$90 & 2.8$\times$2.3   &  0.37   \\

NGC\,1333\,SVS\,13\,C          & 03:29:01.97 & +31:15:38.1 & 50$\pm$10 & 2.2$\times$0.7   &  0.04   \\

NGC\,1333\,IRAS\,4A             & 03:29:10.51 & +31:13:31.3 & 3000$\pm$600 & 2.4$\times$1.7   &   2.43  \\

NGC\,1333\,IRAS\,4B             & 03:29:12.01 & +31:13:08.2 & 920$\pm$180 & 1.4$\times$0.9   &  0.75   \\

NGC\,1333\,IRAS\,4C             & 03:29:12.84 & +31:13:07.0 & 250$\pm$50 & 0.7$\times$0.6   &  0.20   \\

IRAS\,03282       & 03:31:20.94 & +30:45:30.2 & 290$\pm$60 & 1.4$\times$1.3   &  0.24   \\

Per-B1-c             & 03:33:17.87 & +31:09:32.3 & 300$\pm$60 & 3.1$\times$2.4   & 0.24    \\

Per-B1-bN          & 03:33:21.19 & +31:07:43.6 & 240$\pm$50 & 4.6$\times$3.2   &  0.19  \\

Per-B1-bS          & 03:33:21.33 & +31:07:26.3 & 340$\pm$70 & 3.2$\times$1.8   &   0.28  \\

IC348\,MMS1     & 03:43:57.05 & +32:03:05.0 & 260$\pm$50 & 2.5$\times$1.9   &  0.21  \\

IC348\,MMS2     & 03:43:57.73 & +32:03:10.5 & 65$\pm$15 & 4.4$\times$2.6   &  0.053  \\

CB\,17\,IRS        & 04:04:33.76 & +56:56:16.5 & 6.3$\pm$1.3 & 3.4$\times$2.5  &  0.006   \\

CB\,17\,MMS      & 04:04:35.78 & +56:56:03.4 & 3.6$\pm$0.7 & 3.6$\times$1.1  &  0.003   \\

IRAM\,04191\,IRS   & 04:21:56.89 & +15:29:46.1 & 4.5$\pm$1.5 & 2.8$\times$2.7   &  0.001   \\

IRAM\,04191\,MMS   & 04:21:56.36 & +15:29:48.9 & 2.5$\pm$1.0 & 3.7$\times$1.1   &  $<$\,0.001   \\

L1521F IRS      & 04:28:38.91 & +26:51:34.6 & 15$\pm$5 & 5.0$\times$2.1 &  0.004   \\

L1527               & 04:39:53.90 & +26:03:09.8 & 190$\pm$40 & 1.0$\times$0.5 & 0.052    \\

HH\,114\,MMS   & 05:18:15.24 & +07:12:02.7 & 490$\pm$100 & 1.9$\times$1.6 & 1.34   \\

HH\,212\,MMS  & 05:43:51.41 & $-$01:02:53.1 & 76$\pm$15 & 0.6$\times$0.5 &  0.21   \\

CG\,30\,N         & 08:09:33.13 & $-$36:04:58.3 & 260$\pm$50 & 3.3$\times$1.9 &   0.15  \\

CG\,30\,S         & 08:09:32.73 & $-$36:05:19.3 & 100$\pm$20 & 4.6$\times$2.4 &   0.06 \\

B228                & 15:43:02.20 & $-$34:09:07.6 & 150$\pm$30 & 5.5$\times$3.9 &   0.036  \\

B228\,(1.1\,mm)  & 15:43:02.13 & $-$34:09:07.3 & 135$\pm$30 & 5.4$\times$4.3 &  0.032 \\

VLA\,1623\,W    & 16:26:25.64 & $-$24:24:29.3 & 42$\pm$10 & 1.1$\times$0.7 &  0.009   \\

VLA\,1623\,E\,A   & 16:26:26.39 & $-$24:24:30.7 & 79$\pm$15 & 1.0$\times$0.8 &   0.016  \\

VLA\,1623\,E\,B   & 16:26:26.31 & $-$24:24:30.4 & 71$\pm$15 & 0.6$\times$0.6 &  0.014   \\

Oph-MMS\,126   & 16:28:21.58 & $-$24:36:23.6 & 11$\pm$2 & 3.6$\times$1.7 &  0.002   \\

IRAS\,16293\,A   & 16:32:22.87 & $-$24:28:36.6 & 620$\pm$120 & 1.7$\times$1.4   &  0.13  \\

IRAS\,16293\,B   & 16:32:22.62 & $-$24:28:32.6 & 1300$\pm$260 & 1.6$\times$1.4  &  0.26  \\

L723\,VLA2\,MMS1  & 19:17:53.68 & +19:12:19.8 & 59$\pm$12 & 2.7$\times$2.0 &  0.075   \\

L723\,VLA2\,MMS2  & 19:17:53.91 & +19:12:18.4 & 58$\pm$12 & 3.2$\times$1.4 &  0.074   \\

B335                 & 19:37:00.93 & +07:34:09.9 & 145$\pm$30 & 4.5$\times$2.4 &  0.13   \\

L1157               & 20:39:06.18 & +68:02:16.3 & 280$\pm$56 & 3.1$\times$3.0 & 0.42   \\

L1251B\,IRS1   & 22:38:46.68 & +75:11:33.1 & 48$\pm$10 & 4.2$\times$2.3 &  0.061   \\

L1251B\,IRS2   & 22:38:52.94 & +75:11:23.2 & 21$\pm$4 & 3.6$\times$3.3 &  0.027   \\

L1251B\,SMA\,N   & 22:38:48.95 & +75:11:33.4 & 17$\pm$4 & 4.9$\times$4.7  & 0.022   \\

L1251B\,SMA\,S   & 22:38:49.09 & +75:11:25.0 & 28$\pm$6 & 6.1$\times$4.2  &  0.035   \\

\enddata

\tablenotetext{a}{Center position, flux, and FWHM size of the dust continuum sources derived from Gaussian fitting.}%
\tablenotetext{b}{Total gas mass derived from the SMA 1.3\,mm dust continuum observations. Note that the gas mass
is from compact emission only, not the whole envelope.}%
\end{deluxetable}


\begin{deluxetable}{lcccccc}
\tabletypesize{\scriptsize} \tablecaption{SMA 850\,$\mu$m dust continuum results.\label{tbl-5}} \tablewidth{0pt}
\tablehead{\colhead{Source} & \colhead{R.A.$^{a}$} & \colhead{Dec.$^{a}$} & \colhead{$S_{\nu}$$^{a}$} & \colhead{FWHM$^{a}$} & \colhead{$M_{\rm gas}$$^b$} \\
\colhead{} & \colhead{(J2000)} & \colhead{(J2000)} & \colhead{[mJy]} & \colhead{(maj.$''$\,$\times$\,min.$''$)} & \colhead{[$M_{\odot}$]} }

\startdata

L1448C\,N       & 03:25:38.87 & +30:44:05.3 & 410$\pm$80 & 0.6$\times$0.4 & 0.078 \\

L1448C\,S       & 03:25:39.14 & +30:43:60.0 & 40$\pm$10 & 1.2$\times$0.6 &  0.008 \\

NGC\,1333\,IRAS\,2A1  & 03:28:55.57 & +31:14:37.1 & 800$\pm$160 & 1.1$\times$0.8 &  0.15  \\

NGC\,1333\,IRAS\,2A2  & 03:28:55.55 & +31:14:38.5 & 40$\pm$10 & (unresolved) &  0.008  \\

NGC\,1333\,IRAS\,4A1        & 03:29:10.53 & +31:13:31.0 & 3200$\pm$600 & 1.4$\times$0.4 &  0.61  \\

NGC\,1333\,IRAS\,4A2        & 03:29:10.44 & +31:13:32.2 & 1600$\pm$300 & 1.1$\times$0.1 &  0.30   \\

IRAS\,03282\,MMS1     & 03:31:20.94 & +30:45:30.3 & 340$\pm$70 & 0.2$\times$0.1  &  0.065  \\

IRAS\,03282\,MMS2     & 03:31:21.08 & +30:45:29.9 & 30$\pm$10 & (unresolved) &  0.006  \\

HH\,211~SMM1       & 03:43:56.80 & +32:00:50.3 & 80$\pm$20 & 0.1$\times$0.1  &  0.015  \\

HH\,211~SMM2       & 03:43:56.79 & +32:00:50.0 & 25$\pm$10 & $<$\,0.1 &  0.005  \\

HH\,114\,MMS        & 05:18:15.23 & +07:12:02.5 & 1100$\pm$220 & 1.3$\times$1.2  &  0.70  \\

OMC3\,MMS6 SMM1   & 05:35:23.43 & $-$05:01:30.6 & 1350$\pm$200 & 1.0$\times$0.8 &  0.86   \\

OMC3\,MMS6 SMM2   & 05:35:23.52 & $-$05:01:41.2 & 550$\pm$100 & 1.8$\times$1.7 &  0.35  \\

NGC2024~FIR5 SMM1    & 05:41:44.26 & $-$01:55:40.9 & 2640$\pm$500 & 2.9$\times$1.8 & 1.69    \\

NGC2024~FIR5 SMM2    & 05:41:44.51 & $-$01:55:42.4 & 1030$\pm$200 & 3.5$\times$3.0 &  0.66  \\

HH\,212\,MMS1       & 05:43:51.41 & $-$01:02:53.1  & 260$\pm$50  & 2.0$\times$1.0  &  0.17  \\

HH\,212\,MMS2       & 05:43:51.38 & $-$01:02:53.0  & 20$\pm$4  & (unresolved)  &  0.013  \\

HH\,212\,MMS3       & 05:43:51.48 & $-$01:02:53.8  & 27$\pm$5  & 0.5$\times$0.4  &  0.018  \\

HH\,25\,SMM1       & 05:46:07.26 & $-$00:13:30.7 & 510$\pm$100 & 2.8$\times$2.4 &  0.33  \\

HH\,25\,SMM2       & 05:46:07.34 & $-$00:13:43.4 & 430$\pm$80 & 1.3$\times$0.7 &  0.27   \\

HH\,25\,SMM3       & 05:46:07.58 & $-$00:13:53.7 & 150$\pm$30 & 4.8$\times$4.6  &  0.10   \\

IRAS\,16293\,Aa   & 16:32:22.86 & $-$24:28:36.6 & 980$\pm$200 & 0.8$\times$0.7   &  0.05  \\

IRAS\,16293\,Ab   & 16:32:22.88 &  $-$24:28:36.3 & 430$\pm$90 &  0.9$\times$0.6  &  0.02  \\

IRAS\,16293\,B   & 16:32:22.62 & $-$24:28:32.5 & 3130$\pm$630 & 0.6$\times$0.4  &  0.15  \\

L483       & 18:17:29.94 & $-$04:39:39.3 & 140$\pm$28 & 2.2$\times$0.8 & 0.02   \\

\enddata

\tablenotetext{a}{Center position, flux, and FWHM size of the dust continuum sources derived from Gaussian fitting.}%
\tablenotetext{b}{Total gas mass derived from the SMA 850\,$\mu$m dust continuum observations. Note that the gas 
mass is from compact emission only, not the whole envelope.}%
\end{deluxetable}

\clearpage


\section{DISCUSSION}

\subsection{Multiplicity in the Class\,0 Phase}

Before analyzing the binary/multiple frequency in the Class\,0 protostellar phase, it is important 
to note how binary/multiple systems are defined in this work. Strictly speaking, only systems that 
show bound motions can be referred to as binary/multiple systems. However, compared with 
well-studied MS/PMS binaries (for which the orbital motion can be determined by direct imaging 
or spectroscopic monitoring at optical and near-infrared wavelengths; see Duch{\^e}ne et al. 2007), 
these kinds of systems are extremely rare in the protostellar phase (see, e.g., Rodr{\'i}guez 2004).
The main reasons are: (1) the sample of protobinary systems is relatively small, (2) it is difficult to 
directly observe the protostellar phase, due to the limitation of angular resolution obtained at millimeter 
wavelengths, and (3) it is difficult to monitor bound motion in protobinary systems, as most known 
protobinary systems have large separations (100--1000\,AU) and thus have long orbital periods. 
In this work, we use the following criterion to define binary/multiple systems: \noindent {\it the 
separation or the closest separation (if there are more than two components in the systems) is 
less than 5000\,AU.} We explain below why we choose 5000\,AU as the limit.

As introduced in Section~I, through initial fragmentation, large-scale ($\sim$\,0.1--1.0\,pc) molecular 
clumps (or filaments) fragment into individual dense cores, which subsequently collapse to form 
individual stellar systems (i.e., fragmented components from initial clump fragmentation are not bound 
stellar systems but prestellar cores).
The fragmented components from initial fragmentation generally have separations of more than 5000\,AU, 
consistent with the typical Jeans length in large-scale molecular clumps/filaments 
(4000--5000\,AU at $T$ = 10\,\rm K and $n_{\rm H_2}$ = 10$^5$--10$^6$\,cm$^{-3}$). This is also in 
agreement with the results from the high angular resolution survey by Looney et al. (2000) that systems 
with separations $\leq$\,6000\,AU have common envelopes while wider systems have separated envelopes. 
Interestingly, by analyzing observational data in the Taurus region (e.g., Ghez et al. 1993; Gomez
et al. 1993), Larson (1995) also found the existence of an `intrinsic' scale ($\sim$\,0.04\,pc or 8000\,AU) 
in binary star formation process, and this scale is found to be consistent with the Jeans length in large 
molecular clouds.
Therefore, we consider that components with separations $>$\,5000\,AU are likely not relevant for the 
formation of bound stellar systems. This is also the reason why we list several well-known sources, e.g., 
NGC\,1333 IRAS\,2A \& 2B (separation $\sim$\,7500\,AU) and IRAS\,4A \& 4B (separation $\sim$\,7200\,AU), 
as separate targets in this work (see Table~1).

We use the densities of YSOs and dense cores in the Perseus molecular cloud to estimate the likelihood 
of identifying a background or foreground source as a possible binary (or multiple) companion in this study. 
Many of the targets (14 of 33, or 42\%) in this sample are located in the Perseus cloud, and most of the 
rest of the sources are in regions with lower protostellar densities. In general, main sequence and evolved 
stars foreground and background to the cloud have very small amounts of (or no) circumstellar dust, and 
should be undetectable by this interferometric submillimeter and millimeter continuum survey. Therefore, 
the only possible ``contaminating" sources in this study should be young stars with dusty circumstellar 
envelopes associated with the cloud. In Perseus, there are a total of 400 YSOs in an area of 3.86\,deg$^2$ 
(see J{\o}rgensen et al. 2006). Assuming that these YSOs all have detectable dust continuum emission at 
the SMA, the density of potential dust continuum sources is 103.6 YSOs/deg$^2$. We therefore estimate 
the probability of there being a contaminating YSO within a 5000\,AU (21$''$ at the distance of Perseus) 
radius that could mistakenly be identified as a companion to be 0.01.
Another estimate of the contamination likelihood can be obtained using the density of dense cores detected 
in previous single-dish dust continuum surveys. Hatchell et al. (2005; SCUBA 850\,$\mu$m) and Enoch et al. 
(2006; Bolocam 1.1\,mm) found 91 (within an area of 3\,deg$^2$) and 122 (within an area of 7.5\,deg$^2$) 
dense cores in Perseus, respectively. From these two large-scale surveys, the probabilities are estimated to 
be 3.2\,$\times$\,10$^{-3}$ and 1.7\,$\times$\,10$^{-3}$, respectively. Hence, it is very unlikely that any of 
the identified companions in this work are background or foreground sources.

Nevertheless, we want to note that further high angular resolution kinematical data are important for 
complementing the separation criterion used here. For example, for relatively wide protobinary systems 
($>$\,1000\,AU), molecular line observations are needed to verify if the components therein have similar 
systematic velocities, in order to insure that the components resolved therein are physically 
associated\footnote{For several wide protobinary systems (e.g., L1448N, L1448C, IRAM\,04191, CG\,30, 
and L723), high angular resolution molecular line observations have already found that the components 
resolved therein have similar systematic velocities.}. For close systems (on the scale of 100\,AU), high 
angular resolution line (e.g., CO or H$_2$) observations are needed to distinguish if each component 
drives its own outflow, in order to be sure that each component is associated with a protostellar object 
(and not a dense, starless, condensation).

In this work, we adopt the following terminology and definitions to describe the complexities of binary/multiple 
stars:

\begin{equation}
multiplicity~frequency~(MF) = \frac{B+T+Q} {S+B+T+Q},
\end{equation}

\begin{equation}
companion~star~fraction~(CSF) = \frac{B+2T+3Q} {S+B+T+Q},
\end{equation}

\noindent where $S$, $B$, $T$, and $Q$ are the numbers of single, binary, triple, and quadruple systems, respectively.
The `multiplicity frequency', suggested by Reipurth \& Zinnecker (1993), represents the probability that any system is 
a binary/multiple system. The `companion~star~fraction', on the other hand, looks at the individual stars and counts the 
number of companions relative to the sample size. It should be noted that many previous works used 
`companion~star~fraction' rather than `multiplicity frequency' in their analysis (see, e.g., Haisch et al. 2004; Duch{\^e}ne 
et al. 2004; Connelley et al. 2008b). Actually, as noted by Reipurth \& Zinnecker (1993), the most precise description of 
multiplicity consists of explicitly listing the number of singles, binaries, triples, and quadruples, like S:B:T:Q, although this 
description (as well as others) does not account for hierarchicality.

In this SMA survey toward 33 protostars, 22 sources show signatures of binarity or multiplicity (see Appendix~A for detailed 
descriptions of the individual sources). Table~6 lists the names of these sources, as well as their basic observed parameters 
(separation and circumstellar mass ratio). If we adopt 5000\,AU as the limit, source CB\,17 (21\farcs4 or 5350\,AU) is beyond 
this limit and will {\it not} be regarded as a ``bound" binary system in this work. 
Therefore, the numbers of singles\footnote{L1448 IRS2, NGC1333 IRAS\,2B, B1c, CB\,17, L1521F-IRS, L1527\,IRS, 
HH\,114 MMS, B228, Oph MMS\,126, L483, B335, and L1157}, binaries\footnote{L1448C, NGC1333 IRAS\,2A, IRAS\,4A, 
IRAS\,4B, IRAS\,03282, B1b, HH\,211\,MMS, IC348\,MMS, IRAM\,04191, OMC3~MMS6, NGC2024-FIR5, HH\,25\,MMS,
CG\,30, and L723\,VLA2}, triples\footnote{L1448N, PER\,065, HH\,212\,MMS, VLA\,1623, and IRAS\,16293}, and 
quadruples\footnote{NGC\,1333 SVS\,13 and L1251B} are considered to be 12, 14, 5, and 2, respectively, which is 
S:B:T:Q~=~0.86:1.00:0.36:0.14 when normalized to the number of binaries. The overall multiplicity frequency (MF) and 
companion star fraction (CSF) derived in the sample, without correcting for completeness (see below), is 0.64\,$\pm$\,0.08 
and 0.91\,$\pm$\,0.05, respectively, with the separations in the range between 50\,AU (which is the smallest separation 
we detect between two components) and 5000\,AU (the limit set in this work). The statistical standard errors are estimated 
as [MF(1$-$MF)/N]$^{1/2}$ and [CSF(1$-$CSF)/N]$^{1/2}$ (where N is the total number of sources in the sample), respectively. 

It is important to note that the angular resolution is not uniform in this SMA survey (see Tables 2 \& 3). The median angular 
resolution in the survey is 2\farcs5, corresponding to a median linear resolution of 600\,AU in nearby clouds (see Figure~1). 
There are very likely more binary/multiple systems with separations less than 600\,AU unresolved in this survey. Therefore, 
this survey is incomplete across the observed separation range. However, it is difficult to correct for this incompleteness at 
present, because we would have to adopt assumptions regarding to the separation distributions of Class\,0 protobinary 
systems, which are unfortunately unknown yet (see further discussion in $\S$\,4.2). Consequently, the multiplicity frequency 
and companion star fraction derived from this work could be considered as lower limits over the separation range observed here. 

Recently, Maury et al. (2010) argued that binary systems are relatively rare in the Class\,0 phase, as only one object 
shows a possible binary detection in their small PdBI survey toward five Class\,0 objects. However, two protobinary 
systems detected in the SMA survey presented here, L1448C (see Appendix~A.1) and IRAM\,04191 (see Appendix~A.8), 
are missing in their PdBI A-configuration (the most extended configuration) observations due to the low flux-ratio sensitivity 
(only 0.6--0.9; see Maury et al. 2010). Therefore, taking both L1448C and IRAM\,04191 into account, the re-calculated 
multiplicity frequency in their sample is actually 0.60, which is consistent with the multiplicity frequency derived in this SMA 
survey (0.64).

\begin{table*}[hp]
\begin{center}
\caption{Separations and mass ratios of binary/multiple systems observed in this work}\footnotesize
\begin{tabular}{rccc}
\hline \hline
Source                                               & Separation [$''$] & Separation [AU] & $M_{\rm low}/M_{\rm high}$ \\
\hline

L1448N (A, B)                                    & 7.4\,$\pm$\,0.3   & 1800   & 0.21\,$\pm$\,0.06    \\
L1448N (A, NW)                                 & 16.5\,$\pm$\,0.3 & 4000  & 0.50\,$\pm$\,0.15    \\
L1448C (N, S)                                    & 7.0\,$\pm$\,0.5 & 1700 & 0.10\,$\pm$\,0.05    \\
PER\,065 (A, B)                                 & 3.9\,$\pm$\,0.4 & 940 & 0.25\,$\pm$\,0.10   \\
PER\,065 (B, C)                                 & 7.5\,$\pm$\,0.4 & 1800 & 0.73\,$\pm$\,0.22   \\
NGC1333\,IRAS\,2A (A1, A2)            & 1.5\,$\pm$\,0.2 & 360  & 0.05\,$\pm$\,0.02 \\
NGC1333\,SVS\,13 (A, VLA3)            & 5.0\,$\pm$\,0.3  & 1200 & 0.05\,$\pm$\,0.02  \\
NGC1333\,SVS\,13 (VLA3, B)            & 10.4\,$\pm$\,0.3  & 2500 & 0.05\,$\pm$\,0.02  \\
NGC1333\,SVS\,13 (B, C)                  & 19.8\,$\pm$\,0.3  & 4750 & 0.11\,$\pm$\,0.03  \\
NGC1333\,IRAS\,4A (A1, A2)             & 1.4\,$\pm$\,0.2 & 340 & 0.50\,$\pm$\,0.15     \\
NGC1333\,IRAS\,4\,B/C                     &  10.5\,$\pm$\,0.2  & 2500 & 0.27\,$\pm$\,0.10     \\
IRAS03282+3035 (MMS1, MMS2)     & 1.6\,$\pm$\,0.2 &  400 & 0.10\,$\pm$\,0.05   \\
Per-B1-b (N, S)                                   & 17.4\,$\pm$\,0.5 & 4200 & 0.71\,$\pm$\,0.21 \\
HH\,211 (SMM1, SMM2)                    & 0.31\,$\pm$\,0.02 & 74 & 0.30\,$\pm$\,0.10 \\
IC348 (MMS1, MMS2)                       & 9.8\,$\pm$\,0.4 & 2350 & 0.25\,$\pm$\,0.10 \\
$^a$CB\,17 (IRS, MMS)                     & 21.4\,$\pm$\,0.5 & 5350 & 0.57\,$\pm$\,0.17 \\
IRAM\,04191+1522 (IRS, MMS)        & 8.0\,$\pm$\,0.4 & 1100 & 0.56\,$\pm$\,0.17 \\
OMC3\,MMS\,6 (SMM1, SMM2)        & 10.8\,$\pm$\,0.2 & 4750 & 0.41\,$\pm$\,0.12 \\
NGC2024 FIR\,5 (SMM1, SMM2)      & 4.1\,$\pm$\,0.2 & 1800 & 0.39\,$\pm$\,0.12 \\
HH\,212 (SMM1, SMM2)                    & 0.53\,$\pm$\,0.05 & 230 & 0.08\,$\pm$\,0.03\\
HH\,212 (SMM1, SMM3)                    & 1.2\,$\pm$\,0.1 & 530 & 0.10\,$\pm$\,0.03\\
$^b$HH\,25 (SMM1, SMM2)               & 12.7\,$\pm$\,0.3 & 5600 & 0.84\,$\pm$\,0.25 \\
HH\,25 (SMM2, SMM3)                      & 10.9\,$\pm$\,0.3 & 4800 & 0.35\,$\pm$\,0.10\\
CG\,30 (N, S)                                     & 21.8\,$\pm$\,0.5 & 4400 & 0.38\,$\pm$\,0.11 \\
VLA\,1623 (W, E)                               & 9.9\,$\pm$\,0.2   & 1200 & 0.20\,$\pm$\,0.06\\
VLA\,1623 (EA, EB)                            & 1.1\,$\pm$\,0.1 &  130 & 1.0\,$\pm$\,0.3 \\
IRAS16293$-$2422 (A, B)                  & 5.3\,$\pm$\,0.2 &  640 & 0.50\,$\pm$\,0.15  \\
IRAS16293$-$2422 (Aa, Ab)              & 0.42\,$\pm$\,0.05 &  50 & 0.44\,$\pm$\,0.13  \\
L723\,VLA2 (MMS1, MMS2)               & 3.5\,$\pm$\,0.3 &  1050 & 1.0\,$\pm$\,0.3  \\
L1251B (IRS1, SMA-N)                       & 8.9\,$\pm$\,0.4 & 2700  & 0.35\,$\pm$\,0.11 \\
L1251B (SMA-N, SMA-S)                    & 8.4\,$\pm$\,0.4 & 2500 & 0.61\,$\pm$\,0.18 \\
L1251B (SMA-S, IRS2)                       & 15.6\,$\pm$\,0.4 & 4700 & 0.80\,$\pm$\,0.25 \\

\hline \hline
\end{tabular}\\[0.5mm]
{\footnotesize $^a$ CB\,17 is not regarded as a ``bound" binary system in this work.}\\
{\footnotesize $^b$ The separation between sources SMM\,1 and SMM\,2 is beyond
the limit of 5000\,AU, and hence we regard HH\,25\,MMS as a binary system (SMM\,2 and SMM\,3).}
\end{center}
\end{table*}

\subsection{The Distribution of Separations}

Figure~19 shows the distribution of separations for the Class\,0 binary/multiple systems observed in this work 
(see Table~6). It must be noted that these separations, measured in the SMA (sub-)\,millimeter images, are 
the projected separations between the components in individual systems. However, as discussed by Reipurth 
\& Zinnecker (1993), using reasonable statistical assumptions, one could convert the mean projected separation 
($l$) to the mean semi-major axis ($a$) with the ratio of $l$/$a$\,$\sim$\,0.79--0.85 (see Reipurth \& Zinnecker 
1993 and references therein). We therefore consider that there is no significant error by assuming that the overall 
statistical distribution of observed projected binary separations is representative of the actual distribution of the 
semi-major axes. In this work, we do not use these ratios to convert observed projected separations to the 
`semi-major' axes, and instead show the original results from the observations.

As shown in the top of Figure~19, the distribution of separations for Class\,0 protostars shows a general trend 
in which the companion star fraction increases from wide separations to close separations. Quantitatively, the 
companion star fraction is 0.42\,$\pm$\,0.09 for separations from 50\,AU to 1700\,AU, 0.27\,$\pm$\,0.08 from 
1700\,AU to 3350\,AU, and 0.21\,$\pm$\,0.07 from 3350\,AU to 5000\,AU. The median separation in this survey 
is 1800\,AU. Previous large-scale surveys toward MS and PMS stars found that the distributions of separations 
could be well fitted with log-normal functions with peaks at less than 100\,AU (see, e.g., Duquennoy \& Mayor 
1991 and Patience et al. 2002). As shown in the bottom of Figure~19, the distribution of separations for Class\,0 
protostars does not exhibit a log-normal like distribution, and rather shows a trend where most companions are 
found with separations around 1800\,AU. We ascribe this distribution to the limited angular resolution of the SMA 
observations (i.e., the completeness limit of the survey, see below), which allowed us to mostly detect binary 
separations in the order of 1000\,AU. Further higher angular resolution large-scale surveys toward Class\,0 protostars 
are needed, in order to achieve a wider distribution of separations (from $\sim$\,0.1\,AU to $\sim$\,10$^3$\,AU) for 
protobinary systems.

To better understand the evolution of binary properties, it is important to compare the separation distribution 
of Class\,0 protobinary systems with those derived for Class\,I, PMS, and MS binaries. Yet, it must be noted 
in advance that MS and PMS binaries, as well as Class\,I binaries, have been well studied over the past two 
decades thanks to the availability of both large samples and high angular resolution obtained with large 
optical/near-infrared telescopes. The sample size and angular resolution in this SMA survey are an order of 
magnitude less than the optical and near-infrared surveys. Therefore, large uncertainties may remain in our 
comparisons due to the small sample size and limited angular resolution in this work. 

In a large-scale survey toward Class\,I YSOs, Connelley et al. (2008a, b) found the number of single, binary, triple, and 
quadruple stars is S:B:T:Q~=~122:51:12:4, with separations ranging from about 50\,AU to 5000\,AU. Based on the 
large-scale survey of Duquennoy \& Mayor (1991), with a similar separations range, the corresponding number found in 
solar-type MS stars is S:B:T:Q~=~131:32:1:0 (see Connelley et al.  2008b). These two studies, which summarize the binary 
properties for Class\,I YSOs and MS stars, will be mainly used here for comparison, as their separation ranges basically 
match the observed separation range in this SMA survey. From these two large-scale surveys, the multiplicity frequency 
and companion star fraction are derived to be 0.35\,$\pm$\,0.03 and 0.46\,$\pm$\,0.04 for Class\,I YSOs, and 
0.20\,$\pm$\,0.03 and 0.21\,$\pm$\,0.03 for MS stars, respectively. 
Despite the incompleteness in the SMA survey and the difference between the sample sizes, the multiplicity frequency
(0.64\,$\pm$\,0.08) and companion star fraction (0.91\,$\pm$\,0.05) found in Class\,0 protostars are 1.8 (for MF) and 
2.0 (for CSF) times larger than the values found in Class\,I YSOs, and 3.2 (for MF) and 4.3 (for CSF) times larger than 
the values found in MS stars (see Figure~20). Furthermore, the observed fraction of high order multiple systems to binary 
systems in Class\,0 protostars (7:14 or 0.50\,$\pm$\,0.09) is also larger than the fractions found in Class\,I YSOs 
(0.31\,$\pm$\,0.07) and MS stars (0.03\,$\pm$\,0.01)\footnote{The estimated fraction from the whole sample of 
Duquennoy \& Mayor (1991; from $\sim$\,0.1\,AU to $\sim$\,10$^4$\,AU) is about 20\% (see also Tokovinin 2004).}. 
To estimate how different the distribution of separations for Class\,0 protostars is from those distributions for Class\,I YSOs 
and MS stars, we use the equation given by Brandeker et al. (2006) in their Appendix B2 to calculate the probability that the 
results in a given separation range from two studies are consistent with each other. Using this method, the probabilities that 
Class\,0 protostars have the same distribution of separations as Class\,I YSOs and MS stars are estimated to be 
1.9\,$\times$\,10$^{-6}$ and 4.1\,$\times$\,10$^{-15}$, respectively.

The distributions of separations derived for PMS binaries are in general similar to those for Class\,I binaries (see, e.g., 
Connelley et al. 2008b). Therefore, we consider that the comparisons between Class\,0 and PMS binaries will likely hold 
similar results to those derived from the comparisons between Class\,0 and Class\,I binaries. Indeed, for example, in the 
survey toward PMS stars, Ghez et al. (1997) found a companion star fraction of 0.21\,$\pm$\,0.04 with separations from 
150\,AU to 1800\,AU; with the same separation range, the companion star fraction found by Reipurth \& Zinnecker (1993) 
is 0.16\,$\pm$\,0.02. The companion star fraction derived from this survey with the same separation range is 0.39\,$\pm$\,0.08, 
approximately two times higher, comparable to the ratios estimated between the Class\,0 and Class\,I binaries.

As we indicated in $\S$\,4.1, this SMA survey is incomplete across the observed separation range, and the derived multiplicity
frequency and companion star fraction in this work should be considered as lower limits. Therefore, the ratios estimated above 
should also be considered as lower limits. Considering that the median separation in this SMA survey is 1800\,AU (larger than 
the lowest linear resolution in the survey of 1500\,AU), the survey can be regarded to be complete for protobinary systems with 
separations from 1800\,AU to 5000\,AU. 
Since, to our knowledge, this work is the first large-scale observational survey of multiplicity in Class\,0 protostars, the distribution 
of protobinary systems with smaller separations is not yet known. Hence, it is difficult to correct for our incompleteness to derive 
an accurate multiplicity frequency or companion star fraction for Class\,0 protobinary systems with separations between 50\,AU 
and 1800\,AU. 
Nevertheless, we may adopt a simple method to `correct' for completeness in this work. For separations smaller than 1800\,AU, 
we assume that the completeness in a separation range is given by the averaged fraction of the sources in the total sample that 
have linear resolutions intermediate between the higher and lower limits of the range. For example, 55\% of the sample has a 
resolution better than the median resolution of 600\,AU, then we assume that we only detect $\frac{55\% + 100\%}{2}$ of the total 
number of multiple systems in the range between 600\,AU and 1800\,AU. Using this method, the multiplicity frequency and 
companion star fraction are derived to be 0.33\,$\pm$\,0.08 and 0.42\,$\pm$\,0.08 for systems with separations from 1800\,AU to 
5000\,AU (100\% completeness), 0.15\,$\pm$\,0.06 and 0.24\,$\pm$\,0.07 for systems with separations from 600\,AU to 1800\,AU 
(78\% completeness), while 0.15\,$\pm$\,0.06 and 0.24\,$\pm$\,0.07 for systems with separations from 50\,AU to 600\,AU (32\% 
completeness). After correcting for completeness, the overall multiplicity frequency and companion star fraction in this SMA survey 
are 1.0\,$\pm$\,0.1 and 1.5\,$\pm$\,0.2, respectively, over the separation range from 50\,AU to 5000\,AU. Compared to the data 
derived from Class\,I YSOs survey (Connelley et al. 2008a, b), these `corrected' values, result in a difference, in both the multiplicity 
frequency and the companion star fraction, of a factor of three between Class\,0 protostars and Class\,I YSOs, which is indeed larger 
than the factor of two estimated above. However, it must be noted that these `corrected' values are uncertain. Further high angular 
resolution large-scale surveys toward Class\,0 protostars are needed, in order to draw a real statistically significant separation 
distribution for close protobinary systems.

The decrease in both the multiplicity frequency and companion star fraction, as well as the fraction of high order multiple 
systems, suggests that binary properties change as stars evolve from the Class\,0 to Class\,I/PMS, all the way to the MS 
phases. As reviewed by Goodwin et al. (2007), the evolution of binary properties has been ascribed to two main mechanisms: 
(1) the rapid dynamical decay of young small-$N$ systems (2\,$<$\,$N$\,$<$\,12; see, e.g., Reipurth 2000; Reipurth 
\& Clarke 2001; Goodwin \& Kroupa 2005); and (2) dynamical destruction of loose binary systems in dense clusters 
(see, e.g., Kroupa 1995). Here, we consider that small-$N$ decay is likely the main process for the evolution of 
binary properties, because (1) most sources in this sample, as well as those in the Class\,I, PMS, and MS surveys discussed
above, are not in dense cluster environments, and (2) the time scale for the decrease in binary fraction caused by destruction 
in dense cluster (a few 10\,Myr) is much greater than the lifetimes of Class\,0 protostars ($\sim$\,10$^{4}$--10$^{5}$~yr) and 
Class\,I YSOs ($\sim$\,10$^5$~yr). As estimated by Goodwin et al. (2007), a triple system is unstable to decay with a 
general half-life of a few 10$^4$~yr, which is consistent with the lifetime estimated for Class\,0 protostars. Thus, most of 
the decays/ejections are expected to happen during the Class\,0 phase (e.g., Reipurth 2000). This small-$N$ decay will 
modify the binary properties by reducing the overall multiplicity frequency and companion star fraction, and reducing the 
fraction of high order multiple systems to binary systems. All have been evidenced by the comparisons between this work 
and previous large-scale surveys toward Class\,I and MS/PMS binaries. 

In short, we find that approximately two-thirds of Class\,0 protostars are binary or multiple systems in this SMA survey, 
and this could only be a lower limit to the fraction of Class\,0 protostars formed in such systems. The results derived 
from this work, in concert with the comparison between this survey and large surveys toward Class\,I and MS/PMS 
binaries, is consistent with the possibility that most, if not all, stars are formed in binary or multiple systems and that 
single stars result from the decay of multiple systems, as suggested by Larson (1972). 

\subsection{The Distribution of Mass Ratios}

We have derived the circumstellar gas masses for the sources in the sample using the (sub-)\,millimeter
dust continuum emission (see Section\,3 and Tables 4 \& 5). We note that these values are uncertain and 
generally cannot be compared among the sources in the sample, because different sources were observed 
differently (i.e., with different angular resolution, $uv$-coverage, and different resulting sensitivity), which 
probe different size scales and amounts of material. 
Relative masses between components within individual systems are more reliable. Assuming, for simplicity, 
that sources in a multiple system have similar temperatures, then we can obtain the circumstellar mass ratio 
from the continuum emission flux ratio of the sources (see Equations~1-3).
Table~6 lists the circumstellar mass ratios of the binary/multiple systems observed in this work, under the 
assumption that the circumstellar mass is proportional to the (sub)\,millimeter continuum flux. It must be 
noted that this circumstellar mass ratio does not necessarily reflect the mass ratio of the hydrostatic cores 
nor that of the final stars, but still provides a valuable comparison between the components in these systems. 
Figure~21 shows the distribution of circumstellar mass ratios for the protobinary systems in this work. 
The distribution appears to be flat like that of more evolved long-period PMS and MS binary stars (see,
e.g., Reipurth \& Zinnecker 1993; Halbwachs et al. 2003). We find that 67\%\,$\pm$\,8\% of protobinary 
systems have circumstellar mass ratios below 0.5, i.e., unequal masses systems are more common than 
equal masses systems in the sample. This result suggests that unequal-mass systems may be preferred 
in the process of (wide) binary star formation (separations on the order of 1000\,AU).

Figure~22 shows the distribution of circumstellar mass ratios versus separations. The distribution is 
somewhat scattered, similar to the distribution of PMS binaries (see, e.g., Reipurth \& Zinnecker 1993). 
Numerical simulations predict that close binary systems (separations on the order of 1\,AU) are likely to 
have mass ratios near unity (see Bate 2000), which is indirectly supported by statistical studies of MS 
binary systems (see, e.g., Halbwachs et al. 2003). However, we do not find this trend in Figure~22. It is 
probably because (1) the sample size in this work is still small, and (2) we cannot resolve close binary 
systems in which components may prefer similar masses.

\subsection{Sequential Fragmentation in Star Formation}

Different physical processes at different evolutionary stages may trigger different fragmentation processes (e.g., 
Tohline 2002; Machida et al. 2008). At different evolutionary stages, the varying properties of a collapsing core
(e.g., core temperature and density) will result in different fragmentation scales. As introduced in Section~I, 
depending on the evolution of a collapsing core, fragmentation models can be divided into initial fragmentation, 
prompt (isothermal) fragmentation, adiabatic fragmentation and secondary fragmentation. Suggested by these
models, the fragmentation of a core could take place sequentially during the core's different collapsing phases,
which would result in a hierarchical system eventually. However, it is yet unclear whether the fragmentation at 
one specific phase is dominant over other phases, and whether most collapsing cores go though all these 
fragmentation processes.
Based on the extensive single-dish and interferometric observations, we show in Figure~23 a suggested sequential 
fragmentation picture. This is a rough outline and further efforts are needed to collect corresponding samples and 
to derive their kinematical properties, in order to understand the fragmentation mechanisms at different core's collapse 
phases in detail.

The initial fragmentation of a large-scale clump or filament ($\sim$\,0.1--1.0\,pc) occurs before the collapse of individual
prestellar cores, and results in cores with separations of about 10$^3$--10$^4$\,AU. This large-scale fragmentation 
has been frequently observed with single-dish millimeter telescopes  (see, e.g., Kauffmann et al. 2008; Launhardt et al. 
2010), and it is likely controlled by the combined effects of gravity and turbulence. Figure~23 shows a possible example 
of the initial clump fragmentation, CB\,246 (see Launhardt et al. 2010 for more details), where the extended large-scale 
clump has fragmented into separated cores. These cores are invisible in the {\it Spitzer Space Telescope} ($Spitzer$) 
infrared images and may represent the prestellar cores which will further collapse and fragment. 

Most of the protobinary systems that have been observed with current millimeter arrays have separations in scales 
of 1000\,AU (e.g., Looney et al. 2000, and this work). These systems appear to support the scenario of prompt 
fragmentation, which occurs at the end of the isothermal collapse phase (see Tohline 2002). Nevertheless, these 
observed systems are mostly made of Class\,0 objects that have already gone through the fragmentation process, 
and thus these systems do not exhibit the initial conditions of a fragmenting core. An observational example representing 
the prompt fragmentation may be the prestellar core R\,CrA~SMM\,1A (see Figure~23), in which multiple faint condensations 
were discovered with separations between 1000\,AU and 2000\,AU. These condensations are invisible in the deep 
infrared images, have extremely low bolometric luminosities ($<$ 0.1\,$L_{\sun}$) and temperatures ($<$ 20\,K), indicating 
that these are young sources have yet to form protostars, and therefore represent the earliest phase of core fragmentation 
observed (see Chen \& Arce 2010 for more details). More recently, based on the kinematic data collected at different 
interferometric arrays, Chen et al. (2012a) found that most protostellar cores with binary systems (separations scale 
$\sim$\,1000\,AU) formed therein have ratios of rotational energy to gravitational energy $\beta$$_{\rm rot}$ $>$ 1\%, 
which is consistent with theoretical simulations (see, e.g., Boss 1999). This suggests that the level of rotational energy in a 
dense core plays an important role in the prompt fragmentation process.

Unfortunately, there is a lack of observations of protobinary systems with separations between 10\,AU and 100\,AU, which 
could be used to study the adiabatic fragmentation scenario (see Machida et al. 2008). This can be easily explained by the 
fact that current millimeter interferometers (e.g., SMA and PdBI) normally reach 1--2$''$ angular resolution under general 
conditions, and thus mostly resolve protobinary systems with separations of 100--200\,AU or larger in nearby star-forming 
clouds (e.g., Perseus and Ophuichus). In Figure~23, we show the eSMA 850\,$\mu$m image of source IRAS\,16293A, which 
represents the best angular resolution that can be achieved at the SMA. In IRAS\,16293\,A, a close binary system with a 
separation of 50\,AU is revealed (see also Appendix~A.19). We speculate that the close binary system in IRAS\,16293A was 
formed through adiabatic fragmentation, which took place after the initial clump fragmentation (that formed the IRAS\,16293 
and IRAS\,16293\,E cores; see, e.g., Figure~8 in J{\o}rgensen et al. 2008) and the prompt (isothermal) fragmentation (that 
formed the IRAS\,16293 A-B system; see Figure~15d). We note that the protobinary system VLA\,1623 (see Appendix~A.17 
and Figures~15a-b) may show another example of adiabatic fragmentation. Here, the VLA\,1623 system appears to have 
fragmented from the Oph-A core through initial clump fragmentation and have later fragmented into VLA\,1623 West \& East 
(through prompt fragmentation); VLA1623 East seems to have subsequently gone through adiabatic fragmentation to form 
VLA\,1623 East A and B. As we discussed in Appendix~A.17, source VLA\,1623\,B could also be a binary system, which needs 
to be verified by further higher angular resolution observations. 

We believe that routine observations at angular resolution better than 0\farcs1 will reveal more multiple protobinary systems 
with smaller separations, which will then allow us to study in detail a statistically significant number of protostellar binary/multipe 
systems with a wide range of separations (from 1\,AU to 10$^4$\,AU). The Atacama Large Millimeter/submillimeter Array (ALMA), 
capable of attaining an angular resolution down to 10\,mas (when fully completed), will provide a breakthrough in our knowledge 
of binary star formation. It will certainly help us find closer protobinary systems with separations in the scale of 1--10\,AU, and 
will even allow us to test the secondary fragmentation scenario (see, e.g., Machida et al. 2008), by which stellar systems with 
separations of 0.01--0.1\,AU can be formed.

\section{SUMMARY}

We present SMA 1.3\,mm and 850\,$\mu$m dust continuum data toward a sample of 33 Class\,0 protostars in nearby 
clouds (distance $<$\,500\,pc), which thus far represents the largest survey toward protostellar binary/multiple systems. 
The median angular resolution in the survey is 2\farcs5, while the median linear resolution is approximately 600\,AU. 
The main results of this work are summarized below.

(1) Compact millimeter and/or submillimeter dust continuum emission is detected from all sources in the sample, which 
likely originates from the inner envelope and/or circumstellar disk. The measured 1.3\,mm fluxes in the sample range 
from 1.1\,mJy to 3.0\,Jy, with a median value of 100\,mJy, while the 850\,$\mu$m fluxes range from 20\,mJy to 3.2\,Jy, 
with a median value of 430\,mJy. Assuming that the 1.3\,mm and 850\,$\mu$m dust continuum emission is optically thin
and a dust temperature of 30\,K for all sources, the estimates of the total gas masses are in the ranges of 
0.001$-$2.4\,$M_\odot$ at $\lambda$\,1.3\,mm and of 0.005$-$1.7\,$M_\odot$ at $\lambda$\,850\,$\mu$m.

(2) Twenty-one sources in the sample show signatures of binarity/multiplicity, with separations ranging from 50\,AU to
5000\,AU. Of them, four sources are newly-discovered binary/multiple candidates (PER\,065, NCG\,1333 IRAS\,2A, 
IC\,348 MMS, and HH\,25\,SMM). Quantitatively, the numbers of singles, binaries, triples, and quadruples in the sample 
are 12, 14, 5, and 2, respectively. The estimated multiplicity frequency and companion star fraction are 0.64\,$\pm$\,0.08 
and 0.91\,$\pm$\,0.05, respectively, with no correction for completeness. We consider these values as lower limits, as 
it is likely that there are unresolved binary/multiple systems with small ($<$ 600\,AU) separations in this survey.

(3) The derived multiplicity frequency and companion star fraction for Class\,0 protostars are approximately two times 
higher than those values found in the binary surveys toward Class\,I YSOs and pre-main sequence stars, and approximately 
three (for multiplicity frequency) and four (for companion star fraction) times larger than the values found among main 
sequence stars, with a similar range of separations. Furthermore, the observed fraction of high order multiple systems to 
binary systems in Class\,0 protostars (0.50\,$\pm$\,0.09) is also larger than the fractions found in Class\,I YSOs 
(0.31\,$\pm$\,0.07) and main sequence stars ($\leq$\,0.2). The decrease in both the multiplicity frequency/companion 
star fraction and the fraction of high order multiple systems clearly suggests that binary properties evolve as protostars 
evolve, as predicted by numerical simulations.

(4) The distribution of separations for Class\,0 protostars shows a general trend in which the companion star fraction increases 
with decreasing companion separation. Quantitatively, the companion star fraction is 0.42\,$\pm$\,0.09 for separations from 
50\,AU to 1700\,AU, 0.27\,$\pm$\,0.08 from 1700\,AU to 3350\,AU, and 0.21\,$\pm$\,0.07 from 3350\,AU to 5000\,AU, with no
correction for completeness. The median separation in this survey is 1800\,AU (larger than the lowest linear resolution), and 
thus the survey is regarded to be complete for protobinary systems with separations from 1800\,AU from 5000\,AU (over this
separation range the multiplicity frequency and companion star fraction are 0.33\,$\pm$\,0.08 and 0.42\,$\pm$\,0.08, respectively). 
After correcting for incompleteness for separations less than 1800\,AU with a simple method, the overall multiplicity frequency 
and companion star fraction in this survey are 1.0\,$\pm$\,0.1 and 1.5\,$\pm$\,0.2, respectively, over the separation range from 
50\,AU to 5000\,AU.

(5) The distribution of circumstellar mass ratios for the protobinary systems in this survey appears to be flat, like that of 
more evolved long-period main-sequence and pre-main sequence binary stars. We find that 67\%\,$\pm$\,8\% of the 
protobinary systems have circumstellar mass ratios below 0.5 (i.e., unequal masses are much more common than equal 
masses). This implies that unequal-mass systems are preferred in the process of binary star formation (with separations in
the scale of 1000\,AU).

(6) The high angular resolution survey in this work, in concert with previous extensive single-dish observations, strongly 
support that the fragmentation of molecular cloud cores, at different core collapse phases, is the main mechanism for the 
formation of binary/multiple stars. Based on these observations, we suggest an empirical sequential fragmentation picture 
for binary star formation.

\acknowledgments

We thank the anonymous referee for providing many insightful suggestions and comments, which helped us improve 
this work greatly. We thank the SMA staff for technical support during the observations and their maintenance of the 
SMA archival data. X.C. acknowledges the support of the Thousand Young Talents Program of China. This material is 
based on work supported by NSF grant AST-0845619 to H.G.A. JEP has received funding from the European Community's 
Seventh Framework Programme (/FP7/2007-2013/) under grant agreement No 229517.

\clearpage

\appendix

\section{DESCRIPTION OF INDIVIDUAL SOURCES}

In this Appendix, we provide a brief description of the individual sources, and discuss their SMA results in more detail. 
We note that, for sources L1448C, NGC\,1333 IRAS\,2A \& 2B, NGC\,1333 IRAS\,4A \& 4B, L1527, L483, B335, and 
L1157, the SMA results (dust continuum and molecular lines) were presented by J{\o}rgensen et al. (2007). The dust 
continuum results (images and fluxes) derived independently in this work are comparable with those in J{\o}rgensen 
et al. (2007). We refer readers to that paper for more details on the line data of these sources. Other sources for which 
their SMA data have been published elsewhere include: HH\,211~MMS (Lee et al. 2007b; 2009; 2010), NGC\,2024-FIR5 
(Alves et al. 2011), HH\,212~MMS (Lee et al. 2006; 2007a; 2008), L723 VLA2 (Girart et al. 2009), and L1251B (J.~Lee 
et al. 2006; 2007).

\subsection{L1448 Region}

L1448 is a dark globule located on the west end of the Perseus cloud, which has been observed extensively 
in the past two decades (see, e.g., Bally et al. 2008, and references therein). Figure~3a shows the SCUBA 
850\,$\mu$m image of the L1448 region. The region contains three well-known Class\,0 protostars: L1448 
IRS2, L1448\,N, and L1448C (see, e.g., Barsony et al. 1998; O'Linger et al. 2006).

{\bf L1448 IRS2} was suggested to be a protobinary candidate (with a 10$''$ separation) by Wolf-Chase et al. 
(2000) based on the NRAO-12\,m CO\,(1--0) observations that mapped two distinct molecular outflows from 
IRS2. Volgenau et al. (2002) also claim the detection of a binary system in IRS2 using BIMA, but no results 
(e.g., spatial separation or image) have been presented yet. However, the SMA 1.3\,mm dust continuum 
observations with an angular resolution of 3\farcs4\,$\times$\,2\farcs5 (see Figure~3c), as well as $Spitzer$ 
infrared observations (e.g., Chen et al. 2010), do not find any companions around IRS2 (see also O'Linger 
et al. 2006). We therefore conclude that L1448 IRS2 is a single protostar.

{\bf L1448\,N} is also referred to as L1448\,IRS3 by several authors. However, we prefer using the name L1448\,N, 
because the IRAS source IRS3 in fact consists of two protostars, L1448\,N and L1448C (see O'Linger et al. 2006). 
L1448\,N is the brightest source at infrared and millimeter wavelengths in L1448, and has been resolved into three 
distinct sources by previous interferometric observations (e.g., Looney et al. 2000). The three sources are named 
L1448\,NA, L1448\,NB, and L1448\,NW (O'Linger et al. 2006). The SMA 1.3\,mm dust continuum image of L1448\,N 
shows these three distinct sources (see Figure~3b). The dust continuum emission is dominated by source NB, which 
is located 7\farcs4\,$\pm$\,0\farcs3 south of source NA. Source L1448\,NW, located approximately 20$''$ northwest 
of the NA-NB pair, has relatively weaker dust continuum emission in the interferometric observations.

{\bf L1448C}, also known as L1448-mm, is one of the best-studied Class\,0 protostars. L1448C is well-known 
for driving a young, highly collimated, and high-velocity bipolar outflow (e.g., Bachiller et al. 1990; Hirano et al. 
2010). $Spitzer$ infrared observations resolved L1448C into two separated components, with an angular separation 
of about 8$''$ (see J{\o}rgensen et al. 2006). The SMA observations at both 850\,$\mu$m and 1.3\,mm continuum 
bands detect these two components (see Figures~3d \& 3e). The dust continuum emission is dominated by the 
northern component, and only weak dust continuum emission is found from the southern component (5\,$\sigma$ 
detection). The northern component is the driving source of the well-known high-velocity bipolar outflow (J{\o}rgensen 
et al. 2007; Hirano et al. 2010). Maury et al. (2010) argued that the southern component is not a protostar but a 
dusty clump associated with the outflow driven by the northern component. However, we note that a relatively low-velocity 
bipolar outflow associated with the southern component has been found in the SMA CO\,(2--1) observations (see 
J{\o}rgensen et al. 2007) and CO\,(3--2) observations (see Hirano et al. 2010), which clearly indicates that the 
southern component is a protostellar object. 

\subsection{PER\,065}

The protostellar core PER\,065 was discovered by the dust continuum survey in the Perseus cloud, and 
was classified as a Class\,0 protostar by Hatchell et al. (2007) and Enoch et al. (2009). Because of its low 
luminosity (0.03--0.22\,$L_\odot$), it was also cataloged as one of the very low luminosity objects (VeLLOs) 
in Perseus (No.\,065; see Dunham et al. 2008). The SMA 1.3\,mm dust continuum observations resolve, 
for the first time, the faint core into three components, named A, B, and C, respectively (see Figure~4). 
The separations between components A and B and between B and C are 3\farcs9\,$\pm$\,0\farcs4 and 
7\farcs5\,$\pm$\,0\farcs4, respectively. 

\subsection{NGC\,1333 Region}

NCG\,1333 is a well-studied protostellar cluster located in the western part of the Perseus cloud 
(see, e.g., a review by Walawender et al. 2008 and references therein, and Arce et al. 2010). 
Submillimeter and millimeter dust continuum observations have revealed several protostellar cores 
within the region (see, e.g., Sandell \& Knee 2001), of which three cores, SVS\,13, IRAS\,2, and 
IRAS\,4 are the most well-known and studied (see Figure~5a).

Figure~5b shows the SMA 1.3\,mm continuum image of {\bf SVS\,13}, in which three distinct sources,
referred to as A, B, and C, are found. The SMA results are consistent with the results found in other 
high angular resolution observations (e.g., Bachiller et al. 1998; Looney et al. 2000; Chen et al. 2009). 
Another weak continuum source, which is spatially coincident with the radio source VLA\,3 (Rodr\'{i}guez 
et al. 1997) and the 3\,mm dust continuum source detected by Looney et al. (2000) and Chen et al. 
(2009), is found in the SMA 1.3\,mm dust continuum image (see Figure~5c). Also note that, with the 
VLA centimeter observations, Anglada et al. (2000; 2004) resolved source A into a close binary system 
with an angular separation of 0\farcs3.

The SMA continuum images of {\bf IRAS\,2} and {\bf IRAS\,4} are shown in Figures~5d\,\&\,5e and 
Figures~5f\,\&\,5g, respectively. The continuum results of the two sources derived independently in 
this work are comparable with those in J{\o}rgensen et al. (2007). We therefore refer readers to that
work for more details about the SMA results. We note that IRAS\,2A is resolved into two separated
components in the high angular resolution 850\,$\mu$m image (see Figure~5e), with a faint companion 
1\farcs5\,$\pm$\,0\farcs2 to the north of the main dust peak. In the CO observations, a quadrupolar 
outflow centered at IRAS\,2A was detected (see, e.g., J{\o}rgensen et al. 2007). This quadrupolar outflow 
is likely two different outflows driven by the two binary components in IRAS\,2A. Further high angular 
resolution observations are needed to confirm the tentative detection of the faint secondary source in 
IRAS\,2A and to distinguish which source drives which outflow.

\subsection{IRAS 03282+3035}

IRAS\,03282+3035 (IRAS\,03282) is a young Class\,0 protostar located in the western part of the Perseus 
cloud, which has been studied by various groups using different molecular line transitions (see, e.g., Bachiller 
et al. 1991, 1994; Arce \& Sargent 2006; Chen et al. 2007; Tobin et al. 2011).

The OVRO 1.3\,mm dust continuum observations (at 0\farcs9\,$\times$\,0\farcs7 resolution) revealed two 
millimeter sources in IRAS\,03282, with an angular separation of 1\farcs5 (see Launhardt 2004). Although 
the SMA 1.3\,mm observations (at 2\farcs5\,$\times$\,2\farcs2) do not resolve the two components, the 
SMA 850\,$\mu$m observations (at 0\farcs9\,$\times$\,0\farcs8) show two separated components (see 
Figure~6). We refer to the main continuum source as MMS\,1 and the faint source as MMS\,2. In the SMA 
images, the angular separation between the two sources is 1\farcs6\,$\pm$\,0\farcs2, similar to the result 
found in the OVRO 1.3\,mm images. 

\subsection{Perseus B1-c and B1-b}

Barnard~1 (B1) is a dark cloud in the western part of Perseus. Single-dish dust continuum observations have 
revealed four protostellar cores in this cloud (see, e.g., Matthews \& Wilson 2002, Hatchell et al. 2005, and 
Enoch et al. 2006), which are commonly referred to as B1-a, -b, -c, and -d (see Figure~7a). 

The Class\,0 protostar {\bf B1-c} was studied with BIMA by Matthews et al. (2006). The BIMA 3.3\,mm dust 
continuum observations (at 6\farcs4\,$\times$\,4\farcs9 resolution) revealed a protostellar envelope with 
radius of $\sim$\,2400\,AU and mass of 2.1--2.9\,$M_\odot$. The SMA 1.3\,mm dust continuum observations,
with higher angular resolution (3\farcs4\,$\times$\,2\farcs9), resolve an inner envelope with radius of 
700\,$\pm$\,100\,AU and mass of 0.24\,$\pm$\,0.05\,$M_\odot$ (see Figure 7b). 

The {\bf B1-b} protostellar core is elongated in the north-south direction in the SCUBA images (see Figure~7a). 
The 3\,mm dust continuum observations at the Nobeyama Millimeter Array (NMA) found two separate sources 
in the B1-b core, referred to as B1-bN and B1-bS (Hirano et al. 1999). The SMA 1.3\,mm observations show 
similar results to those found by Hirano et al. (1999): the two sources are separated by 17\farcs4\,$\pm$\,0\farcs5 
and have similar circumstellar masses (mass ratio 0.7\,$\pm$\,0.2).

\subsection{IC\,348 MMS and HH\,211 MMS}

{\bf IC\,348~MMS} is a Class\,0 protostar discovered in the southwestern part of the IC\,348 cluster 
(see Eisl{\"o}ffel et al. 2003). It drives a collimated molecular outflow, traced by H$_2$ emission 
(Eisl{\"o}ffel et al. 2003) and CO emission (Tafalla et al. 2006).
The SMA 1.3\,mm dust continuum images of IC\,348~MMS reveal, for the first time, two sources 
embedded in the common envelope detected by single-dish dust continuum observations (see 
Figure~8b). The angular separation between the two sources is 9\farcs8\,$\pm$\,0\farcs4, and the 
mass ratio is 0.25\,$\pm$\,0.10. The main continuum source MMS\,1 is the driving source of the 
collimated molecular outflow. 

The Class\,0 protostar {\bf HH\,211~MMS} is well-known for driving a highly collimated bipolar outflow, 
which has a narrow bipolar jet-like component in the high-velocity CO/SiO emission and a relatively 
wide-angle bipolar cavity in the low-velocity CO emission (see, e.g., Gueth \& Guilloteau 1999; Hirano 
et al. 2006; Lee et al. 2007b). 

Based on very high angular resolution SMA observations (at 0\farcs20\,$\times$\,0\farcs15), Lee et al. 
(2009) found that HH\,211 MMS actually contains two dust continuum sources, SMM\,1 and SMM\,2, 
with an angular separation of 0\farcs31\,$\pm$\,0\farcs02 (see Figure~8c). 
Source SMM\,1 appears to be the driving source of the collimated jet/outflow, while the nature of source 
SMM\,2, which has a mass of only 1.5--4\,$M_{\rm Jup}$, is still uncertain. More interestingly, based on 
the reflection-symmetric wiggle of the HH\,211 jet, Lee et al. (2010) suggested that source SMM\,1 itself 
could be a very low mass protobinary with a separation of $\sim$\,4.6\,AU (the putative 4.6\,AU companion 
was not directly detected, but inferred from the morphology of the outflow). Further higher angular resolution 
observations are needed to confirm this interesting tentative detection. More details about HH\,211 MMS 
and its surrounding envelope can be found in Gueth \& Guilloteau (1999), Lee et al. (2007b; 2009; 2010), 
and Tanner \& Arce (2011).

\subsection{CB\,17}

CB\,17 (also known as L1389) is a small, isolated, and slightly cometary-shaped dark cloud, 
which was classified as a Bok globule by Clemens \& Barvainis (1988). CB\,17 has been studied 
by various groups using different molecular line transitions (e.g.,Pavlyuchenkov et al. 2006) and 
multi-wavelenghth dust continuum emission (e.g., Launhardt et al. 2010).

The SMA 1.3\,mm dust continuum observations reveal two sources within CB\,17, which are 
separated by 21\farcs4\,$\pm$\,0\farcs5 in the northwest-southeast direction (see Figure~9a). 
The northwestern dust continuum source is associated with an infrared source seen in the 
$Spitzer$ images and is referred to as CB\,17\,IRS (see Chen et al. 2012b). No compact infrared 
emission is detected from the southeastern source in the $Spitzer$ bands from 3.6 to 70\,$\mu$m, 
and we refer to this source as CB\,17\,MMS. Interestingly, the SMA CO\,(2--1) observations suggest 
that source CB\,17\,MMS, an object with an extremely low bolometric luminosity ($\leq$\,0.04\,$L_\odot$), 
is driving a low velocity ($\sim$\,2.5\,km\,s$^{-1}$) molecular outflow. These characteristics have lent 
this source to be classified as a candidate of the so-called first hydrostatic cores (see Chen et al. 2012b 
for more details).

\subsection{IRAM\,04191+1522}

IRAM\,04191+1522 (IRAM\,04191) is a well-studied Class\,0 protostar located in the southern 
part of the Taurus molecular cloud (Andr{\'e} et al. 1999). IRAM\,04191 drives a large bipolar molecular 
outflow, features bright molecular line and dust continuum emission, and shows evidences of supersonic 
infall, fast rotation, depletion, and deuteration (see, e.g., Andr\'{e} et al. 1999; Belloche et al. 2002; 
Belloche \& Andr\'{e} 2004; Lee et al. 2005). 

The SMA 1.3\,mm dust continuum observations reveal two distinct continuum sources with an angular 
separation of 8\farcs0\,$\pm$\,0\farcs4, embedded within the elongated core of IRAM\,04191 (see 
Figure~9b). The discovery of this binary system was reported by Chen et al. (2012a). The southeastern 
source, associated with an infrared source seen in the $Spitzer$ images (see Dunham et al. 2006), 
represents the well-known Class\,0 protostar and is referred to as IRAM\,04191\,IRS here. In contrast, 
the northwestern source has no infrared emission detected in the $Spitzer$ images, and is referred to 
as IRAM\,04191\,MMS (see Chen et al. 2012a for more details).

\subsection{L1521F-IRS and L1527-IRS}

L1521F is a well-studied dense core in Taurus. This core shows high central density, infall asymmetry, 
molecular depletion, and enhanced deuterium fractionation, and was suggested as a highly evolved 
starless core by Crapsi et al. (2004). However, $Spitzer$ observations found that L1521F harbors a very 
low luminosity protostar, refereed to as {\bf L1521F-IRS} (see Bourke et al. 2006), which is in the early
stage of the Class\,0 phase (Shinnaga et al. 2009; Terebey et al. 2009).
The SMA 1.3\,mm dust continuum observations of L1521F-IRS show a faint source (see Figure~9c). More 
than 95\% of the flux around L1521F-IRS is resolved out by the SMA observations, compared to the flux 
detected in the IRAM-30m 1.2\,mm maps (600\,$\pm$\,150\,mJy; Bourke et al. 2006).
We also note that L1521F-IRS was observed at higher angular resolution with the IRAM-PdBI by Maury 
et al. (2010). In the PdBI 1.3\,mm dust continuum images, there is also only one object detected. 

{\bf L1527-IRS} is another protostar in Taurus, which is associated with a far-infrared source and is probably 
more evolved than L1521F-IRS (see, e.g., Terebey et al. 2009). Based on the JCMT 800\,$\mu$m continuum 
observations, Fuller et al. (1996) suggested L1527-IRS is a binary system with the secondary at about 20$''$ 
from the main protostar. However, no evidence is found of this secondary source in the SMA 1.3\,mm dust 
continuum image (see Figure~9d). We therefore conclude that L1527-IRS is a single protostar (see also 
Maury et al. 2010).
 
\subsection{HH\,114 MMS}

HH\,114 MMS is a Class\,0 protostar located in the western part of the $\lambda$ Orionis molecular shell,
which drives a highly collimated bipolar outflow (see Arce \& Sargent 2006). In the SMA 1.3\,mm and 
880\,$\mu$m dust continuum images (see Figure~10), source HH\,114 MMS remains as a single object at 
the angular resolution of 2\farcs2\,$\times$\,1\farcs9, which is consistent with the OVRO 2.7\,mm and 3.4\,mm 
continuum observations (Arce \& Sargent 2006). 

\subsection{OMC3 MMS\,6}

The MMS\,6 core is the brightest source in submillimeter and millimeter wavelengths in the Orion 
molecular cloud-3 (OMC-3) region (Chini et al. 1997; Lis et al. 1998; Johnstone \& Bally 1999). The 
SMA 850\,$\mu$m continuum images show two sources in the MMS\,6 core, referred to as SMM\,1 
and SMM\,2, with an angular separation of 10\farcs8\,$\pm$\,0\farcs2 (see Figure~11a).  
The northern continuum source SMM\,1 is spatially coincident with a continuum source detected in 
early interferometric observations (see Matthews et al. 2005; Takahashi et al. 2009). The southern 
continuum source SMM\,2 is associated with a continuum source detected in the BIMA 1.3\,mm 
observations by Matthews et al. (2005), but it was not detected in the multi-wavelength observations 
by Takahashi et al. (2009). We note that source MMS\,6 was also observed in the SMA 0.9\,mm 
continuum by Takahashi et al. (2009), but the sensitivity in their image was about 2 times worse than 
that in the SMA 850\,$\mu$m image presented here. 

\subsection{NGC\,2024 FIR\,5}

FIR\,5 is a Class\,0 protostar located in the NGC\,2024 cloud, the most active star-forming region 
in the Orion B cloud (see, e.g., a review by Meyer et al. 2008 and references therein). With a high
velocity, collimated molecular outflow extending over 5$'$ south of the core (Richer et al. 1992), 
FIR\,5 is the brightest and also probably most evolved protostellar core among the dense cores in 
NGC\,2024 (Mezger et al. 1992). 

In the SMA 850\,$\mu$m continuum images, the FIR\,5 core is resolved into two main components,
referred to as SMM\,1 and SMM\,2, with an angular separation of 4\farcs1\,$\pm$\,0\farcs2 (see 
Figure~11b). The SMA results are consistent with the results derived from the PdBI 3\,mm continuum 
observations, in which the FIR\,5 core was also resolved into two main components (see Wiesemeyer 
et al. 1997). Nevertheless, in the BIMA 1.3\,mm deep observations, Lai et al. (2002) also found several 
weaker dust condensations around the two main components, which are not detected in the SMA images. 
We note that the SMA results of FIR\,5 (including also dust polarization and CO emission results) were 
independently published by Alves et al. (2011), and similar dust continuum results were derived there.

\subsection{HH\,212 MMS}

HH\,212 is a remarkable jet in the Orion L1630 cloud, discovered by Zinnecker et al. (1998). The jet is 
associated with a bipolar molecular outflow seen in CO lines (see, e.g., Lee et al. 2000; 2006; 2007a) 
and SiO lines (see, e.g., Codella et al. 2007; Lee et al. 2008). The driving source of the jet/outflow, 
referred to as HH\,212\,MMS, is a low-luminosity Class\,0 protostar associated with a cold IRAS source 
(IRAS\,05413-0104; Zinnecker et al 1992). 

The SMA 1.3\,mm and 850\,$\mu$m dust continuum images of HH\,212 MMS were shown by Lee 
et al. (2006; 2007a; 2008). By re-analyzing the high angular resolution 850\,$\mu$m images in Lee 
et al. (2008; angular resolution 0\farcs35\,$\times$\,0\farcs32), a faint continuum component is 
found to the west of source HH\,212\,MMS (see Figure~12c). Hereafter, we refer to the main dust 
continuum component as MMS\,1 and to the faint continuum component as MMS\,2. The separation 
between the two components is measured to be 0\farcs53\,$\pm$\,0\farcs05.
We also retrieved the SMA 230\,GHz data of HH\,212 MMS obtained in the Extended and Very 
Extended configurations (the 230\,GHz data published by Lee et al. 2006 were obtained only 
in the Compact configuration). The high angular resolution 1.3\,mm dust continuum images (angular 
resolution 0\farcs8\,$\times$\,0\farcs7) show that source HH\,212 MMS is extended in the east-west 
direction and a component is tentatively detected in the 1.3\,mm images, which is spatially associated 
with the new component MMS\,2 seen in the 850\,$\mu$m images (see Figure~12d).

We also note that another dust continuum source was tentatively detected in the PdBI 1.4\,mm continuum 
observations, which is about 1\farcs7 to the southeast of the main dust continuum source MMS\,1 (see 
Codella et al. 2007). In the SMA 850\,$\mu$m images, a weak dust continuum source ($\sim$\,7\,$\sigma$ 
detection) is seen near this PdBI 1.4\,mm source but 0\farcs6 closer to the main continuum source MMS\,1. 
We referred to this weak continuum source as MMS\,3 (see Figure~12c). Altogether, there may be three 
dust continuum sources found within the elongated envelope of HH\,212 MMS. Further high angular 
observations are needed to confirm this tentative multiple system detection. 

\subsection{HH\,25\,MMS}

The protostellar jet HH\,25 is located in the northern part of the Orion~B dark cloud. The jet is associated 
with a compact but collimated molecular outflow (e.g., Gibb \& Davis 1998; Gibb et al. 2004). The driving 
source of the jet/outflow, referred to as HH\,25\,MMS, was revealed by a VLA 3.6\,cm survey in this region 
(Bontemps et al. 1995), and was classified as a Class\,0 protostar with a luminosity of $\sim$\,6\,$L_\odot$ 
(Gibb \& Davis 1998).

In the single-dish dust continuum maps (IRAM-30m, Lis et al. 1999; JCMT/SCUBA, Phillips et al. 2001), 
HH\,25\,MMS is elongated in the north-south direction, with the VLA source located at the northern tip of the 
extension. In the SMA 870\,$\mu$m dust continuum image, three distinct, previously unknown sources are 
revealed within the large-scale elongated core, approximately aligned in the north-south direction (see 
Figure~13a). The three continuum sources, from north to south, are referred to as SMM\,1, SMM\,2, and 
SMM\,3. The angular separations between sources SMM\,1 and SMM\,2 and between sources SMM\,2 and 
SMM\,3 are measured to be 12\farcs7\,$\pm$\,0\farcs3 and 10\farcs9\,$\pm$\,0\farcs3, respectively. Source 
SMM\,1 is spatially coincident with the VLA 3.6\,cm source, and is the driving source of the HH\,25 jet/outflow.

\subsection{CG\,30}

CG\,30 (also known as BHR\,12; Bourke et al. 1995) is a bright-rimmed cometary globule located in the 
Gum Nebula region. In the SCUBA submillimeter (Henning et al. 2001) and ATCA 3\,mm (Chen et al. 
2008a) continuum observations, this globule was resolved into a wide protobinary system with an angular 
separation of $\sim$\,22$''$. 

The SMA 1.3\,mm dust continuum image of CG\,30 shows two compact sources (see Figure~13b). 
The angular separation between the two sources is measured to be 21\farcs8\,$\pm$\,0\farcs5, 
consistent with the ATCA result. Following Chen et al. (2008a), we refer to the northern source as 
CG\,30N (which is spatially coincident with IRAS\,08076$-$3556) and to the southern source as 
CG\,30S. The two sources each drive their own bipolar protostellar jets/outflows, which are roughly 
perpendicular to each other (see Chen et al. 2008a, b). Further discussion of the gas kinematics 
(from ATCA N$_2$H$^+$ and SMA CO observations) and infrared emission (from $Spitzer$ 
observations) in CG\,30 can be found in Chen et al. (2008a, b).

\subsection{B228}

B228 is a Class\,0 protostar located in the Lupus\,I cloud (Shirley et al. 2000). It drives a compact bipolar 
molecular outflow that extends in the northeast-southwest direction (see van~Kempen et al. 2009). In the 
SMA dust continuum images, an elongated source is seen at both 1.3\,mm and 1.1\,mm (see Figure~14). 
The source is extended in the southwest direction, roughly following the direction of the blueshifted outflow
lobe, suggesting that the outflow has a strong impact on the dust envelope, similar to L1157 (see, e.g., 
Gueth et al. 2003). 

\subsection{VLA\,1623}

VLA\,1623 is the prototypical Class\,0 protostar (Andr{\'e} et al. 1993), which drives a large scale
bipolar outflow (Dent et al. 1995; Yu \& Chernin 1997). With VLA centimeter continuum observations,
Bontemps \& Andr{\'e} (1997) found a series of emission clumps, named A, B, and C. At high angular 
resolution, clump A was further resolved into two components (Bontemps \& Andr{\'e} 1997; Looney 
et al. 2000). 

In the SMA high angular resolution images (resolution 0\farcs6\,$\times$\,0\farcs3; observed in 2007), 
three distinct dust continuum sources are detected (see Figure~15a). The three sources, aligning 
roughly from east to west, are referred to as VLA\,1623\,A, VLA\,1623\,B, and VLA\,1623 West,
respectively. Sources A and B, the east pair with an angular separation of 1\farcs1\,$\pm$\,0\farcs1 
(see Figure~15b), are spatially coincident with the BIMA 2.7\,mm sources A and B in Looney et al. 
(2000) and the two VLA centimeter sources in clump A in Bontemps \& Andr{\'e} (1997). The mass 
ratio between sources A and B is close to 1.0\,$\pm$\,0.3, similar to the result found by Looney et al. 
(2000). Source VLA\,1623 West, located approximately 10$''$ to the east of the A-B pair, is spatially 
coincident with clump B in Bontemps \& Andr{\'e} (1997). 

More recently, Maury et al. (2012) presented high angular resolution SMA data of VLA\,1623 
(observed in 2009) and argued that sources VLA\,1623\,B and VLA\,1623\,West are two dusty 
knots due to shocks along the outflow cavity wall of source VLA\,1623\,A (although they did not 
show any evidence that VLA\,1623\,A is the driving source of the VLA\,1623 outflow). However, 
in the $Spitzer$ mid-infrared images, source VLA\,1623\,West is associated with a point-like 
infrared source, strongly suggesting that it is a young stellar object (X.~Chen et al. in preparation). 
Furthermore, the SMA CO\,(2--1) data show a quadrupolar outflow centered at source VLA\,1623\,B, 
indicating that this source is not only a protostellar object but also a potential close binary system 
unresolved in the current SMA dust continuum observations (X.~Chen et al., in preparation).

\subsection{Oph-MMS\,126 (IRAS\,16253--2429)}

Oph-MMS\,126 is a Class\,0 protostar discovered in the SEST 1.2\,mm continuum survey toward 
the $\rho$ Ophiuchi molecular cloud, which is associated with a faint, cold IRAS source 
(IRAS\,16253--2429) and drives a low-velocity bipolar molecular outflow (see Stanke et al. 2006). 
In the SMA images, a faint dust continuum source is seen (about 5\,$\sigma$ detection; see Figure~15c), 
which is associated with an infrared source seen in the $Spitzer$ images (see Barsony et al. 2010). The 
total flux was measured to be 11\,$\pm$\,2\,mJy in the SMA 1.3\,mm dust continuum images. Compared 
with the total flux obtained in the SEST observations (411\,mJy; Stanke et al. 2006), more than 97\% of 
the dust continuum emission flux from the large envelope was resolved out by the SMA.

\subsection{IRAS\,16293--2422}

IRAS\,16293--2422 (IRAS\,16293) is a well-studied Class\,0 protostar located in the $\rho$ 
Ophiuchi cloud. Millimeter interferometric observations resolved this source into a binary system,
referred to as IRAS\,16293\,A and B, with a separation of 5\farcs2 (see, e.g., Mundy et al. 1992; Loony 
et al. 2000). In the high angular resolution VLA radio continuum observations, source A is further resolved 
into a chain of radio components (see Loinard et al. 2007a; Pech et al. 2010), which are suggested to be 
related to a jet from source A. Recent ALMA Science verification data revealed an inverse P-Cygni profile 
towards source B, which is direct evidence of infalling material close to the central protostar (Pineda et al. 2012).

Figure~15d shows the SMA 1.3\,mm continuum image of IRAS\,16293, in which two continuum sources, 
A and B, are detected with an angular separation of 5\farcs3\,$\pm$\,0\farcs2. In the higher angular 
resolution eSMA 850\,$\mu$m images, source A is further resolved into two close continuum sources 
(see Figure~15e). The separation between the two sources is 0\farcs42\,$\pm$\,0\farcs05. Following 
Chandler et al. (2005), the two sources are referred to as Aa and Ab. 

\subsection{L483}

L483 is an isolated dense core, associated with a strong infrared source (IRAS\,18148$-$0440). 
It was classified as a Class\,0/I transition object by Tafalla et al. (2000). High angular resolution 
molecular line observations found a cold, quiescent core in L483 (see J{\o}rgensen 2004). In the 
SMA 850\,$\mu$m dust continuum observations, source L483 remains as single object at the angular
resolution of 2\farcs6\,$\times$\,1\farcs9 (see Figure~16), which is consistent with previous interferometric 
millimeter continuum observations (see, e.g., J{\o}rgensen 2004). 

\subsection{L723 VLA\,2}

L723 is a small isolated dark cloud. At the center of the cloud core is an IRAS point source (IRAS\,19156+1906), 
which was classified as a Class\,0 protostar with a bolometric luminosity $L_{\rm bol}$ $\sim$ 3.0\,$L{_\odot}$
and a circumstellar mass $M_{\rm env}$\,$\sim$\,1.2 $M_{\odot}$ (Shirley et al. 2000). 
Anglada et al. (1991) detected two radio continuum sources (VLA\,1 and VLA\,2) with 15$''$ separation, 
both located within the error ellipse of the IRAS position. However, only VLA\,2 was found to be associated 
with dense gas in the cloud core (Girart et al. 1997). 

Figure~17a shows the SMA 1.3\,mm dust continuum images of L723 VLA\,2, in which two separated compact 
sources (labeled MMS\,1 and MMS\,2) are revealed. The angular separation (3\farcs5\,$\pm$\,0\farcs3) and 
mass ratio (1.0\,$\pm$\,0.3) between the two sources derived in the SMA images are consistent with the results 
derived earlier in the OVRO 3\,mm continuum observations (see Launhardt 2004). The two close compact 
sources are embedded in a common envelope seen in the SCUBA 850\,$\mu$m observations. We note that the 
SMA data of L723 VLA\,2 (including molecular lines results) have been published by Girart et al. (2009). 
Also note that VLA 7\,mm continuum observations discovered two components in source MMS\,2 with an 
angular separation of $\sim$\,1$''$ (see Carrasco-Gonz{\'a}lez et al. 2008), which would be unresolved by 
the SMA observations presented here.

\subsection{B335 and L1157}

{\bf B335} is a well-studied, isolated Class\,0 protostar, and is widely accepted as one of the best 
protostellar collapse candidates (see, e.g., Zhou et al. 1993; Evans et al. 2005). It drives a large-scale 
bipolar outflow extending in the east-west direction (see, e.g., J{\o}rgensen et al. 2007). 
{\bf L1157} is also a well-studied, isolated Class\,0 protostar, and is well-known for its prominent outflow.
This outflow shows rich chemistry in various molecular lines observations (see, e.g., Bachiller et al. 2001; 
Arce et al. 2008), and is commonly considered the prototypical chemically active outflow. 
In the SMA 1.3\,mm dust continuum observations, both B335 and L1157 remain as single sources (see 
Figure~17b \& 17c), which is consistent with previous high angular resolution continuum observations 
toward the two sources (see, e.g., Harvey et al. 2003 for B335 and Chiang et al. 2010 for L1157). 

\subsection{L1251B}

L1251B is a small protostellar group located at 300\,$\pm$\,50\,pc (Kun \& Prusti 1993). It is
associated with an IRAS point source (IRAS\,22376+7455), and was found to be associated 
with a CO outflow by Sato \& Fukui (1989). Using OVRO and SMA (as well as single-dish 
telescopes), J.~Lee et al. (2006; 2007) presented multiple-line observations toward L1251B, 
which shows very complex kinematics, including infall, rotation, and outflow motions.

The SMA 1.3\,mm dust continuum observations show four distinct sources embedded in the
common envelope seen in the single-dish observations. Following J.~Lee et al. (2007), the four 
sources are named L1251B~IRS1, IRS2, SMA-N, and SMA-S (see Figure~17d). Both IRS1 and 
IRS2 are associated with the infrared sources seen in the $Spitzer$ images, and were classified 
as Class\,0/I transition objects by J.~Lee et al. (2006). Sources SMA-N and SMA-S have weak 
1.3\,mm dust continuum emission (see Figure~17d) and are not detected in the $Spitzer$ images,
and are likely still in the early stages of the Class\,0 phase (see J.~Lee et al. 2007 for more details).

\clearpage




\begin{figure*}
\begin{center}
\includegraphics[width=14cm, angle=0]{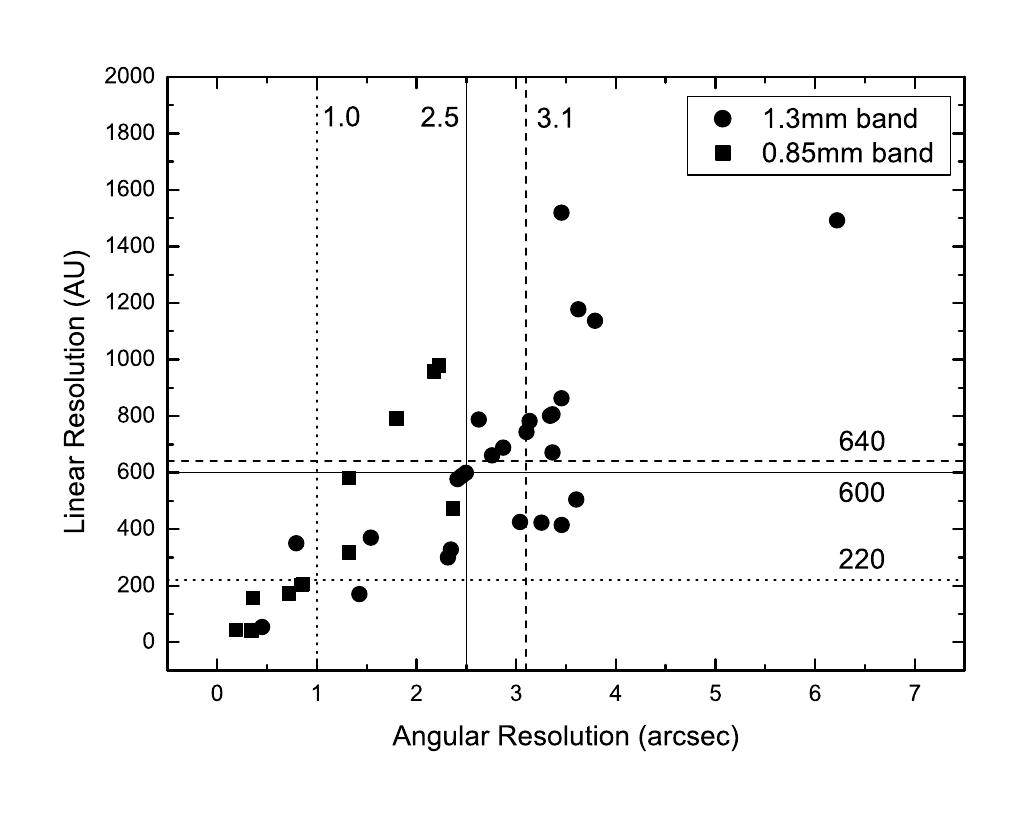}
\caption{Distribution of angular resolutions and corresponding linear resolutions in this SMA survey. 
The dashed lines show the median values in the 1.3\,mm continuum observations (3\farcs1 and 640\,AU), 
the dotted lines show the median values in the 850\,$\mu$m continuum observations (1\farcs0 and 220\,AU),
while the solid lines show the median values for all observations (2\farcs5 and 600\,AU).\label{resolution}}
\end{center}
\end{figure*}

\begin{figure*}
\begin{center}
\includegraphics[width=14cm,angle=0]{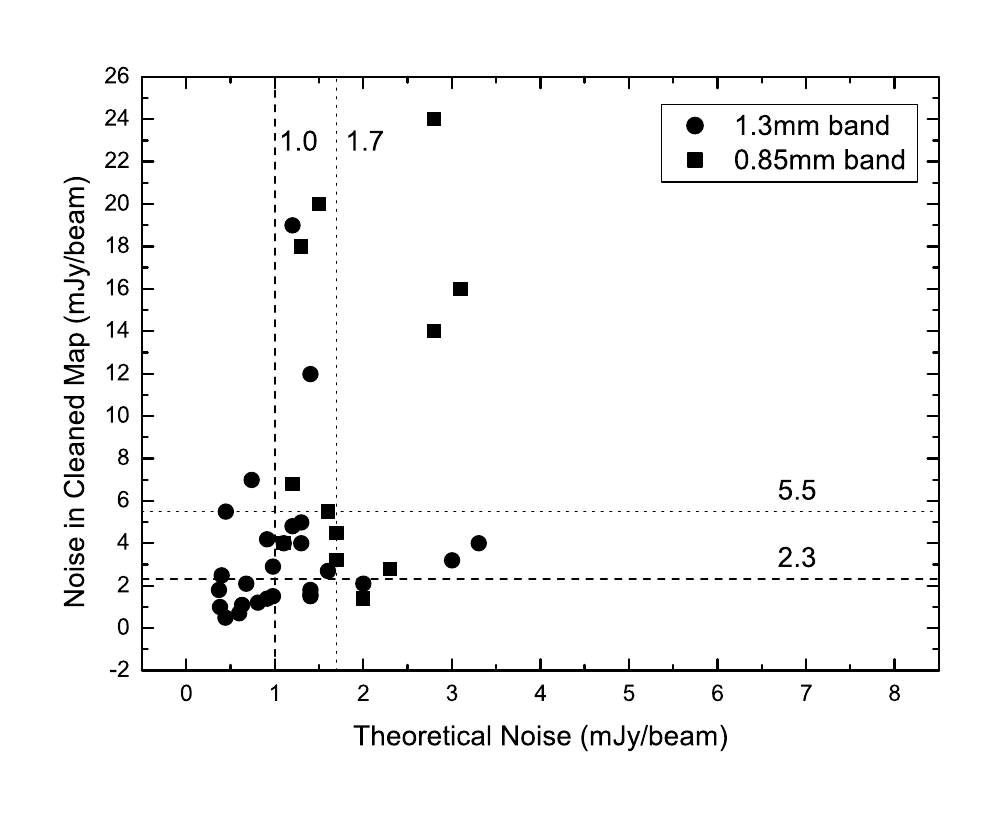}
\caption{Distribution of 1\,$\sigma$ theoretical noises and 1\,$\sigma$ measured noises in cleaned maps in this 
SMA survey. The dashed lines show the median values in the 1.3\,mm continuum observations (1.0 and 2.3\,mJy\,beam$^{-1}$), 
and the dotted lines show the median values in the 850\,$\mu$m continuum observations (1.7 and 
5.5\,mJy\,beam$^{-1}$).\label{noise}}
\end{center}
\end{figure*}

\begin{figure*}
\begin{center}
\includegraphics[width=15cm,angle=0]{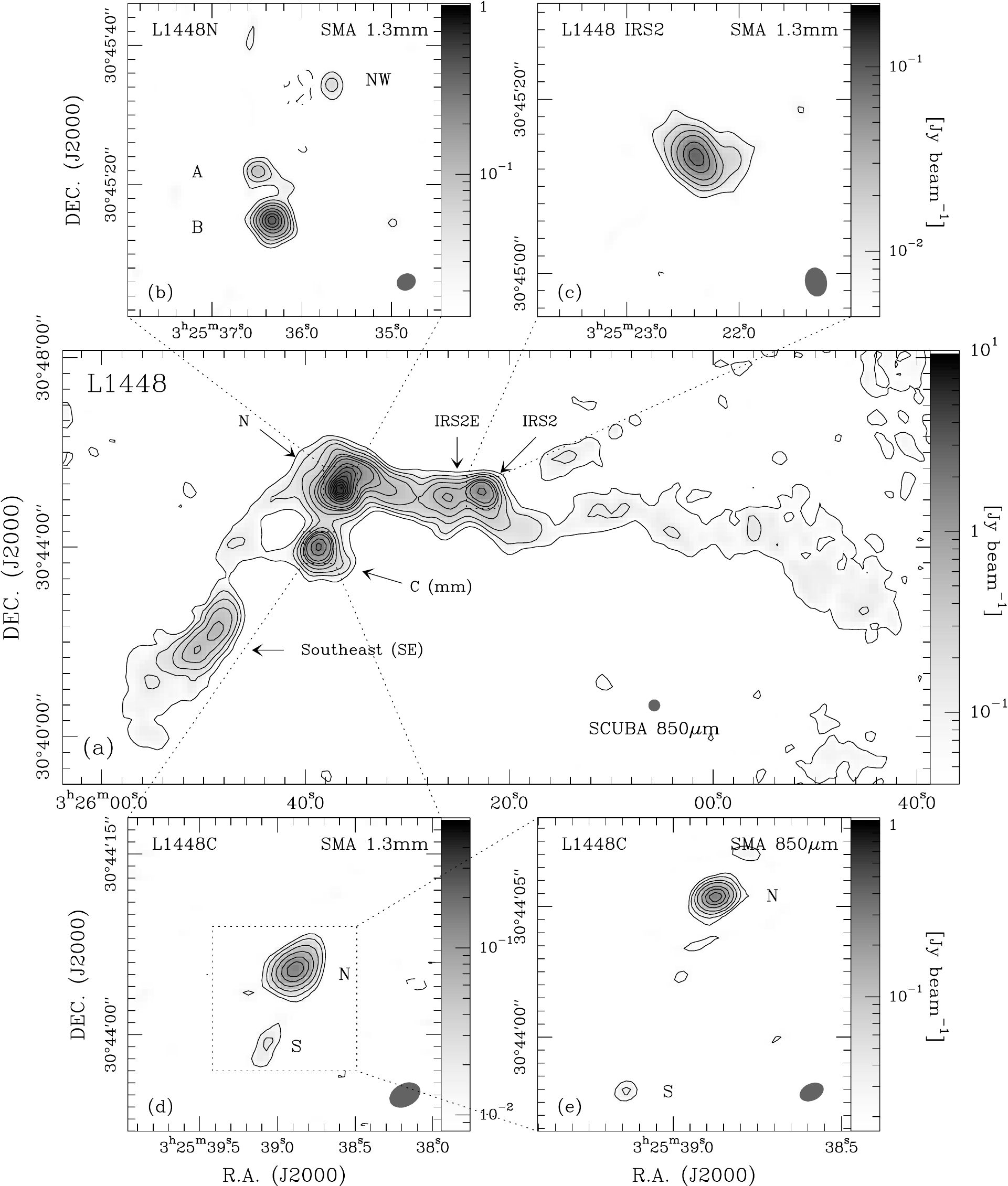}
\caption{\footnotesize (a) JCMT/SCUBA 850\,$\mu$m dust continuum image of L1448. 
Contour levels correspond to 1, 3, 5, 8, 12, 15, and 20\,$\sigma$, then increase in 
steps of 15\,$\sigma$ (1\,$\sigma$ $\sim$\,40\,mJy\,beam$^{-1}$). The SCUBA FWHM 
beam ($\sim$\,14$''$) is shown as a grey oval in the image. (b) SMA 1.3\,mm dust 
continuum image of L1448N. Contour levels correspond to $-$3, 3, 6, 10, and 15\,$\sigma$, 
then increase in steps of 10\,$\sigma$ (1\,$\sigma$ $\sim$\,7.0\,mJy\,beam$^{-1}$). (c) 
SMA 1.3\,mm dust continuum image of L1448 IRS2. Contour levels correspond to $-$3, 
3, 6, 10, and 16\,$\sigma$, then increase in steps of 10\,$\sigma$ (1\,$\sigma$ $\sim$\,2.1\,mJy\,beam$^{-1}$). 
(d) SMA 1.3\,mm dust continuum image of L1448C. Contour levels correspond to $-$3, 3, 5, 9, 
and 15\,$\sigma$, then increase in steps of 8\,$\sigma$ (1\,$\sigma$ $\sim$\,4.0\,mJy\,beam$^{-1}$). 
(e) SMA 850\,$\mu$m image of L1448C. Contour levels correspond to $-$3, 3, 6, 10, and 15\,$\sigma$, 
then increase in steps of 10\,$\sigma$ (1\,$\sigma$ $\sim$\,5.5\,mJy\,beam$^{-1}$). The synthesized 
SMA beam is shown as a grey oval in the bottom right corner of each SMA dust continuum 
image.\label{fig1_l1448_scuba_sma}}
\end{center}
\end{figure*}

\begin{figure*}
\begin{center}
\includegraphics[width=10cm,angle=0]{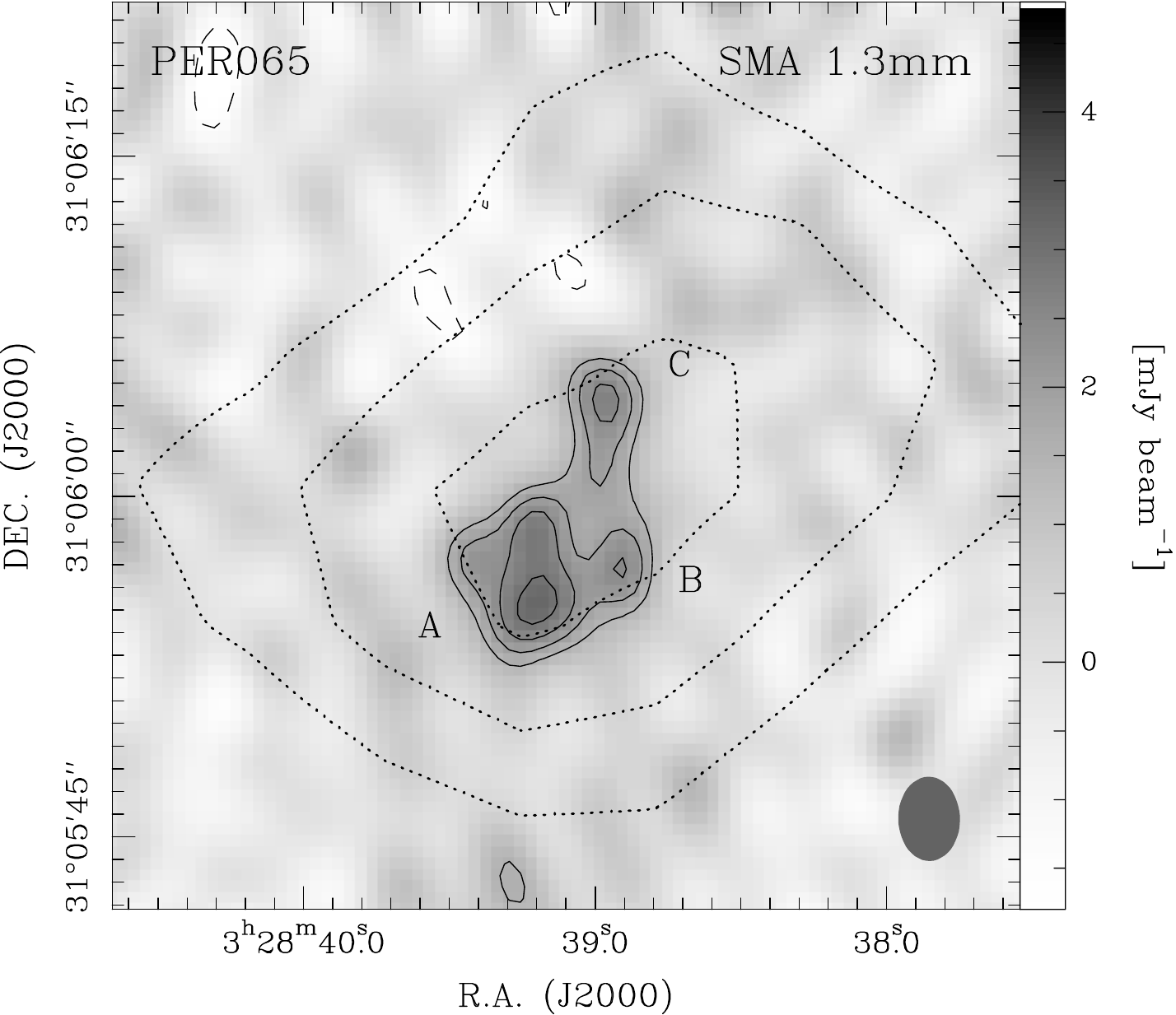}
\caption{SMA 1.3\,mm dust continuum image of source PER\,065, overlapped with the SCUBA
850\,$\mu$m contours. The SMA contours (solid lines) start from $-$3, 3\,$\sigma$, then increase 
in steps of 1\,$\sigma$ (1\,$\sigma$ $\sim$\,5.0\,mJy\,beam$^{-1}$). The synthesized SMA beam 
is shown as a grey oval in the bottom right corner. The SCUBA 850\,\micron\ contours (dotted lines) 
correspond to 75\%, 85\%, and 95\% of the peak emission ($\sim$\,0.3\,Jy\,beam$^{-1}$).\label{PER065_sma}}
\end{center}
\end{figure*}

\begin{figure*}
\begin{center}
\includegraphics[width=15cm,angle=0]{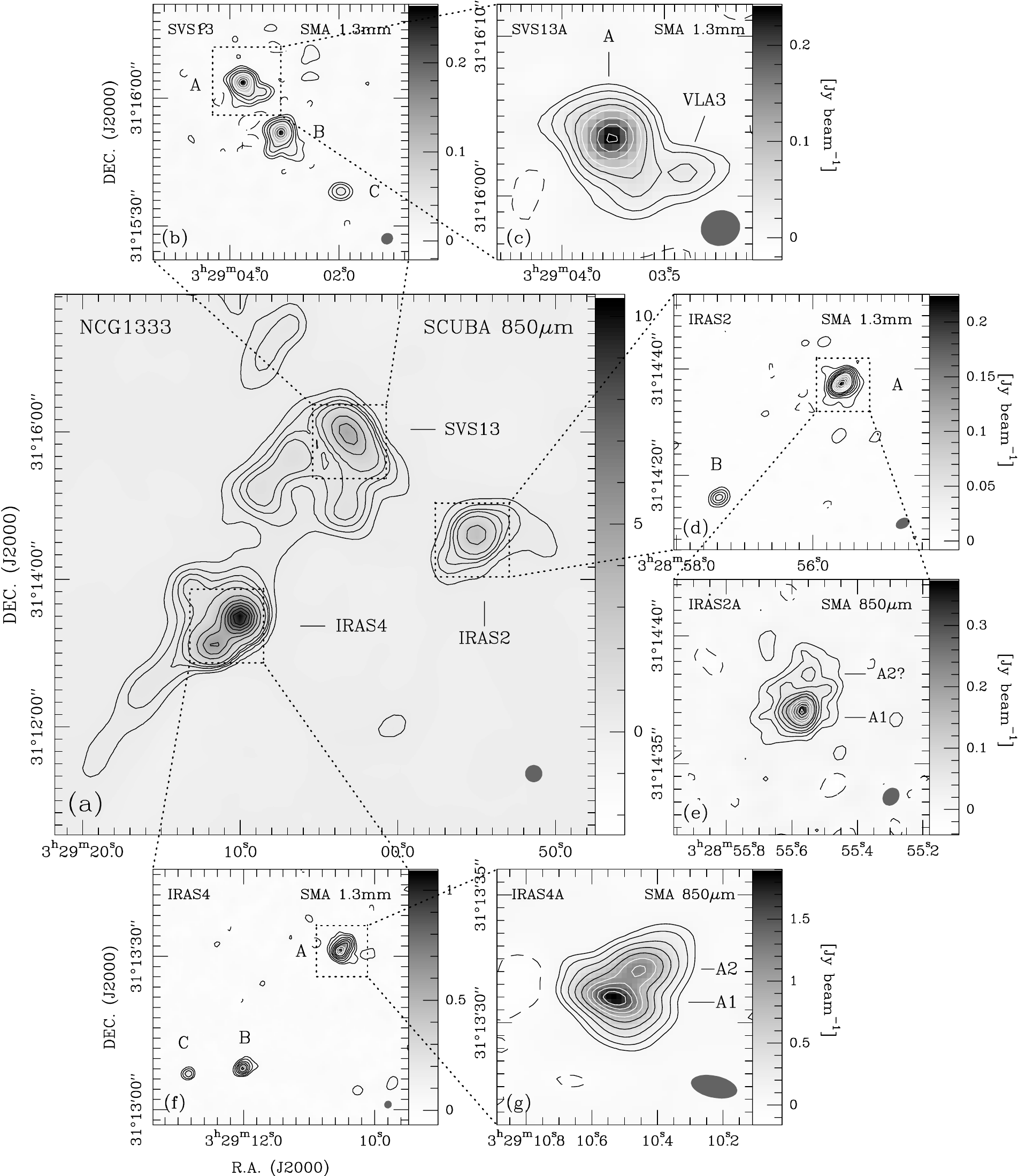}
\caption{\footnotesize (a) JCMT/SCUBA 850\,$\mu$m dust continuum image of  the NCG\,1333 central region.
Contour levels correspond to 5, 10, 15, 20 and 30\,$\sigma$, then increase in steps of 30\,$\sigma$ (1\,$\sigma$ 
$\sim$\,42\,mJy\,beam$^{-1}$). The SCUBA FWHM beam is shown as a grey circle in the bottom right corner of the 
image. (b) SMA 1.3\,mm dust continuum image of SVS\,13. Contour levels correspond to $-$3, 3, 6, 12, 20, and 30\,$\sigma$, 
then increase in steps of 20\,$\sigma$ (1\,$\sigma$ $\sim$\,2.5\,mJy\,beam$^{-1}$). (c) Enlarged view of source 
SVS\,13A. (d) SMA 1.3\,mm dust continuum image of IRAS\,2. Contour levels correspond to $-$3, 3, 6, 10, 15, 22, 30, 
and 40\,$\sigma$, then increase in steps of 15\,$\sigma$ (1\,$\sigma$ $\sim$\,2.0\,mJy\,beam$^{-1}$). (e) SMA 
850\,$\mu$m dust continuum image of IRAS\,2A. Contour levels correspond to $-$3, 3, 6, 10, 15, 22, and 
30\,$\sigma$, then increase in steps of 10\,$\sigma$ (1\,$\sigma$ $\sim$\,4.5\,mJy\,beam$^{-1}$). (f) SMA 
1.3\,mm dust continuum image of IRAS\,4. Contour levels correspond to $-$3, 3, 6, 12, 20, and 30\,$\sigma$, 
then increase in steps of 15\,$\sigma$ (1\,$\sigma$ $\sim$\,12\,mJy\,beam$^{-1}$). (g) SMA 850\,$\mu$m dust 
continuum image of IRAS\,4A. Contour levels correspond to $-$3, 3, 6, 12, and 20\,$\sigma$, then increase in 
steps of 10\,$\sigma$ (1\,$\sigma$ $\sim$\,24\,mJy\,beam$^{-1}$). The synthesized SMA beam is shown 
as a grey oval in the bottom right corner of each SMA dust continuum image.\label{NGC1333_scuba_sma}}
\end{center}
\end{figure*}

\begin{figure*}
\begin{center}
\includegraphics[width=16cm,angle=0]{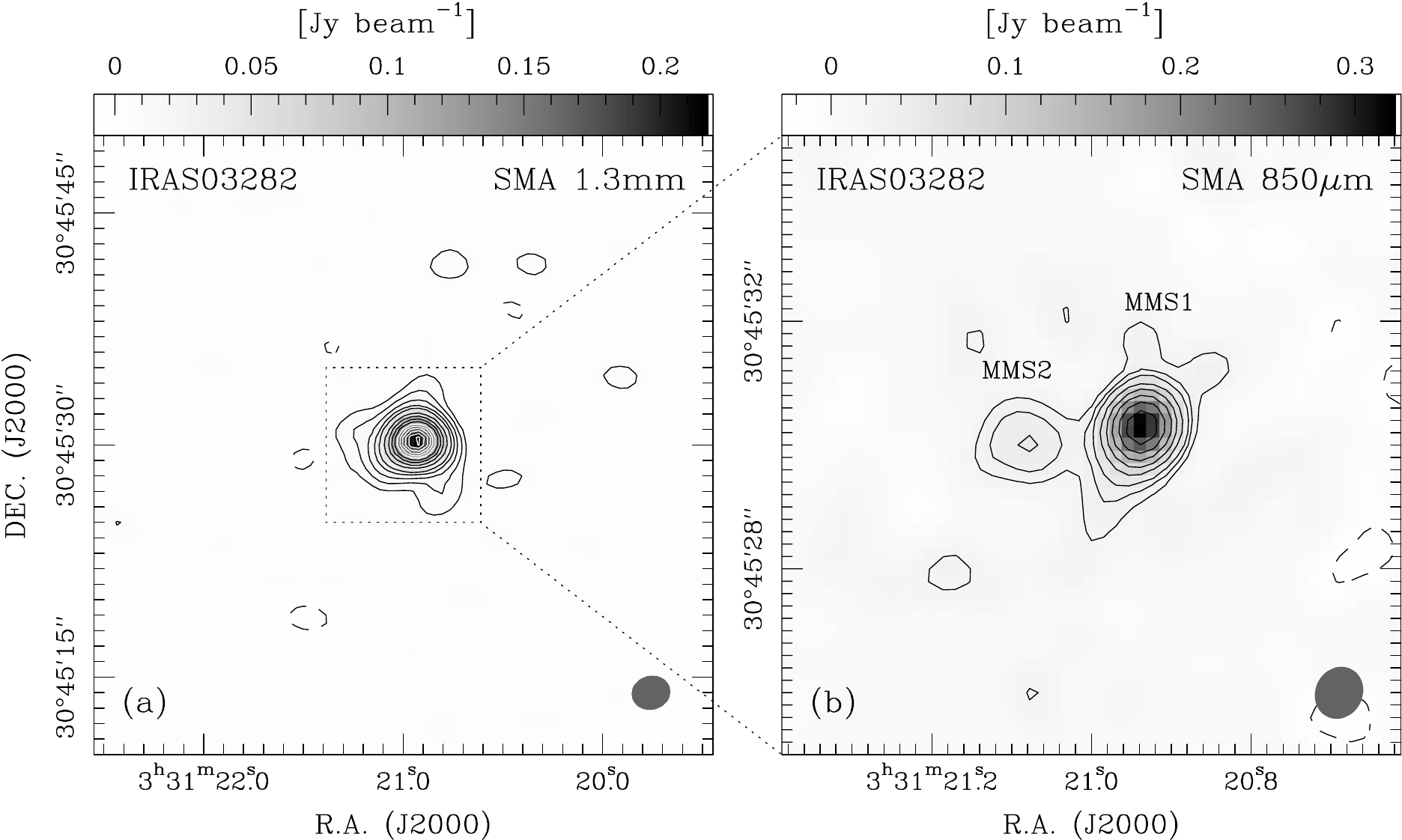}
\caption{(a) SMA 1.3\,mm dust continuum image of IRAS\,03282. The SMA contour levels correspond 
to $-$3, 3, 6, 10, 15, and 20\,$\sigma$, then increase in steps of 10\,$\sigma$ 
(1\,$\sigma$ $\sim$\,1.8\,mJy\,beam$^{-1}$). (b) SMA 850\,$\mu$m  dust continuum image of IRAS\,03282. 
Contour levels correspond to $-$3, 3, 5, 7, 11, 16, 22, and 30\,$\sigma$, then increase in steps of 
10\,$\sigma$ (1\,$\sigma$ $\sim$\,4.0\,mJy\,beam$^{-1}$). The synthesized SMA beam is shown as a grey 
oval in the bottom right corner of the two SMA dust continuum images.\label{sma_IRAS03282}}
\end{center}
\end{figure*}

\begin{figure*}
\begin{center}
\includegraphics[width=16cm,angle=0]{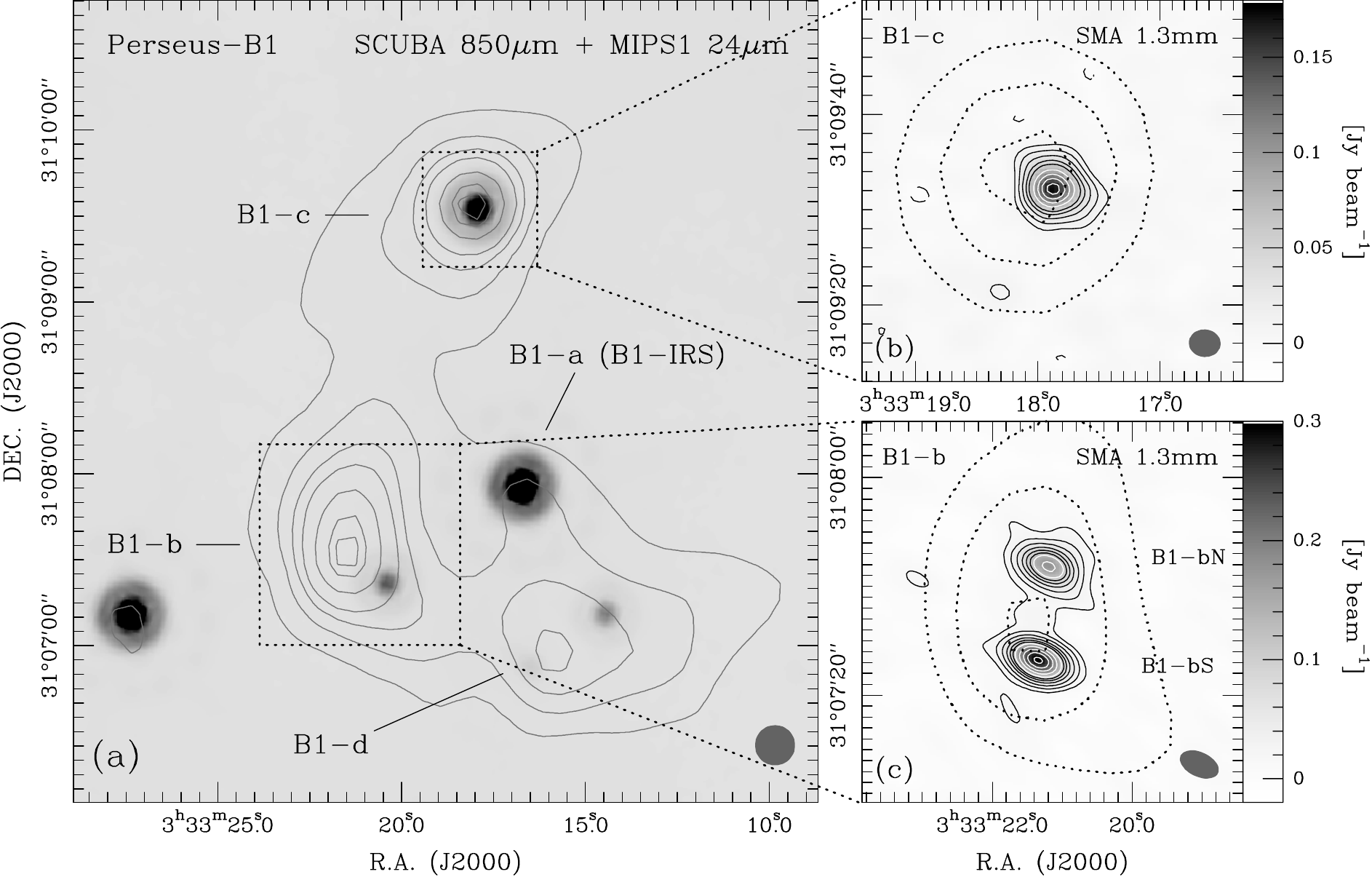}
\caption{(a) JCMT/SCUBA 850\,$\mu$m dust continuum image of the Per-B1 central region, plotted on 
the $Spitzer$ 24\,$\mu$m image. The SCUBA contour levels correspond to 10\%, 20\%, 30\%, 
40\%, 60\%, 80\%, and 90\% of the peak emission ($\sim$\,2.5\,Jy\,beam$^{-1}$). The SCUBA FWHM 
beam is shown as a grey circle in the bottom right corner of the image. (b) SMA 1.3\,mm dust continuum 
image of Per-B1c, overlaid with the SCUBA 850\,$\mu$m contours. The SMA contours (solid lines) correspond 
to $-$3, 3, 6, 10, 15, and 20\,$\sigma$, then increase in steps of 10\,$\sigma$ 
(1\,$\sigma$ $\sim$\,2.7\,mJy\,beam$^{-1}$). The SCUBA 850\,\micron\ contours (dotted lines) correspond 
to 50\%, 70\%, and 90\% of the peak emission. (c) SMA 1.3\,mm dust continuum image of Per-B1b, 
overlapped with the SCUBA 850\,$\mu$m contours. The SMA contours (solid lines) correspond to $-$3, 3, 
6, 10, 15, 20, 25, and 30\,$\sigma$, then increase in steps of 10\,$\sigma$ (1\,$\sigma$ $\sim$\,4.2\,mJy\,beam$^{-1}$). 
The SCUBA 850\,\micron\ contours (dotted lines) correspond to 30\%, 50\%, and 90\% of the peak emission. 
The synthesized SMA beam is shown as a grey oval in the bottom right corner of the two SMA dust continuum 
images.\label{B1_scuba_sma}}
\end{center}
\end{figure*}

\begin{figure*}
\begin{center}
\includegraphics[width=16cm,angle=0]{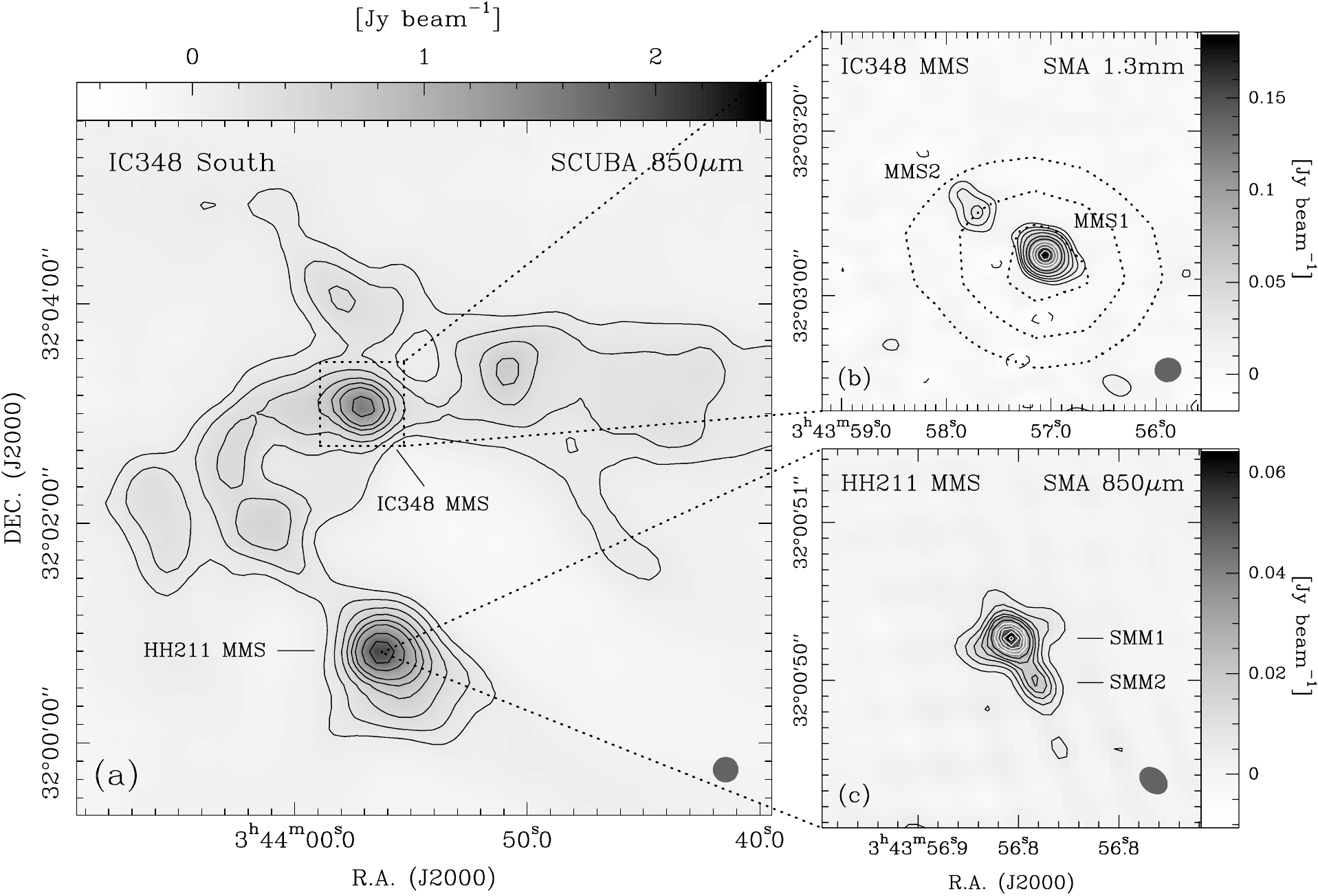}
\caption{(a) SCUBA 850\,$\mu$m dust continuum image of the IC\,348 southern region. The SCUBA contour 
levels correspond to 3, 6, 10, 15, 20 and 25\,$\sigma$, then increase in steps of 10\,$\sigma$ 
(1\,$\sigma$ $\sim$\,40\,mJy\,beam$^{-1}$). The SCUBA FWHM beam is shown as a grey circle in the bottom 
right corner of the image. (b) SMA 1.3\,mm dust continuum image of IC\,348\,MMS, overlapped with the SCUBA 
850\,$\mu$m contours. The SMA contours (solid lines) correspond to $-$3, 3, 5, 7, 10, 15, 20, and 25\,$\sigma$, 
then increase in steps of 6\,$\sigma$ (1\,$\sigma$ $\sim$\,4.0\,mJy\,beam$^{-1}$). The SCUBA 850\,\micron\ 
contours (dotted lines) correspond to 50\%, 70\%, and 90\% of the peak emission of the IC\,348\,MMS core 
($\sim$\,1.6\,Jy\,beam$^{-1}$). (c) SMA 850\,$\mu$m image of HH211\,MMS. The SMA contours correspond 
to $-$3, 3, 6, 9, 12, 15, 20, and 25\,$\sigma$, then increase in steps of 5\,$\sigma$ (1\,$\sigma$ $\sim$\,1.4\,mJy\,beam$^{-1}$). 
The synthesized SMA beam is shown as a grey oval in the bottom right corner of the two SMA dust continuum 
images.\label{HH211_IC348_sma}}
\end{center}
\end{figure*}

\begin{figure*}
\begin{center}
\includegraphics[width=14cm,angle=0]{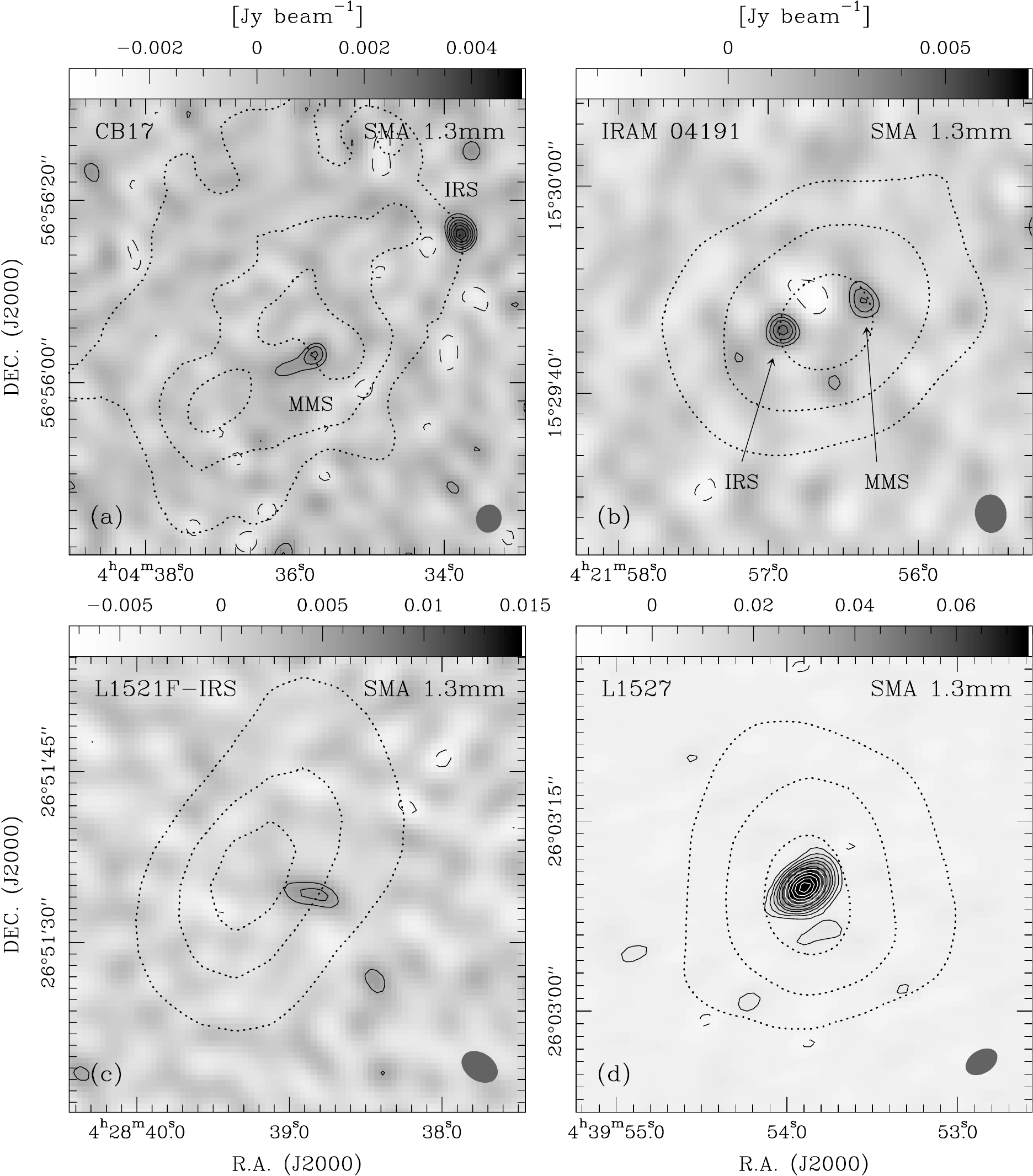}
\caption{\footnotesize (a) SMA 1.3\,mm dust continuum image of CB\,17, overlapped with the SCUBA 850\,$\mu$m 
contours. The SMA contours (solid lines) start at $-$3, 3\,$\sigma$, then increase in steps of 1\,$\sigma$ 
(1\,$\sigma$\,$\sim$\,0.5\,mJy\,beam$^{-1}$). The SCUBA contours (dotted lines) correspond to 50\%, 70\%, and 
90\% of the peak emission ($\sim$\,170\,mJy\,beam$^{-1}$). (b) The same as Figure~a, but for source IRAM\,04191. 
The SMA contours start at $-$3, 3\,$\sigma$, then increase in steps of 1\,$\sigma$ (1\,$\sigma$\,$\sim$\,7\,mJy\,beam$^{-1}$). 
The SCUBA 850\,$\mu$m contours correspond to 50\%, 70\%, and 90\% of the peak emission ($\sim$\,0.37\,Jy\,beam$^{-1}$). 
(c) The same as Figure~a, but for source L1521-IRS. The SMA contours correspond to $-$3, 3, and 5\,$\sigma$ 
(1\,$\sigma$\,$\sim$\,1.5\,mJy\,beam$^{-1}$). The SCUBA 850\,$\mu$m contours correspond to 70\%, 85\%, and 95\% 
of the peak emission ($\sim$\,0.44\,Jy\,beam$^{-1}$). (d) The same as Figure~a, but for source L1527. The SMA 
contours correspond to $-$3, 3, 6, 10, 15, 20, 30, and 40\,$\sigma$, then increase in steps of 15\,$\sigma$ 
(1\,$\sigma$ $\sim$\,1.5\,mJy\,beam$^{-1}$). The SCUBA 850\,$\mu$m contours correspond to 50\%, 70\%, and 90\% 
of the peak emission ($\sim$\,0.5\,Jy\,beam$^{-1}$). The synthesized SMA beam is shown as a grey oval in the bottom 
right corner of the each SMA image.\label{CB17_Taurus_sma}}
\end{center}
\end{figure*}

\begin{figure*}
\begin{center}
\includegraphics[width=14cm,angle=0]{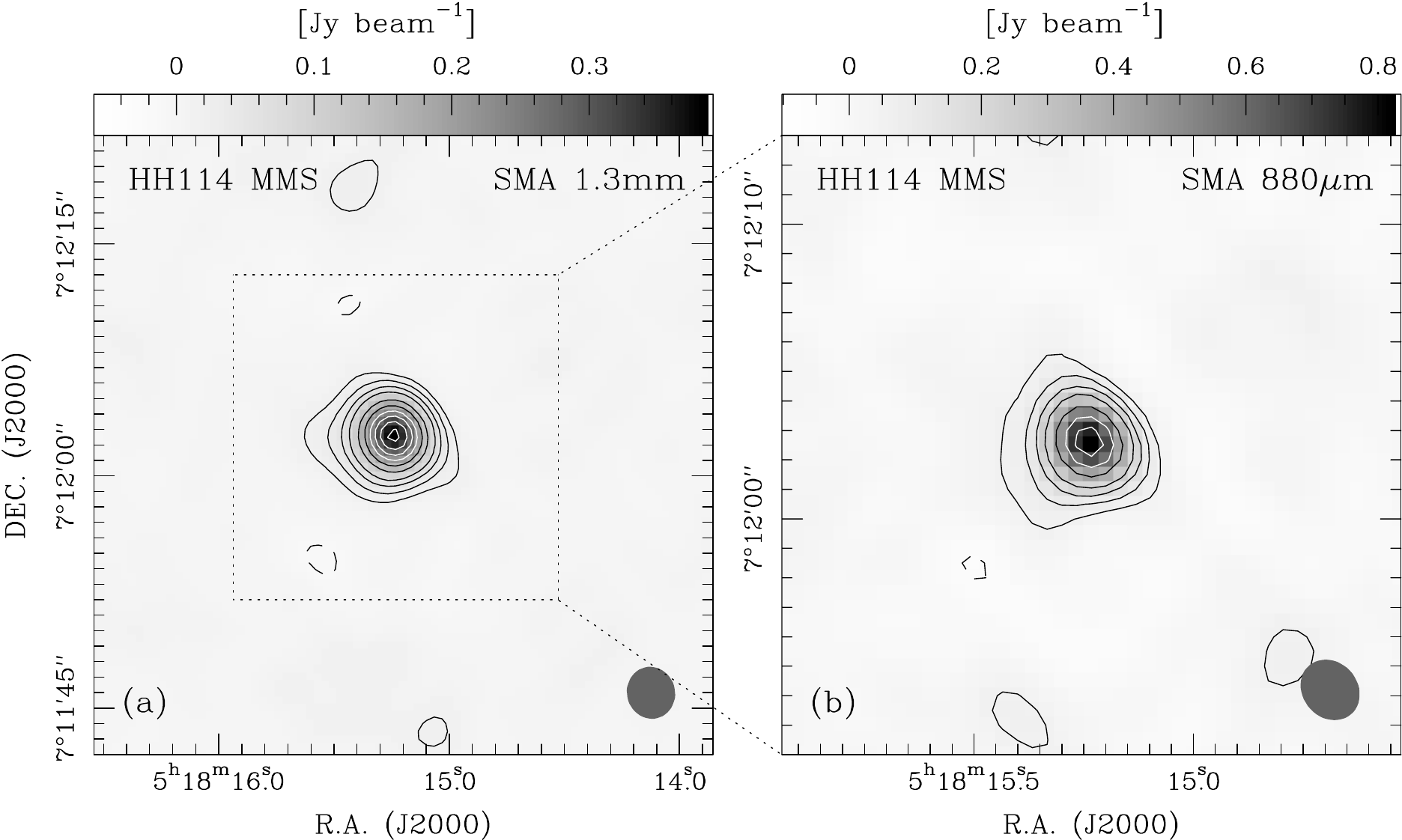}
\caption{(a) SMA 1.3\,mm dust continuum image of HH\,114\,MMS. The contour levels correspond to $-$3, 3, 6, 10, and 
15\,$\sigma$, then increase in steps of 10\,$\sigma$ (1\,$\sigma$ $\sim$\,5.0\,mJy\,beam$^{-1}$). (b) SMA 880\,$\mu$m 
dust continuum image of HH\,114 MMS. The contour levels correspond to $-$3, 3, 6, 10, and 15\,$\sigma$, then increase 
in steps of 10\,$\sigma$ (1\,$\sigma$ $\sim$\,16\,mJy\,beam$^{-1}$). The synthesized SMA beam is shown as a grey oval 
in the bottom right corner of the two SMA dust continuum images.\label{HH114_sma}}
\end{center}
\end{figure*}

\begin{figure*}
\begin{center}
\includegraphics[width=16cm,angle=0]{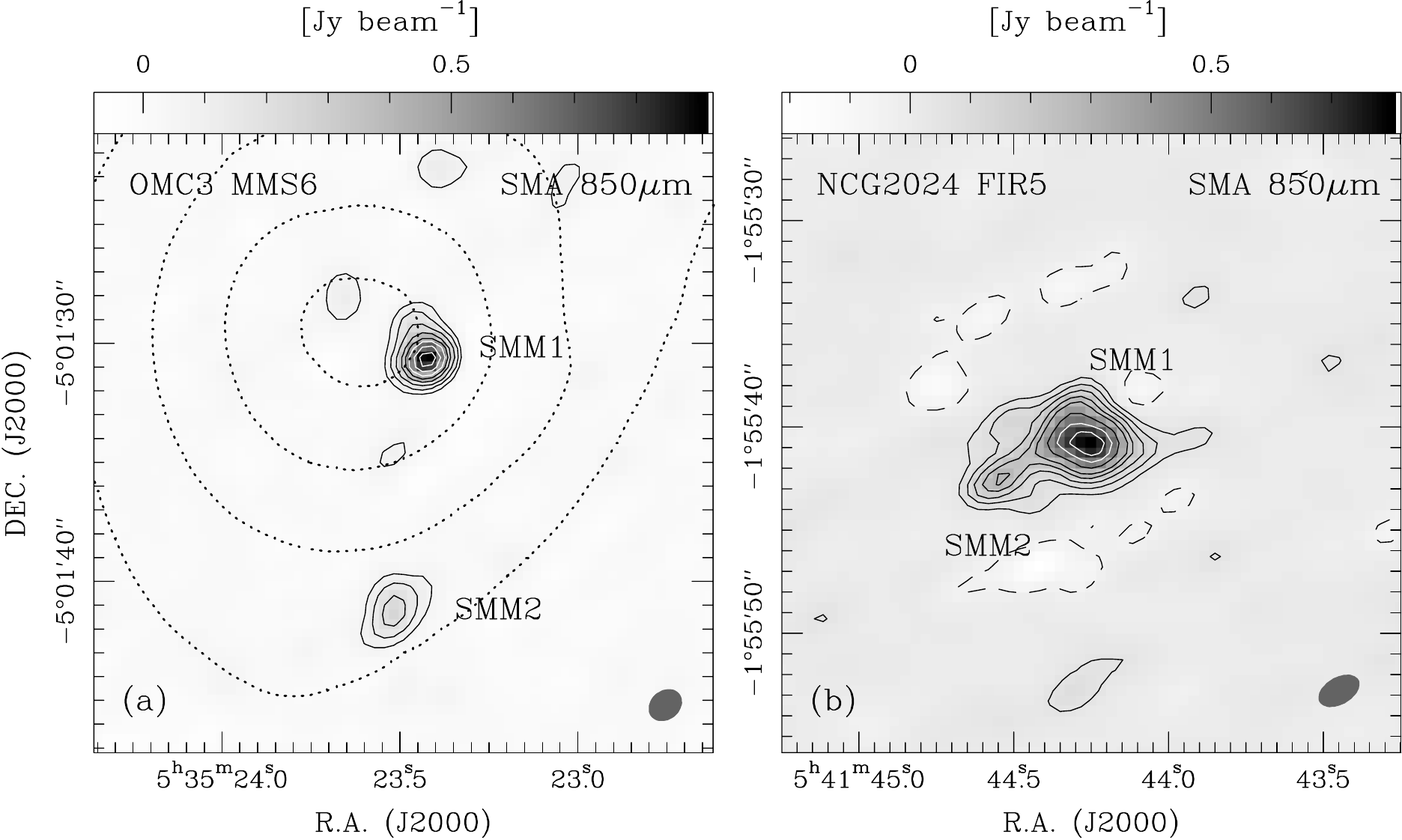}
\caption{(a) SMA 850\,$\mu$m dust continuum image of OMC3 MMS\,6, overlapped with the SCUBA 850\,$\mu$m contours. 
The SMA contours (solid lines) correspond to $-$3, 3, 6, 10, 15, 20, and 25\,$\sigma$, then increase in steps of 8\,$\sigma$ 
(1\,$\sigma$ $\sim$\,20\,mJy\,beam$^{-1}$). The SCUBA 850\,$\mu$m contours (dotted lines) correspond to 25\%, 50\%, 
75\%, and 95\% of the peak emission ($\sim$\,27\,Jy\,beam$^{-1}$). (b) SMA 850\,$\mu$m dust continuum image of 
NGC\,2024\,FIR5. The SMA contours correspond to $-$3, 3, 6, 10, 15, and 20\,$\sigma$, then increase in steps of 10\,$\sigma$ 
(1\,$\sigma$ $\sim$\,14\,mJy\,beam$^{-1}$). The synthesized SMA beam is shown as a grey oval in the bottom right corner of 
the two SMA dust continuum images.\label{OMC3_MMS6_sma}}
\end{center}
\end{figure*}

\begin{figure}
\begin{center}
\includegraphics[width=15cm,angle=0]{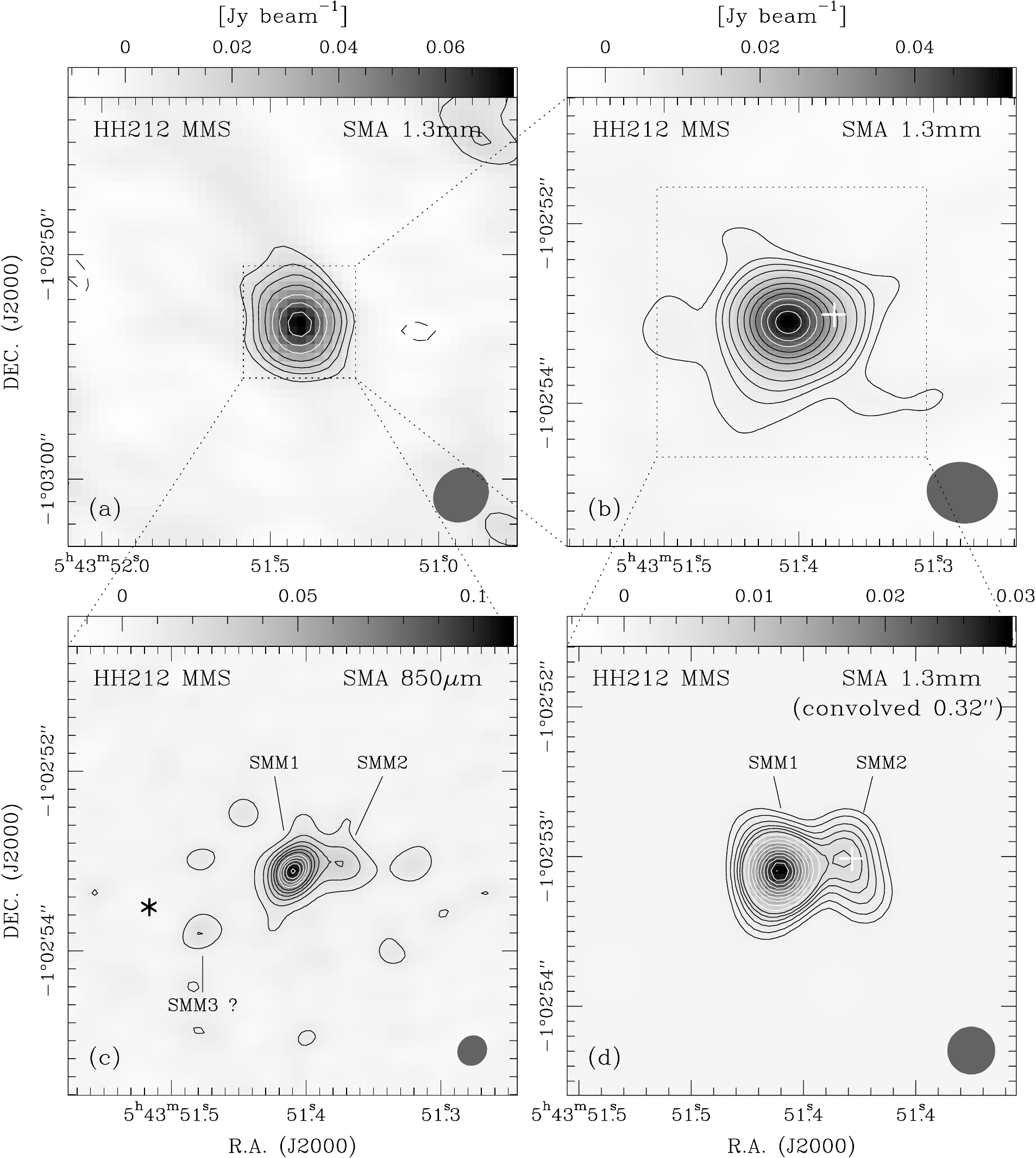}
\caption{\footnotesize (a) SMA 1.3\,mm dust continuum image of HH\,212 MMS (resolution 2\farcs6\,$\times$\,2\farcs3). 
The SMA contours correspond to $-$3, 3, 5, 7, and 10\,$\sigma$, then increase in steps of 4\,$\sigma$ (1\,$\sigma$ 
$\sim$\,3.0\,mJy\,beam$^{-1}$). (b) SMA 1.3\,mm dust continuum image (resolution 0\farcs8\,$\times$\,0\farcs7); the 
contours correspond to $-$3, 3, 5, 7, and 10\,$\sigma$, then increase in steps of 5\,$\sigma$ (1\,$\sigma$ 
$\sim$\,1.1\,mJy\,beam$^{-1}$). (c) The SMA 850\,$\mu$m dust continuum image (resolution 0\farcs35\,$\times$\,0\farcs32); 
the contours correspond to $-$3, 3, 7, and 11\,$\sigma$, then increase in steps of 4\,$\sigma$ (1\,$\sigma$ 
$\sim$\,2.8\,mJy\,beam$^{-1}$). (d) The restored SMA 1.3\,mm image with angular resolution of 0\farcs32\,$\times$\,0\farcs32;
the contours correspond to $-$3, 5, 8, 12, 16, 20, 25, and 30\,$\sigma$, then increase in steps of 10\,$\sigma$ (1\,$\sigma$ 
$\sim$\,0.25\,mJy\,beam$^{-1}$). The white crosses in Figures~b~\&~d show the position of the secondary continuum 
source derived from Figure~c, while the stellar symbol in Figure~c shows the position of the 1.4\,mm dust continuum 
source detected by Codella et al. (2007).\label{HH212_sma}}
\end{center}
\end{figure}

\begin{figure}
\begin{center}
\includegraphics[width=16cm,angle=0]{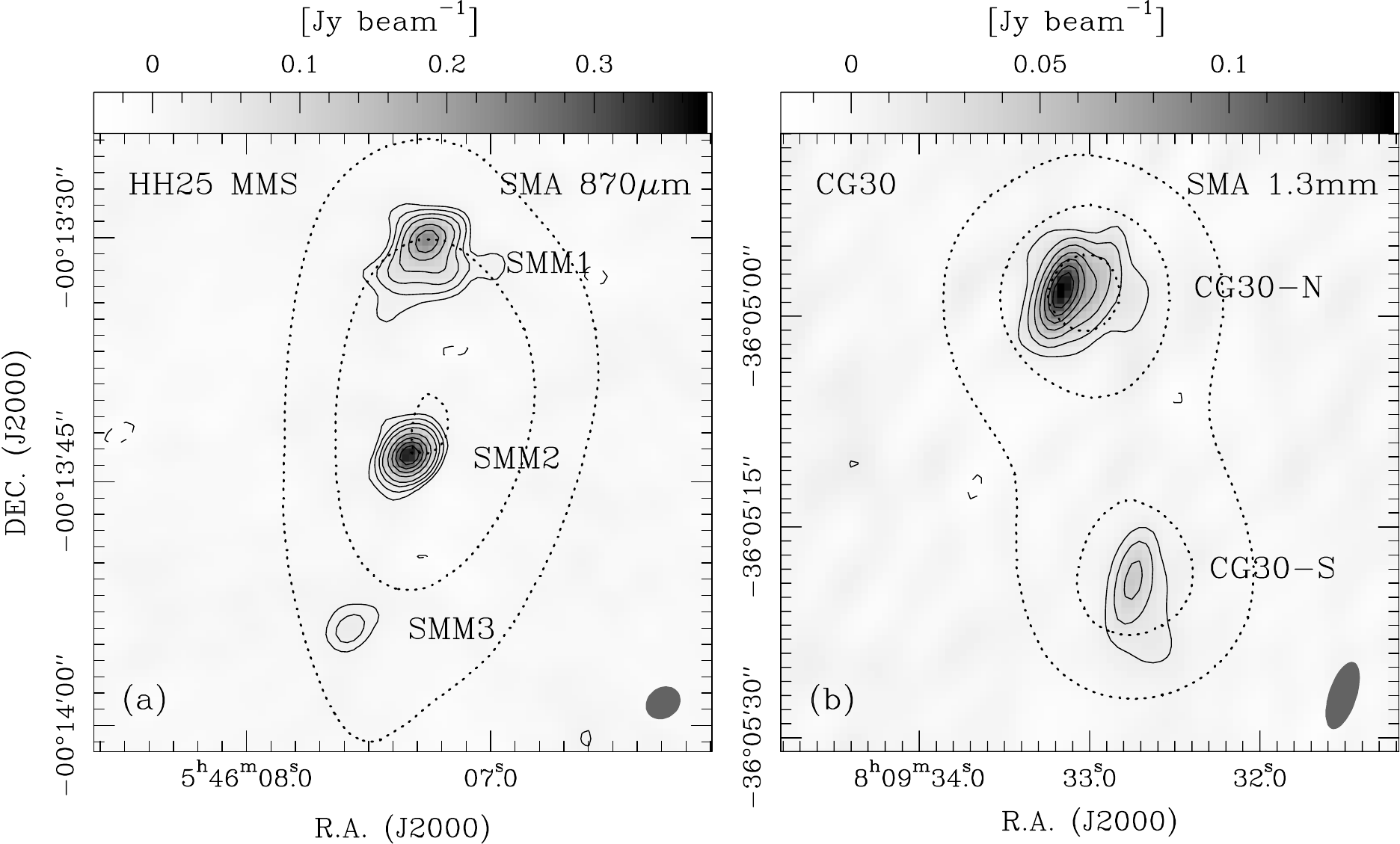}
\caption{(a) SMA 870\,$\mu$m dust continuum image of HH\,25 MMS, overlapped with the SCUBA 850\,$\mu$m contours.
The SMA contour levels (solid lines) correspond to $-$3, 3, 5, 8, and 12\,$\sigma$, then increase in steps of 8\,$\sigma$ 
(1\,$\sigma$ $\sim$\,6.8\,mJy\,beam$^{-1}$). The SCUBA contours (dotted lines) represent the 50\%, 75\%, and 99\% levels 
of the peak emission ($\sim$\,1.6\,Jy\,beam$^{-1}$). (b) SMA 1.3\,mm dust continuum image of CG\,30, overlapped with the 
SCUBA 850\,$\mu$m contours. The SMA contours (solid lines) correspond to $-$3, 3, 6, 10, 15, 20, 25, and 30\,$\sigma$, 
then increase in steps of 10\,$\sigma$ (1\,$\sigma$ $\sim$\,4.0\,mJy\,beam$^{-1}$). The SCUBA contours (dotted lines) 
represent the 50\%, 75\%, and 95\% levels of the peak emission ($\sim$\,1.2\,Jy\,beam$^{-1}$). The synthesized SMA beam 
is shown as a grey oval in the bottom right corner of the two SMA dust continuum images.\label{HH25_CG30}}
\end{center}
\end{figure}

\begin{figure}
\begin{center}
\includegraphics[width=16cm,angle=0]{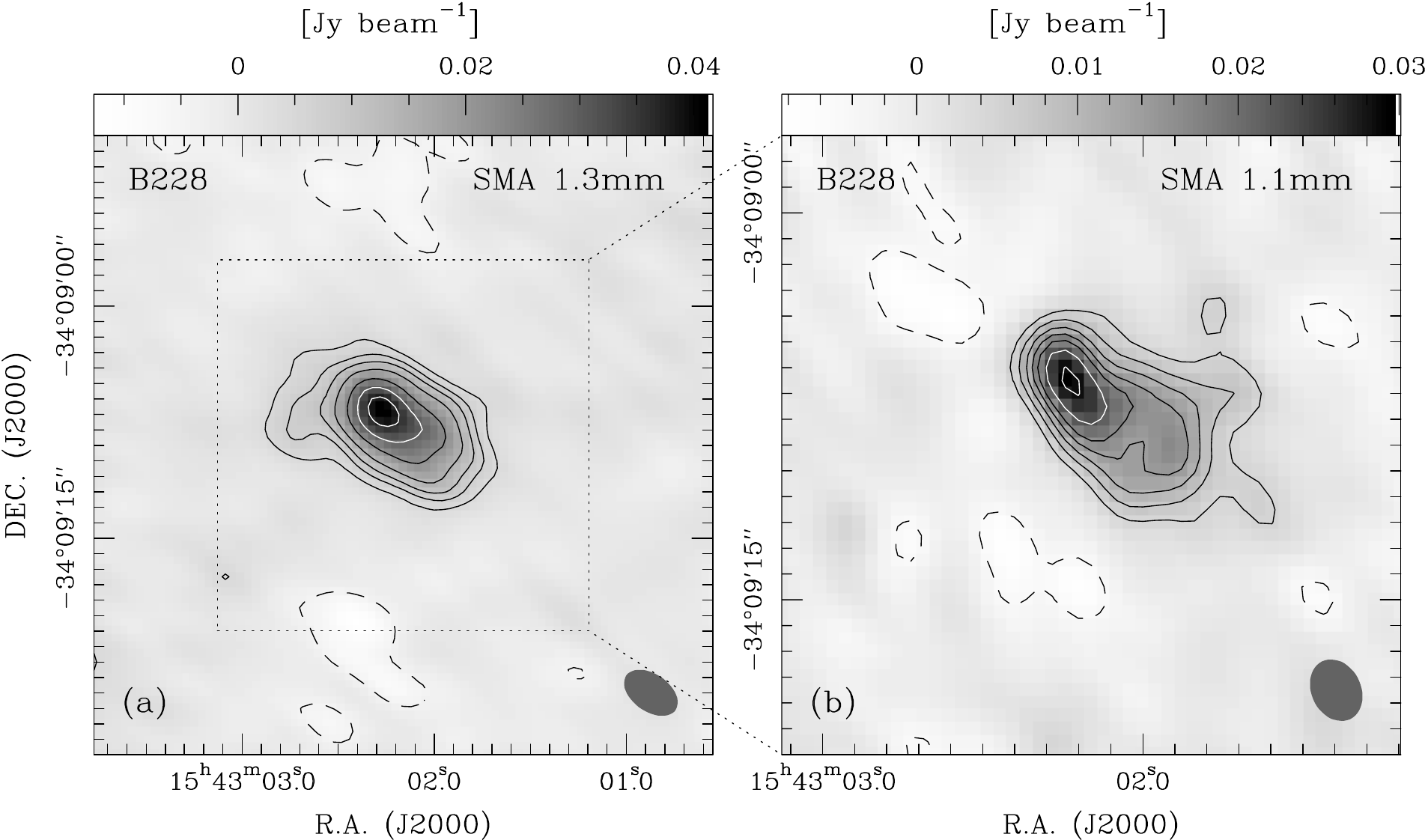}
\caption{(a) SMA 1.3\,mm dust continuum image of B\,228. The contour levels correspond to $-$3, 3, 5, 7, and 10\,$\sigma$, 
then increase in steps of 5\,$\sigma$ (1\,$\sigma$ $\sim$\,1.5\,mJy\,beam$^{-1}$). (b) SMA 1.1\,mm dust continuum 
image of B\,228. The contour levels correspond to $-$3, 3, 5, 7, 9, 11, 14, and 18\,$\sigma$ 
(1\,$\sigma$ $\sim$\,1.6\,mJy\,beam$^{-1}$). The synthesized SMA beam is shown as a grey oval in the bottom right 
corner of the two SMA images.\label{B228_SMA}}
\end{center}
\end{figure}

\begin{figure*}
\begin{center}
\includegraphics[width=15cm,angle=0]{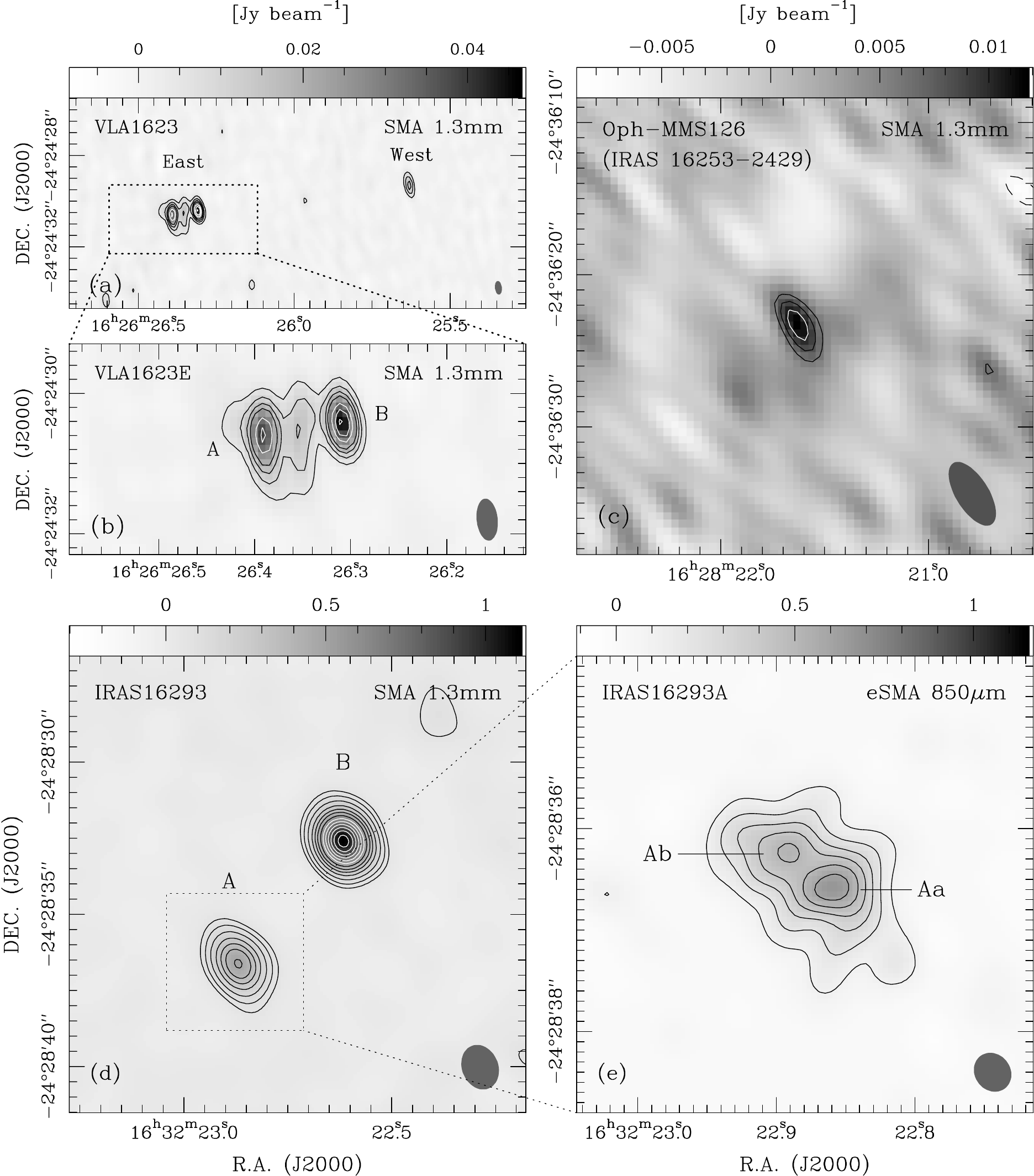}
\caption{(a) SMA 1.3\,mm dust continuum image of VLA\,1623. The contour levels correspond to 5, 10, 15, 20, and 30\,$\sigma$, 
then increase in steps of 8\,$\sigma$ (1\,$\sigma$ $\sim$\,1.0\,mJy\,beam$^{-1}$). (b) The enlarged view of the VLA\,1623\,East
pair. (c) SMA 1.3\,mm dust continuum image of Oph-MMS\,126. The contour levels correspond to $-$3, 3, 4, and 5\,$\sigma$ 
(1\,$\sigma$ $\sim$\,2.1\,mJy\,beam$^{-1}$). (d) SMA 1.3\,mm dust continuum image of IRAS\,16293. The contour levels 
correspond to $-$3, 3, 5, 8, 12, 16, and 20\,$\sigma$, then increase in steps of 5\,$\sigma$ (1\,$\sigma$ $\sim$\,19\,mJy\,beam$^{-1}$). 
(e) The eSMA 850\,$\mu$m continuum image of source IRAS\,16293\,A. The contour levels correspond to $-$3, 5, and 10\,$\sigma$, 
then increase in steps of 5\,$\sigma$ (1\,$\sigma$ $\sim$\,20\,mJy\,beam$^{-1}$). The synthesized SMA beam is shown as a grey 
oval in the bottom right corner of each SMA image.\label{Ophiucus_sma}}
\end{center}
\end{figure*}

\begin{figure*}
\begin{center}
\includegraphics[width=10cm,angle=0]{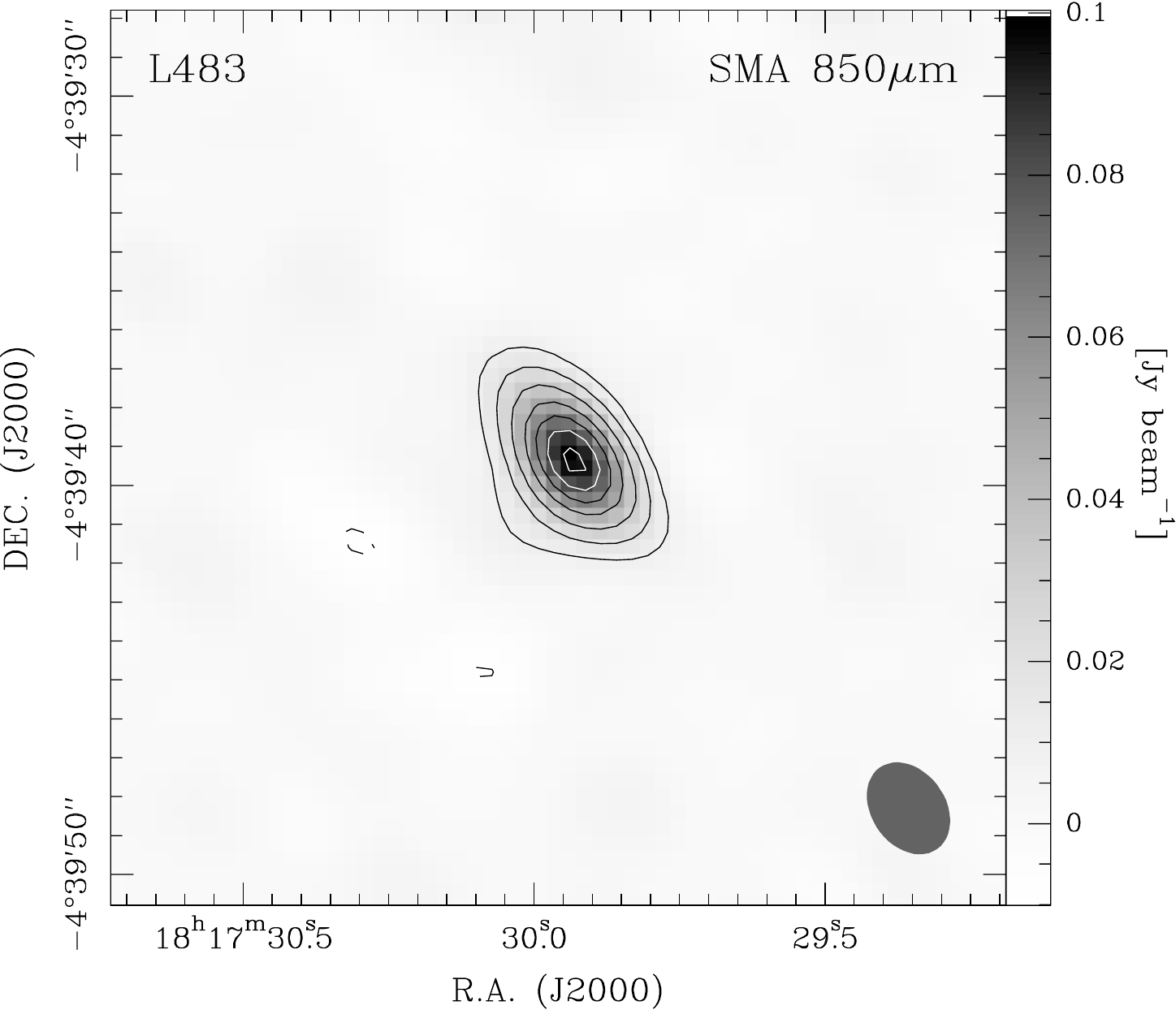}
\caption{SMA 850\,$\mu$m dust continuum image of L483. The contour levels correspond to $-$3, 3, 6, and 10\,$\sigma$, then 
increase in steps of 5\,$\sigma$ (1\,$\sigma$ $\sim$\,3.2\,mJy\,beam$^{-1}$). The synthesized SMA beam is shown as a grey 
oval in the bottom right corner.\label{L483_sma}}
\end{center}
\end{figure*}

\begin{figure}
\begin{center}
\includegraphics[width=15cm,angle=0]{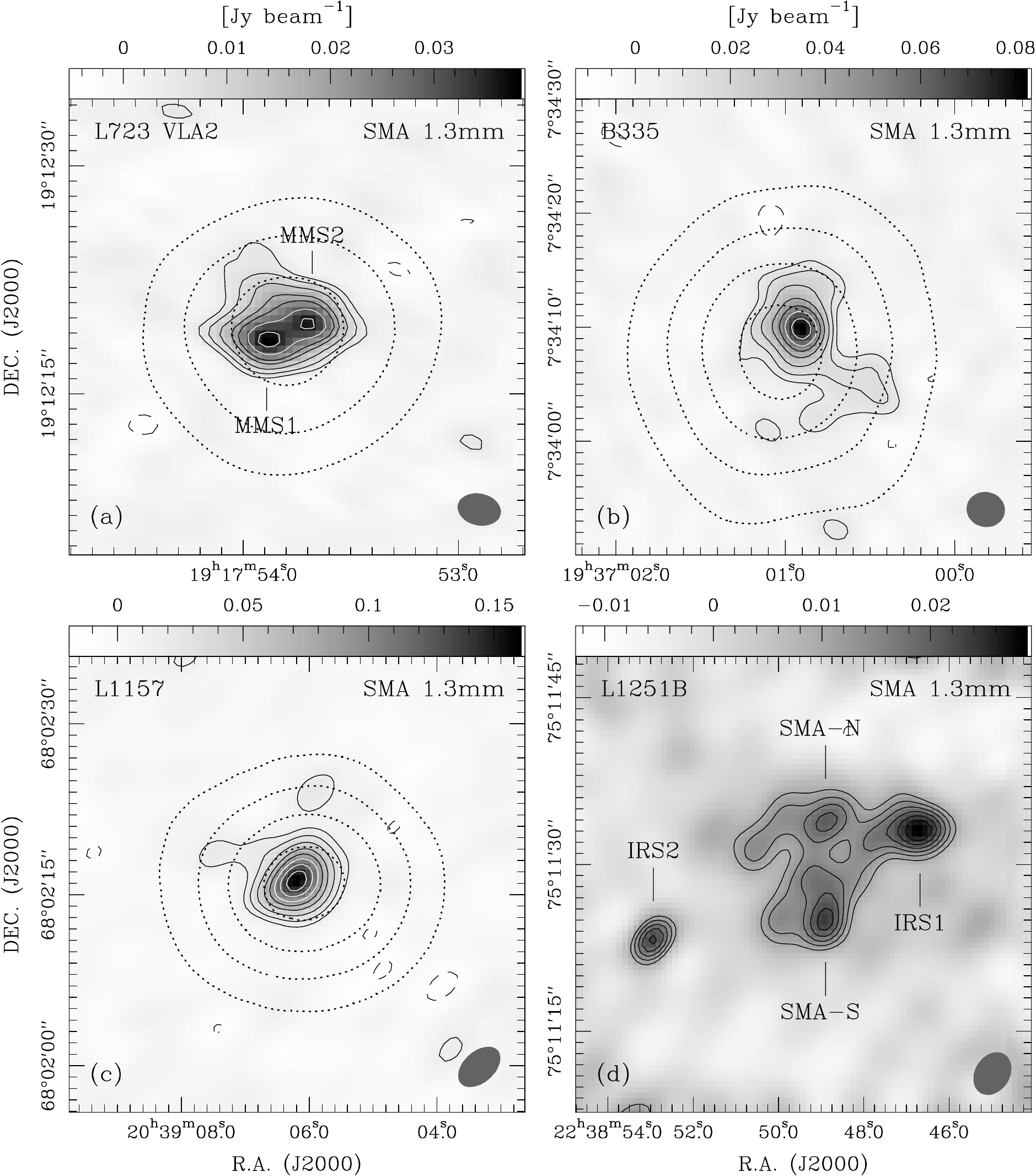}
\caption{\footnotesize (a) SMA 1.3\,mm dust continuum image of L723, overlapped with the SCUBA 850\,$\mu$m contours. 
The SMA contours (solid lines) correspond to $-$3, 3, 5, 8, 12, and 15\,$\sigma$ levels and then increase in steps of 
5\,$\sigma$ (1\,$\sigma$\,$\sim$\,1.4\,mJy\,beam$^{-1}$). The SCUBA 850\,\micron\ contours (dotted lines) represent 
30\%, 50\%, and 90\% of the peak emission ($\sim$\,0.9\,Jy\,beam$^{-1}$). (b) Same as Figure~a, but for source B335. 
The SMA contours correspond to $-$3, 3, 5, 8, 12, 16, and 20\,$\sigma$ levels and then increase in steps of 5\,$\sigma$ 
(1\,$\sigma$\,$\sim$\,2.9\,mJy\,beam$^{-1}$). The SCUBA 850\,\micron\ contours represent 30\%, 50\%, 70\%, and 90\% 
of the peak emission ($\sim$\,1.2\,Jy\,beam$^{-1}$). (c) Same as Figure~b, but for source L1157, where 1\,$\sigma$ noise 
in the SMA image is $\sim$\,4.8\,mJy\,beam$^{-1}$ and the peak value in the SCUBA 850\,\micron\ image is 
$\sim$\,1.4\,Jy\,beam$^{-1}$. (d) SMA 1.3\,mm dust continuum image of L1251B. The SMA contours start at $-$3, 3\,$\sigma$ 
and then increase in steps of 1\,$\sigma$ (1\,$\sigma$\,$\sim$\,3.2\,mJy\,beam$^{-1}$).\label{L723_B335_L1157_L1251B}}
\end{center}
\end{figure}

\newpage

\begin{figure}
\begin{center}
\includegraphics[width=16cm,angle=0]{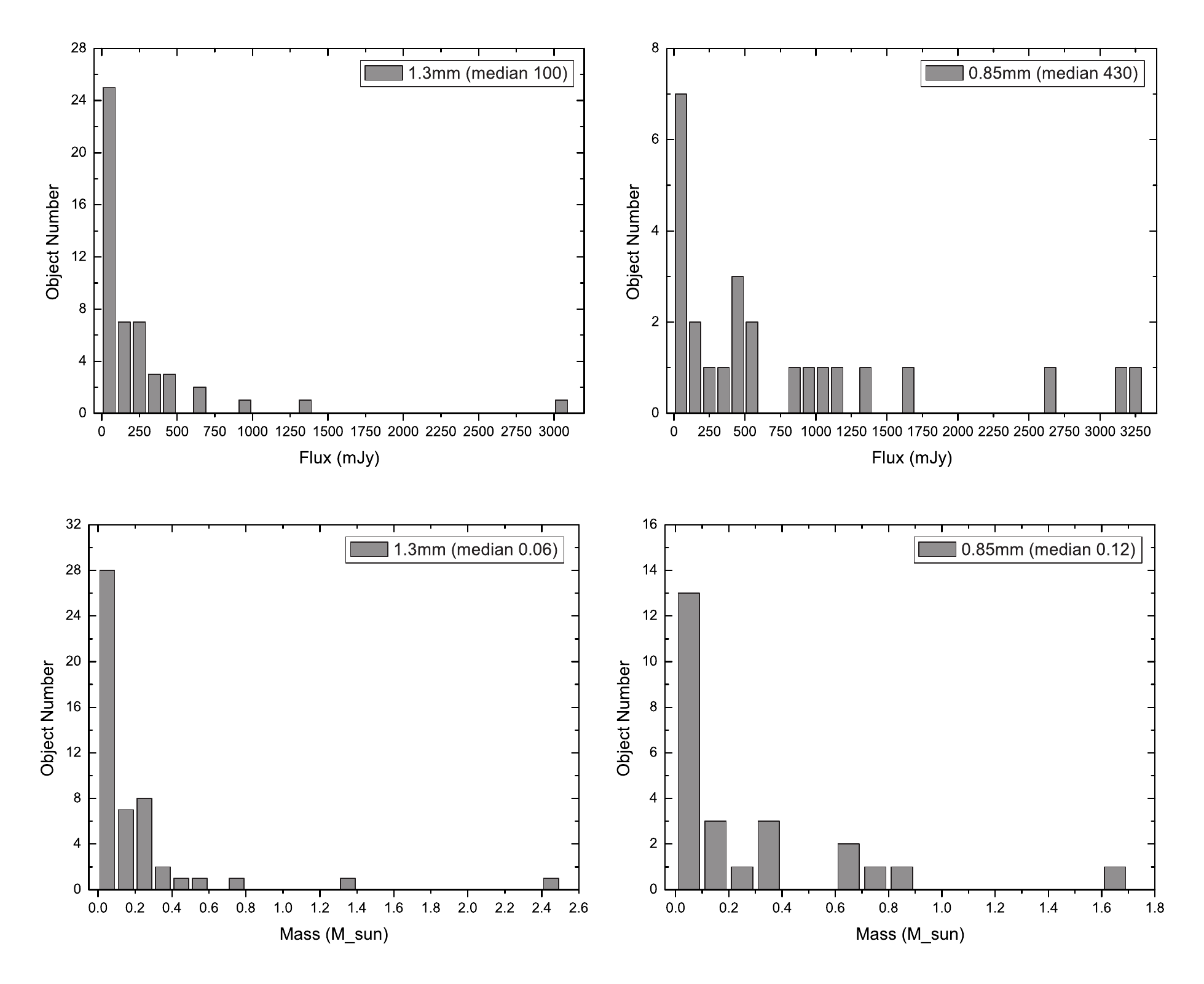}
\caption{{\it Top Left:} Distribution of the 1.3\,mm fluxes derived in this SMA survey. The median value
is 100\,mJy. {\it Top Right:} Distribution of the 850\,$\mu$m fluxes (median value 430\,mJy).
{\it Bottom Left:} Distribution of the circumstellar gas masses derived from the 1.3\,mm dust continuum 
observations (median value 0.06\,$M_\odot$). {\it Bottom Right:} Distribution of the circumstellar gas masses 
derived from the 850\,$\mu$m dust continuum observations (median value 0.12\,$M_\odot$).\label{flux_mass}}
\end{center}
\end{figure}

\begin{figure}
\begin{center}
\includegraphics[width=17cm,angle=0]{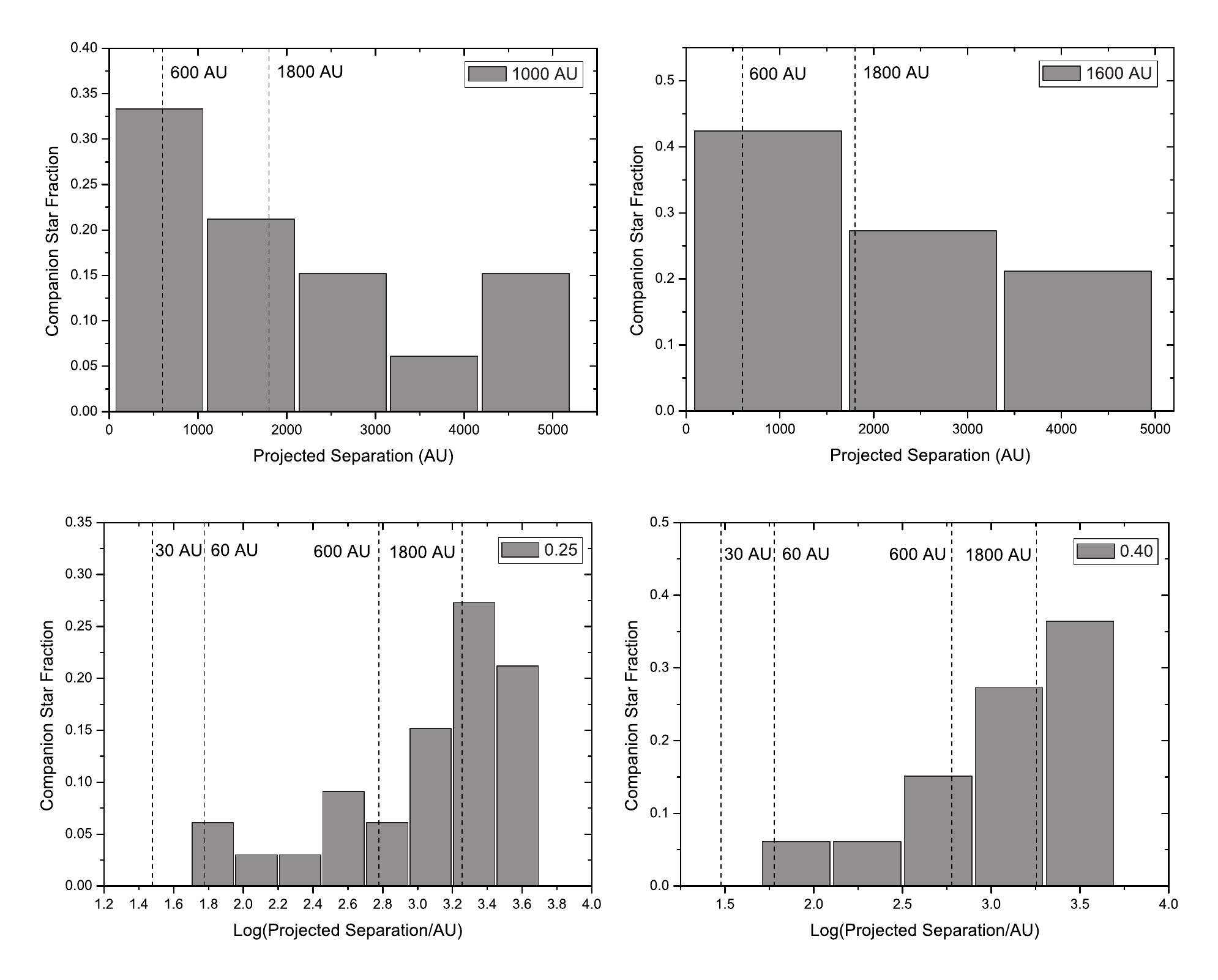}
\caption{Distribution of separations for protobinary systems in this SMA survey (see Table~6), binned in intervals 
of linear values of 1000\,AU ($top~left$) and 1600\,AU ($top~right$), and binned in intervals of log values of 0.25
($bottom~left$) and 0.40 ($bottom~right$). The two dashed lines shown in the top figures mark the median linear 
resolution (600\,AU) and median (projected) separation (1800\,AU) in this survey, while the four dashed lines shown 
in the bottom figures mark the peak values in the distributions of separations for main sequence stars (30\,AU; see
Duquennoy \& Mayor 1991) and pre-main sequence stars (60\,AU; see Patience et al. 2002), and the median linear 
resolution and separation in this survey. We note that the `separations' shown here are observed projected separations.
\label{separation_v1}}
\end{center}
\end{figure}

\begin{figure}
\begin{center}
\includegraphics[width=17cm,angle=0]{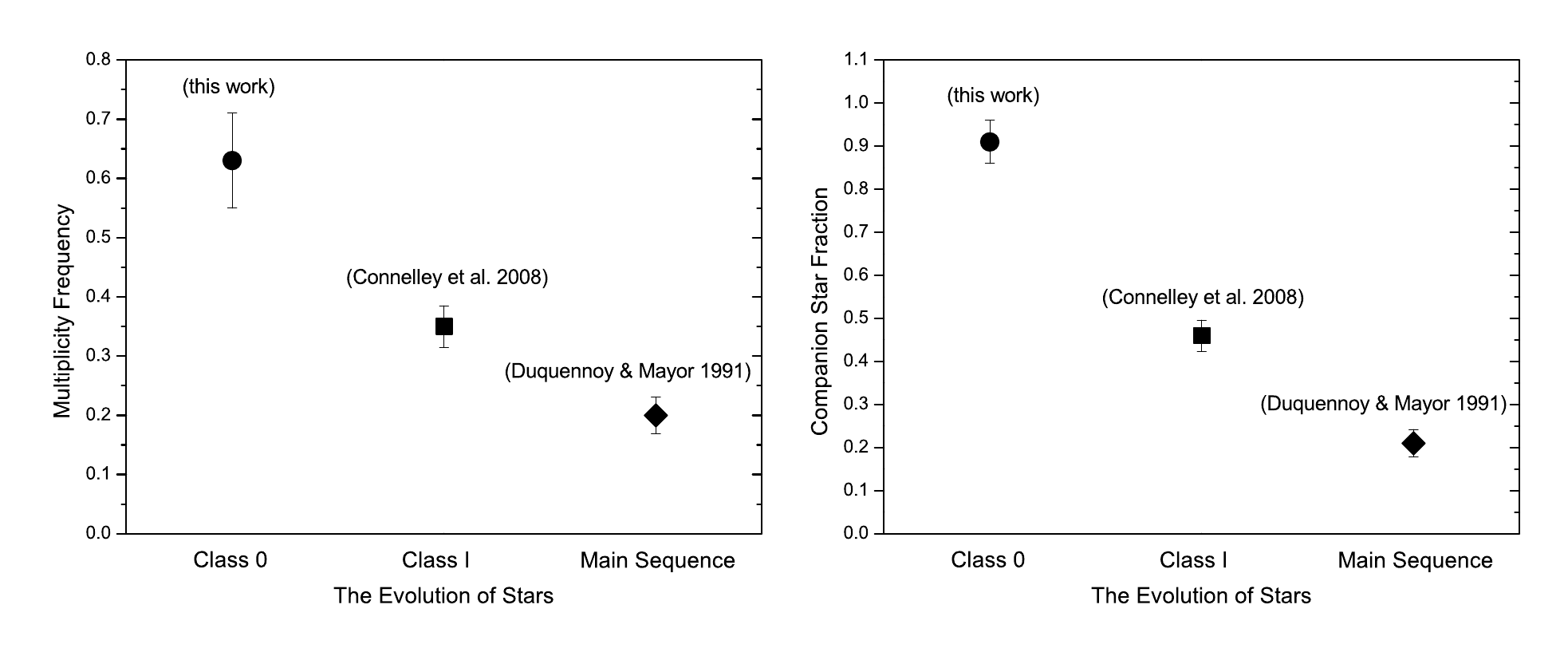}
\caption{Observed multiplicity frequencies ($left$) and companion star fractions ($right$) with separations from 50\,AU 
to 5000\,AU for Class\,0 protostars (this work), Class\,I young stellar objects (data from Connelley et al. 2008a, b), and 
main sequence stars (data from Dquennoy \& Mayor 1991). It must be noted that the SMA survey for Class\,0 protobinary 
systems is incomplete across the observed separations range, and the derived values should be considered as lower 
limits.\label{evolution_v1}}
\end{center}
\end{figure}

\begin{figure}
\begin{center}
\includegraphics[width=17cm,angle=0]{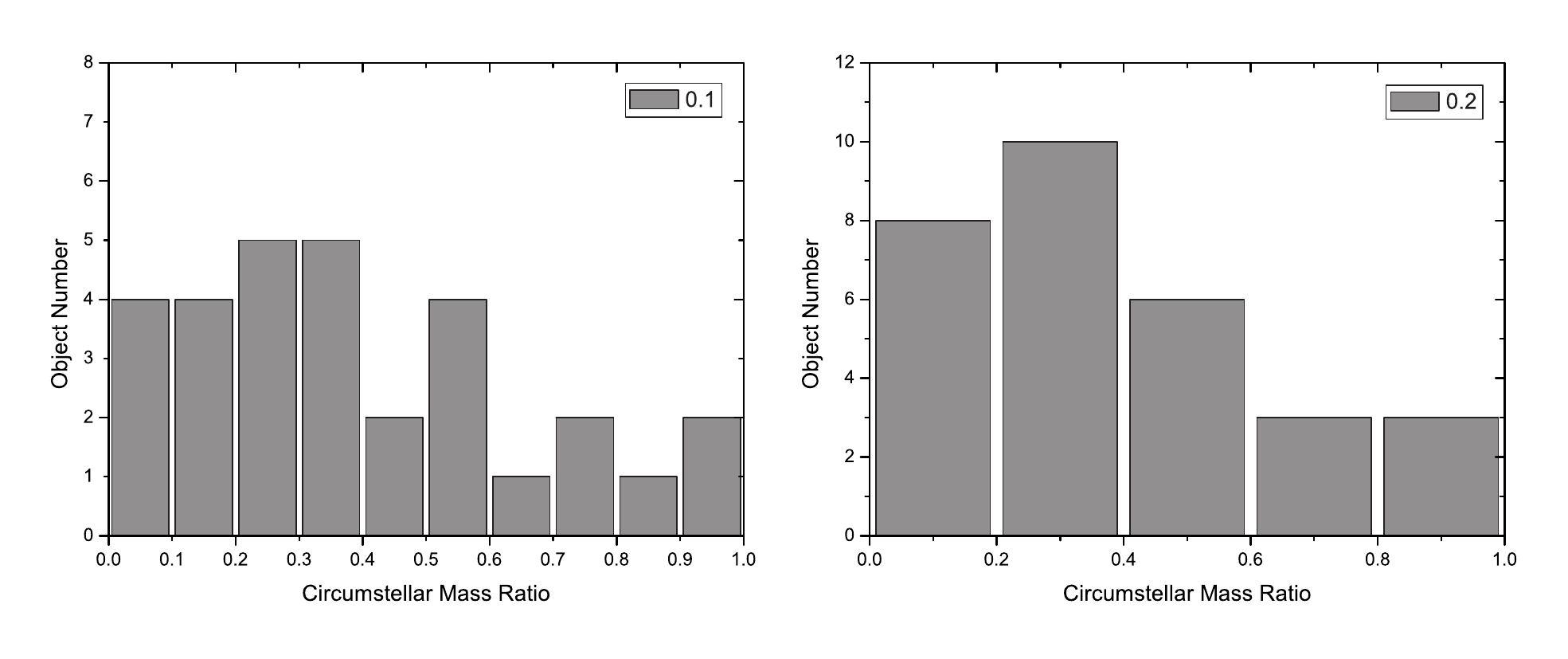}
\caption{Histograms of circumstellar mass ratios for Class\,0 protobinary systems in this work (see Table~6), 
binned in intervals of 0.1 ($left$) and 0.2 ($right$).\label{massratio_v1}}
\end{center}
\end{figure}

\begin{figure}
\begin{center}
\includegraphics[width=17cm,angle=0]{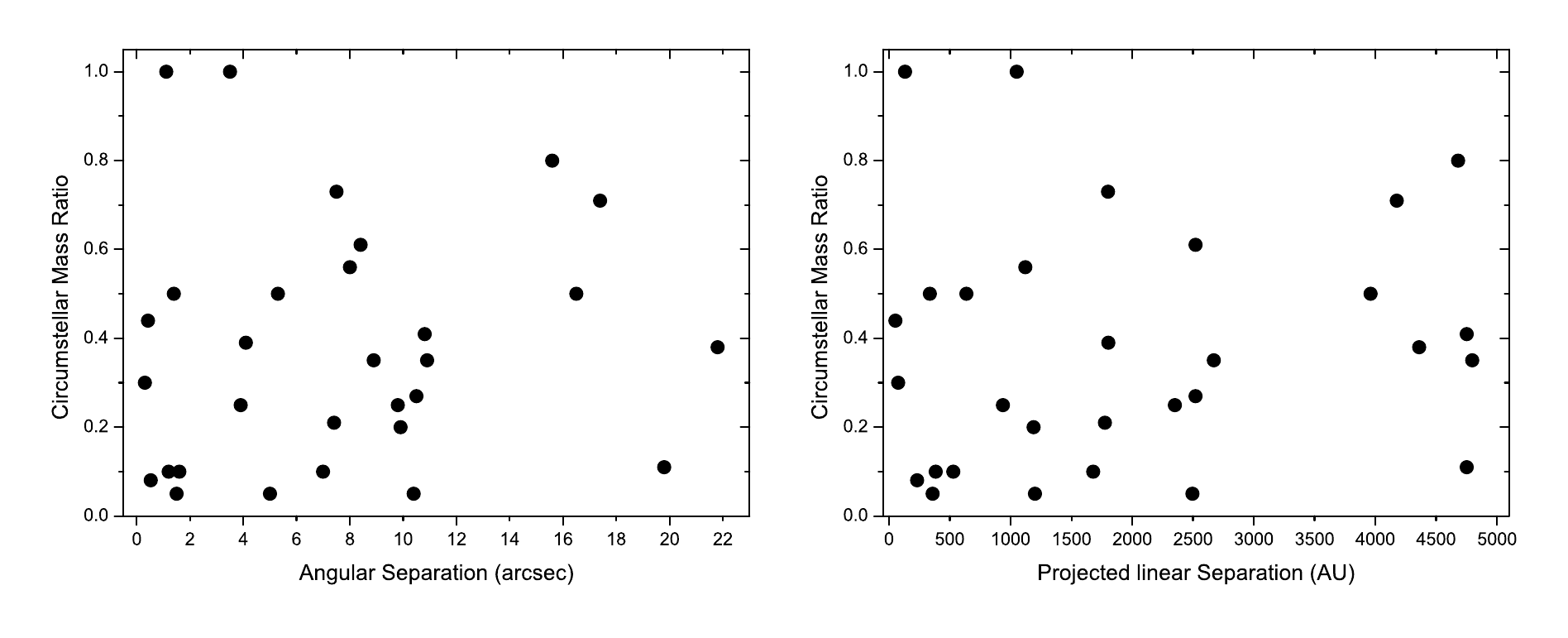}
\caption{The circumstellar mass ratios plotted as a function of angular separations ($left$) and projected linear separations
($right$) for the Class\,0 protobinary systems in this work (see Table~6). The diagrams show that there is no obvious relation 
between mass ratios and separations for the protobinary systems in this sample.\label{separation_massratio}}
\end{center}
\end{figure}

\begin{figure}
\begin{center}
\includegraphics[width=16cm,angle=0]{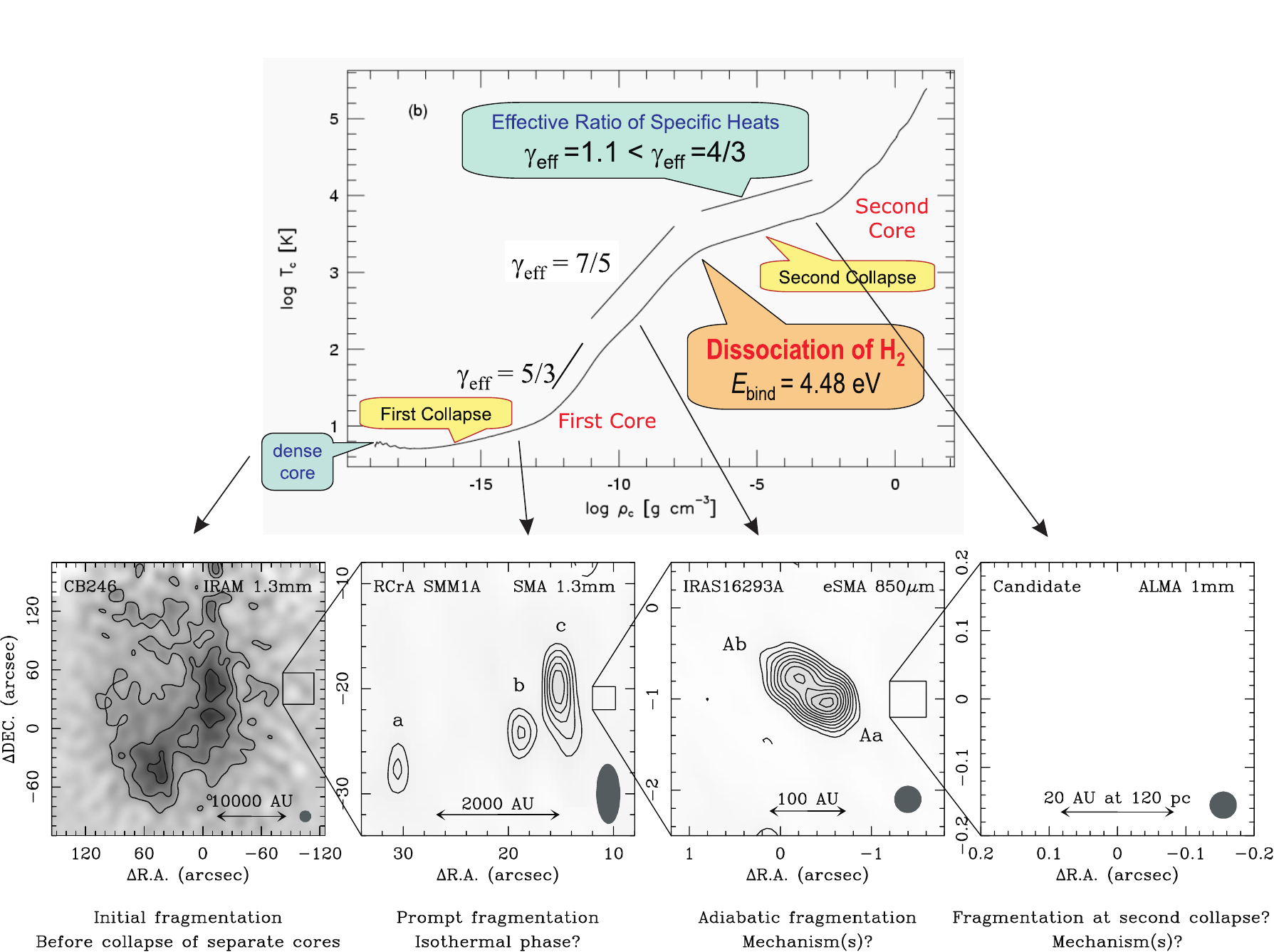}
\caption{{\it Top:} Temperature and density evolution at the center of a gravitationally collapsing cloud 
core obtained by Masunaga \& Inutsuka (2000) in their radiation hydrodynamical calculation of spherically symmetric
protostellar collapse. Each of these phases in the temperature evolution is characterized by a distinct value of the 
effective ratio of specific heats, $\gamma_{\rm eff}$. The image is adopted from Andr{\'e} et al. (2009). {\it Bottom:} A 
suggested picture of the sequential fragmentation for binary stars. From left to right: the IRAM-30m 1.3\,mm dust 
continuum image of Bok globule CB\,246 (data from Launhardt et al. 2010), the SMA 1.3\,mm dust continuum image 
of R\,CrA SMM\,1A (data from Chen \& Arce 2010), and the eSMA 850\,$\mu$m image of IRAS\,16293\,A (this work),
while the right blank panel shows a possible view with an angular resolution of 0\farcs04 attained at the ALMA.
\label{evolution_v2}}
\end{center}
\end{figure}

\end{document}